\documentclass[fleqn,10pt,twoside]{report}
\usepackage{graphicx}
\usepackage[english,dutch]{babel}
\usepackage[ps2pdf,bookmarks,breaklinks,pdfborder={0 0 0},pdfpagescrop={92 75 546 750}]{hyperref}
\usepackage{placeins} 
\usepackage[usenames]{color}
\usepackage{booktabs}
\usepackage{textcomp} 
\usepackage{pifont}   
\usepackage{textpos}
\usepackage{makeidx}
\usepackage{amsfonts}
\usepackage{amssymb}
\usepackage{wrapfig}

\makeindex
\definecolor{gray}{rgb}{0.600,0.600,0.600}

\def\dbar{{\mathchar'26\mkern-12mu {\rm d}}}

\usepackage{microtype}
\usepackage{lmodern}
\usepackage{sectsty}
\allsectionsfont{\sffamily}
\DeclareMathAlphabet{\mathpzc}{OT1}{pzc}{m}{it}

\usepackage[titles]{tocloft}

\def\startchapter{
  \ifodd\count0
    \clearpage\thispagestyle{myheadings}\vspace*{\fill}\clearpage
  \else
    \vspace*{\fill}\clearpage\fi}

\usepackage{fancyhdr}
\fancypagestyle{plain}{
\fancyhf{}
\fancyfoot[C]{\scriptsize\sf\thepage}

}

\makeatletter
\def\cleardoublepage{\clearpage\if@twoside \ifodd\c@page\else%
\hbox{}%
\thispagestyle{empty}%
\newpage%
\if@twocolumn\hbox{}\newpage\fi\fi\fi}
\makeatother

\def\clap#1{\hbox to 0pt{\hss#1\hss}}

\def\mathrlap{\mathpalette\mathrlapinternal}

\def\mathrlapinternal#1#2{\rlap{$\mathsurround=0pt#1{#2}$}}

\def\XXint#1#2#3{{\setbox0=\hbox{$#1{#2#3}{\int}$}
\vcenter{\hbox{$#2#3$}}\kern-.5\wd0}}

\def\dtilde#1{\mathrlap{\rlap{\raise1.2pt\hbox{$\mathsurround=0pt%
\tilde{\phantom{\hbox{$#1$}}}$}}\tilde{\phantom{\hbox{$#1$}}}}#1}

\begin{document}
\thispagestyle{empty} 

\newpage
\thispagestyle{empty} \mbox{}
\begin{textblock}{120}(0,17.5)
\textblockcolour{}
\vspace{-\parskip}

 \Large\sffamily\textbf{STRUCTURES OF NONEQUILIBRIUM\\ FLUCTUATIONS}

\end{textblock}
\begin{textblock}{120}(0,32.5)
\textblockcolour{}
\vspace{-\parskip}
\large\sffamily{DISSIPATION AND ACTIVITY}
\end{textblock}
\begin{textblock}{120}(0,52.5)
\textblockcolour{}
\vspace{-\parskip}
\large\sffamily\textbf{Bram WYNANTS}
\end{textblock}
\begin{textblock}{120}(-2,105)
\textblockcolour{}
\vspace{-\parskip}
\begin{tabular}{p{8cm}l}

Supervisor: 							&\\
Prof. Dr. C. Maes, K.U.Leuven 						&\\
								&  \\
Board of examiners:                                           	&     \\
 Prof.\ Dr. D. Boll\'e, K.U.Leuven, president           			& Dissertation presented in  \\
 Prof.\ Dr. J. Indekeu, K.U.Leuven, secretary          			& partial fulfilment of the \\
 Prof.\ Dr. W. Troost, K.U.Leuven                   		        & requirements for the \\
 Prof.\ Dr. M. Fannes, K.U.Leuven                         			& degree of Doctor of \\
Prof.\ Dr. H. Touchette,  					& Science \\
 \ \ \ \ \ \ \ Queen Mary, University of London  & \\
Prof.\ Dr. T. Bodineau, 						&  \\
 \ \ \ \ \ \ \ Ecole Normale Sup\'erieure, Paris & \\

\end{tabular}
\end{textblock}
\vfill
\begin{center} {May 2010} \end{center}
\pagenumbering{roman}

\selectlanguage{english}
\pagenumbering{roman}

\chapter*{Abstract}

We discuss research done in two important areas of nonequilibrium statistical
mechanics: fluctuation dissipation relations 
and dynamical fluctuations. The work discussed here was reported before in 
\cite{bbmw09,bmw09a,bmw09b,mnw08b,mnw08a}.\\
In equilibrium systems the fluctuation-dissipation theorem
gives a simple relation between the response of observables to a perturation
and correlation functions in the unperturbed system. 
Our contribution here is an investigation of the form of the response function for systems out of equilibrium.
We found that the response function can generally be written as the sum of two
correlation functions. One correlation function is linked to entropy exchange
with the environment, and thus to heat dissipation. The other correlation
function has to do with a quantity which we call traffic and which describes in a sense
the activity of the system. The
results are applied to several explicit examples for which simulations
have provided some visualization.\\
Furthermore, we use the theory
of large deviations to examine dynamical fluctuations in systems
out of equilibrium.  In dynamical fluctuation theory we consider two kinds of observables:
 occupations (describing
the fraction of time the system spends in each configuration)
and currents (describing the changes of configuration
the system makes). We explain how
to compute the rate functions of the large deviations, and what the physical
quantities are that govern their form. As for fluctuation-dissipation relations,
entropy and traffic are the main ingredients. Moreover,
the rate function that governs the joint probabilities of occupations and currents is explicitly computed for
the classes of models considered and is expressed in terms
of entropy and traffic.
The rate function for the occupations can be expressed entirely in terms
of traffic. We also show that this traffic can be seen as a thermodynamic
potential for currents. Finally, for the 
close-to-equilibrium regime, known variational principles as the 
minimum entropy production principle are recovered.

\cleardoublepage
\chapter*{Nomenclature}

\begin{tabular}{ll}
 $ k_B$    	& Boltzmann's constant, usually we work in units in which $k_B= 1$,\\
 $ \beta = \frac{1}{k_BT} $		& inverse temperature,\\
 $ U,V$					& potentials,\\
 $ F,f,g $				& forces,\\
 $ W, Q $ 					& work and heat,\\
 $ \mu $ 				& chemical potential.\\
\end{tabular}

\paragraph{Configurations and trajectories\\}
\begin{tabular}{ll}
 $ \Omega $					& configuration/state space,\\
 $x,y,\ldots$ 					& configurations/states,\\
 $t,s,T$ 					& times,\\
 $x_t$						& configuration at time $t$,\\
 $\lambda_t$					& protocol,\\
 $\omega = (x_t)_{0\leq t \leq T}$		& trajectory/path during an interval $[0,T]$,\\
 $ \pi  $   					& kinematical time-reversal operator\\ 
						& changing the signs of velocities,\\

 $ \theta\omega = (\pi x_{T-t})_{0\leq t \leq T} $ & time-reversal of trajectories,\\
 $d\mathcal{P}_{x_0}(\omega)$			& path-probability measure for paths given $x_0$,\\
 $\mu_0$					& probability distribution of initial configuration,\\
 $d\mathcal{P}_{\mu_0}(\omega)$			& path-probability measure with\\
						& initial state sampled from $\mu_0$,\\
 $ d\mathcal{P}^R $				& path-probability measure with\\
						& reversed protocol,\\
 $\left<f(\omega)\right>_{x_0} = \int d\mathcal{P}_{x_0}(\omega)f(\omega)$			& expectation value of a function $f$,\\
 $\left<f(\omega)\right>_{\mu_0} = \int d\mathcal{P}_{\mu_0}(\omega)f(\omega)$			& expectation value of a function $f$,\\
 $ \frac{d\mathcal{P}_{x_0}^*}{d\mathcal{P}_{x_0}}(\omega)$ 				& Radon-Nikodym derivative,\\
 $A(\omega) = -\log \frac{d\mathcal{P}_{x_0}^*}{d\mathcal{P}_{x_0}}(\omega)	$		& the action,\\
 
 $\mu_t(x)$ 					& time-evolved probability distribution,\\
 $\rho(x)$					& stationary distribution,\\
 $ j_{\mu}(x)	$				& probability current when in $\mu$.\\
\end{tabular}

\paragraph{Entropy and traffic\\}
\begin{tabular}{ll}
 $S(\omega) = \log \frac{d\mathcal{P}_{x_0}}{d\mathcal{P}^R_{x_T}\theta}(\omega) $	& entropy flux into environment,\\
 $ S_{\mu}(\omega) = \log \frac{d\mathcal{P}_{\mu_0}}{d\mathcal{P}^R_{\mu_T}\theta}(\omega) $ & measure of irreversibility,\\
 $s(\mu)$ & Shannon/Gibbs entropy of $\mu$,\\
 $ \sigma(\mu) $   & expected entropy production rate when in $\mu$,\\
 $ S_{ex}(\omega) = A^R(\theta\omega) - A(\omega) $   & excess entropy flux,\\
 $ \mathcal{T}_{ex}(\omega) = A^R(\theta\omega) + A(\omega) $    & excess traffic,\\
 $ \tau(\mu) $    &  expected traffic rate when in $\mu$.
\end{tabular}

\paragraph{Markov jump processes\\}
\begin{tabular}{ll}
 $x,y,\ldots$    & configurations,\\
 $ k_t(x,y) $    & transition rates,\\
 $ \lambda_t(x) = \sum_yk_t(x,y) $   & escape rate.\\
\end{tabular}

\paragraph{Diffusions\\}
\begin{tabular}{ll}
 $x$			& position,\\
 $x^i=x_i$ 		& components of the position,\\
 $v$			& velocity,\\
 $m$			& mass,\\
 $\gamma$		& friction coefficient,\\
 $\chi$ 		& mobility,\\
 $D$			& diffusion coefficient,\\
 $B_t$			& Wiener process.\\
\end{tabular}

\paragraph{Fluctuation-dissipation\\}
\begin{tabular}{ll}
  $h_t$    			& time-dependent amplitude of the perturbation,\\
  $V$				& perturbing potential,\\
 $Q$				& observable, not to be confused with heat,\\
 $R_{QV}(t,s)$			& response function,\\
 $\tau(\omega,s)$		& functional derivative of excess traffic w.r.t. $h_s$.\\
\end{tabular}

\paragraph{Dynamical fluctuations\\}
\begin{tabular}{ll}
 $p_{\omega}$			& empirical occupation vector/density,\\
 $J_{\omega}$ 			& empirical current,\\
 $\mu$				& fluctuation of the occupations,\\
 $j$				& fluctuation of the current,\\
 $I$				& rate function or fluctuation functional,\\
 $K_f$				& a quantity $K$ computed in a dynamics determined by a force $f$,\\
 $\mu_1$			& small deviation of $\mu$ from $\rho$,\\
 $j_1$				& small deviation of $j$ from $j_{\rho}$.\\
\end{tabular}

\cleardoublepage
\chapter*{List of figures}

\begin{tabular*}{\textwidth}{@{\extracolsep{\fill}}llr}
 & &\textbf{Page}\\
Figure \ref{carnot}. & Sadi Carnot, 1796 - 1832.  & \pageref{carnot}\\
& From \textit{http://www.gap-system.org/$\sim$history/} & \\
& \textit{Pict Display/Carnot\_Sadi. html}, & \\
 & 01-03-2010 & \\
& & \\
Figure \ref{joule}. & The experiment with which J.P. Joule showed that heat  & \pageref{joule}\\
& and mechanical work are both forms of energy transfer. & \\
& From \textit{http://www.kutl.kyushu-u.ac.jp/seminar/} & \\
& \textit{MicroWorld1\_E/Part3\_E/P31\_E/heat\_E.htm}, & \\
& 01-03-2010 & \\
& & \\
Figure \ref{boltz}. & Boltzmann's gravestone.  & \pageref{boltz}\\
& From \textit{http://nl.wikipedia.org/wiki/Ludwig\_ Boltzmann}, & \\
& 01-03-2010 & \\
& & \\
Figure \ref{fig:catalyst}. & A catalytic reaction cycle.  & \pageref{fig:catalyst}\\
&  Based on \textit{http://en.wikipedia.org/ wiki/Catalytic\_cycle}, & \\
& 02-03-2010 & \\
& & \\
Figure \ref{fig:markovjump}. & A realization of a Markov jump process.  & \pageref{fig:markovjump}\\
& & \\
Figure \ref{fig:exclusion}. & A visualization of an exclusion process.  & \pageref{fig:exclusion}\\
& & \\
Figure \ref{fig:1}. & Plot of the quantities involved in
Eq.~(\ref{chit}).  & \pageref{fig:1}\\
&  From \cite{bmw09b} & \\
\end{tabular*}

\begin{tabular*}{\textwidth}{@{\extracolsep{\fill}}llr}
Figure \ref{fig:overd}. & Response and fluctuations of the overdamped particle & \pageref{fig:overd}\\
& in a tilted periodic potential.  & \\
&  From \cite{bmw09b} & \\
& & \\
Figure \ref{fig:Lang}. & Integrated correlation functions with various friction  & \pageref{fig:Lang}\\
 &  coefficients.  & \\
&  From \cite{bbmw09} & \\
& & \\
Figure \ref{fig:Lang2}. & Integrated correlation functions with various forces.  & \pageref{fig:Lang2}\\
&  From \cite{bbmw09} & \\
& & \\
Figure \ref{fig:co}. & Visualization of the fluctuation-response relation for  & \pageref{fig:co}\\
&  coupled oscillators.  & \\
&  From \cite{bbmw09} & \\
\end{tabular*}

\cleardoublepage
\phantomsection
\markboth{{Contents}}{{Contents}}
\tableofcontents

\cleardoublepage
\pagenumbering{arabic}
\cleardoublepage
\chapter{Introduction}

\textit{In this chapter we quickly
review some important concepts in thermodynamics
and statistical mechanics that are relevant
for the rest of this thesis. For a more thorough
introduction we refer to standard textbooks. After this
we introduce the reader to the realm of nonequilibrium
phenomena and motivate the research discussed in this thesis.}

\section{Thermodynamics}
\index{thermodynamics}
\begin{wrapfigure}[16]{r}{0.3\textwidth}
\vspace{-20pt}
  \begin{center}
    \includegraphics[width=0.28\textwidth]{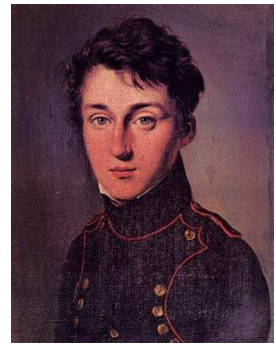}
  \end{center}
\vspace{-20pt}
  \caption{Sadi Carnot, 1796 - 1832.
\label{carnot}}
\end{wrapfigure}
Thermodynamics studies energy conversions
between mechanical work and heat. Although several results in thermodynamics date back
to the seventeenth century, the theory as we now know it 
had its major breakthrough in the nineteenth
century, starting with groundbreaking theoretical considerations by Sadi Carnot.
It is no coincidence that this happened in the century of the industrial revolution. 
Thermodynamics
provided a deep and indispensable understanding of the principles
by which engines (and refrigerators) operate, and the fundamental limits they
must obey. It is also a logical starting point for our discussion. 
In the following pages we quickly review some aspects
of thermodynamics that are relevant to the rest of this thesis; it
is certainly not our goal to review the whole theory of thermodynamics. 

Classical thermodynamics as developed in the nineteenth century is a theory
of macroscopic systems. 
As we now know, a macroscopic system  consists of a large ($10^{23}$)
number of particles (atoms/molecules). 
Thermodynamics thus describes systems with 
variables such as temperature, pressure,
volume, etc, ignoring the microscopic details of the system (on the level of the molecules).
When these variables change in time we speak of a thermodynamic process\index{thermodynamic process}.

\paragraph{Equilibrium}
As this text is situated in nonequilibrium thermal physics, it is
useful to describe what it means to be \emph{in} equilibrium.
With equilibrium we mean the following\index{equilibrium!thermodynamic}:
\begin{itemize}
 \item Two systems are said to be in mechanical equilibrium with each other when the pressures ($P$)
they exert on each other are equal. If the pressures were different, one system
would do work on the other, causing a change in the volumes ($V$).
 \item Two systems that can exchange particles are in diffusive equilibrium when their chemical
potentials ($\mu$) are the same. If they are not then there will be a net current of particles
from one system to another, causing the particle numbers ($N$) of the system to change.
\item Two systems are in thermal equilibrium when, after being brought into thermal contact with each other,
they do not exchange heat. In this case their temperatures ($T$) are the same.
If the temperatures are not the same, there will be heat flow from one system to the other,
causing a change in a new quantity, named entropy ($S$). (We will come back to this later).
\item Two systems are in thermodynamic equilibrium if they are in mechanical, diffusive and thermal
equilibrium.
\end{itemize}
Usually we do not speak about two systems but about one system and its environment,
and say that a system is in equilibrium if it is in equilibrium with its environment.

These definitions already tell us that we distinguish three ways of exchanging energy
between systems: work, particle exchange and heat exchange.
Furthermore six variables are introduced, grouped in three pairs: $P,V$ and $\mu,N$ and $T,S$.
These variables are not independent. Simple systems, such as gases (or more generally pure fluids)  can be described 
by taking one variable of each group,
depending on the interaction of the system with its environment. 
This collection of variables is then called the (equilibrium) state of the system.
For example, if the system is
mechanically isolated from the environment, then its volume is fixed and can be used to
describe the system. One can also fix the pressure and let the volume vary, thus allowing exchange of work, and so on. 
For example a system that can exchange work and heat but not particles
with its environment is best described by the variables $P,N,T$. 

Note that $V,N,S$ are extensive\index{extensive variable}
variables, i.e. they scale with the `size' of the system (e.g. 
if we take two copies of the same system, it has twice the volume, twice the number of particles
and twice the entropy). In contrast $P,\mu,T$ are intensive\index{intensive variable}
variables, i.e. they are independent of the size of the system.

There is a special class of thermodynamic processes which have the following property:
if the process is run backwards, eventually the system and its environment
return to the same equilibrium state they had before the original process.
We call this a reversible process\index{thermodynamic process!reversible}.
Actually, for this to be true in real systems,
the process should go infinitely slowly, and can in that case be described by
a sequence of equilibrium states. As it turns out, many processes in real life
are slow enough for this description to be a reasonable approximation.

Of course, not all quantities mentioned here have a clear intuitive meaning. Especially the
quantities heat and entropy that have to do with thermal equilibrium are vague. 
The laws of thermodynamics give a further specification of these concepts. 
We discuss them here shortly, as the understanding of entropy and its role in nonequilibrium systems is 
very important throughout this thesis.

\paragraph{The first law}\index{thermodynamics!first law}
The internal energy $U$\index{internal energy} of a system corresponds microscopically to the sum of all kinetic energies of the particles
and all their interaction potentials.
The first law of thermodynamics then dictates that a small change of the energy $dU$ of a system
can only be caused by work, heat or a particle flow:
\[  dU = \dbar Q + \dbar W + \mu  dN \]
where $\dbar Q$ is an infinitesimal amount of heat flow into the system and $\dbar W$ is 
an infinitesimal amount of work done on the system. The notation with $\dbar$ stems from the fact that
heat and work can't generally be expressed as the difference of a state function
(a function only depending on the state of the system, not on its history).
If the system contains different
species of particles with chemical potentials $\mu_i$ then the last term should be replaced by 
$\sum_i\mu_idN_i$.

\begin{figure}[ht]
\begin{center}
\includegraphics[width=7cm]{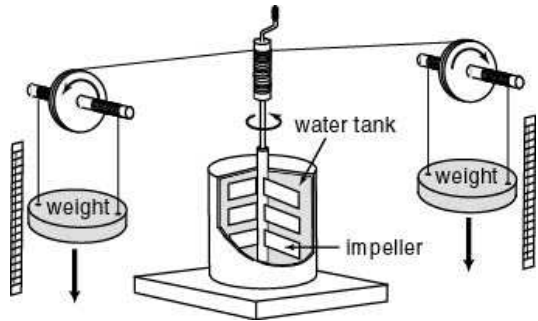}
\end{center}
\caption{The experiment with which J.P. Joule showed that heat and mechanical
work are both forms of energy transfer.\label{joule}
}
\end{figure}

For a system with a fixed number of particles and a fixed volume,
the first law tells us that the energy of a system can only change through
heat flow. In a sense this is a definition of heat\index{heat}: it is an energy transfer
that is not work and not a particle current.

To have a better understanding
of the concept of heat let us make a little thought experiment. Think
of a box containing a gas (system). We imagine the box standing in a room filled with air (environment).
The box is closed and has a fixed volume.
Going down to the microscopic level (leaving thermodynamics for a moment),
we can imagine that particles outside the box collide with the box,
thus exchanging work with the box. The particles in the box also exchange
work with the box. In this way energy can be transferred between the system
and the environment. However when we zoom out to the macroscopic
level, we can no longer see the individual collisions of all the particles,
and if we do not see forces, we can't compute their work. We only see that
energy is exchanged. In this way heat is defined as energy exchange due to work
of forces we can't see from our macroscopic point of view. More generally
heat can also be particle exchange of particles we can't see. 
Note that this definition
is arbitrary in that it depends on the level of description of the system.

\paragraph{The second law}\index{second law} This law is used to define the quantity entropy\index{entropy!thermodynamic}. 
It states that there exists 
a state function,
called the entropy $S$, such that
\[ \Delta S \geq \int \frac{\dbar Q}{T} \]
where $\Delta S = S_f - S_i$ is the change in entropy of the system from the initial
value $S_i$ to the final value $S_f$, and the integral is over a thermodynamic process. 
Moreover, only for a reversible process the inequality becomes an equality, thereby
exactly defining entropy changes.
Apart from defining the entropy, this second law is also a restriction
on the heat flow during a process. For example this law predicts that
heat never spontaneously flows from a cold to a hot object.
It also places a bound on engines that extract work by utilizing
two reservoirs at different temperatures.

For an isolated system (i.e. $\dbar Q = 0$), what does it mean to be `in equilibrium'? 
From the second law we see that any process the system undergoes will
increase its entropy (or leave it unchanged). Only when the system reaches its equilibrium
state, the entropy does not change. One can thus characterize
equilibrium for an isolated system as the state with maximum entropy.

\paragraph{Thermodynamic potentials}\index{thermodynamic potential}
Thermodynamic potentials are (scalar) functions
that describe the thermodynamic state of a system.
From them, many relations between thermodynamic quantities can be derived.
Therefore they play a role in thermodynamics comparable to
the Lagrangian or Hamiltonian in mechanics. Part of this thesis
discusses the use of thermodynamical potentials in nonequilibrium systems.
Because of this, we provide here a quick introduction to these
potentials in thermodynamics. 

The first and most intuitive thermodynamic potential is the
internal energy of the system: $U$. We can use the first law for a reversible process to write
\[ dU = TdS -PdV + \mu  dN \]
where $PdV$ is the work done by the system on its environment.
From this we can see that the natural variables of $U$ are $S,V$ and $N$, while
the variables $T,P,\mu$ all depend on them through
\[ \frac{\partial U}{\partial S} = T,\ \ \ \ \frac{\partial U}{\partial V} = -P,\ \ \ \ \frac{\partial U}{\partial N} = \mu \]
The internal energy is thus fixed in a system
for which $S,V,N$ are fixed and is thus the most natural potential here. Equivalently we can write the entropy $S = S(U,V,N)$
which is most natural in a system with $U,V,N$ fixed, i.e. a totally isolated system.
As we already argued, the entropy characterizes the equilibrium state of such a system
because it is maximal then. Such a characterization of equilibrium is a key feature
of thermodynamic potentials.

For a system with a fixed volume and particle number, but which can exchange heat with 
an environment which is in equilibrium at a fixed temperature, the natural variables are $T,V,N$.
What characterizes equilibrium for such a system? To see this, we assume that 
 the total of system plus environment is isolated. The energy of an isolated
system is constant. So $dU+ dU_e =0$, where $dU$ is the change of energy of the
system, and $dU_e$ of the environment. From the first law of thermodynamics,
we see that for the environment (which is at equilibrium): $dS_e = \frac{1}{T}dU_e = -\frac{1}{T}dU$. 
The system is in equilibrium with its environment if
the total of the two is in equilibrium. In that case the total entropy
is maximal:
\[ d(S + S_e) = 0,\ \ \  d^2(S+S_e)<0 \] 
Written in terms of the system, this gives:
\[ d(S - \frac{U}{T}) = 0, \ \ \ d^2(S - \frac{U}{T}) <  0 \]
This defines a thermodynamic potential $S-U/T$, but a slightly
different potential is
more commonly used: the Helmholtz\index{Helmholtz free energy}\index{free energy}
free energy $F = U - TS$. We see that it is minimal in equilibrium.
This free energy for an equilibrium system is thus defined by
\[ F(T,V,N) = \inf_{S}[U(S,V,N) - TS] \]
which means that it is a Legendre transform of the internal energy $U$. 

Similarly the enthalpy \index{enthalpy}
\[H=U+PV\]
is defined for a system that is thermally isolated but can change its volume
with an environment at constant pressure. In equilibrium we have
\[ H(S,P,N) = \inf_{V}[U(S,V,N) + PV ] \]
The Gibbs free energy \index{Gibbs free energy}\index{free energy}
\[G(T,P,N) = U+PV-TS\]
is defined for a system in contact with an environment at a constant temperature and pressure.
More potentials can be defined when particle exchange with the environment is allowed.

The collection of thermodynamic potentials, all connected through
Legendre transforms, form a very powerful and useful theoretical formalism
for thermodynamics of equilibrium systems.

\section{Equilibrium statistical mechanics}\label{intro-esm}
\index{equilibrium statistical mechanics}
Statistical mechanics, also called statistical thermodynamics, is the theoretical framework that 
explains thermodynamics as a set of macroscopic properties of materials,
 starting from the microscopic properties of the individual atoms or molecules.
But it does more: it gives a more accurate description because it also describes 
the fluctuations from the expected macroscopic behaviour.
It uses statistics/probability theory to be able to study systems that consist 
of a large number ($10^{23}$) particles. Historically, it was
founded in the second half of the nineteenth century, mainly by Ludwig Boltzmann
and James Clerk Maxwell, but also by Gibbs, Einstein and Planck.
Their statistical mechanics is a theory of equilibrium systems. We give a quick overview of
some relevant aspects:

\paragraph{Micro and macro}
In statistical mechanics one makes the difference between microstates and
macrostates. A microstate (or microscopic configuration) \index{microstate}
of a system contains all microscopic information about the system,
such as the positions and velocities of all its particles. 
A macrostate (or macroscopic configuration) gives a description\index{macrostate}
of the system in terms of a few macroscopic properties, such
as temperature and volume. Many microstates may correspond to the
same macrostate.

\paragraph{Entropy}\index{entropy!Boltzmann}
\begin{wrapfigure}[24]{r}{0.35\textwidth}
\vspace{-20pt}
  \begin{center}
    \includegraphics[width=0.32\textwidth]{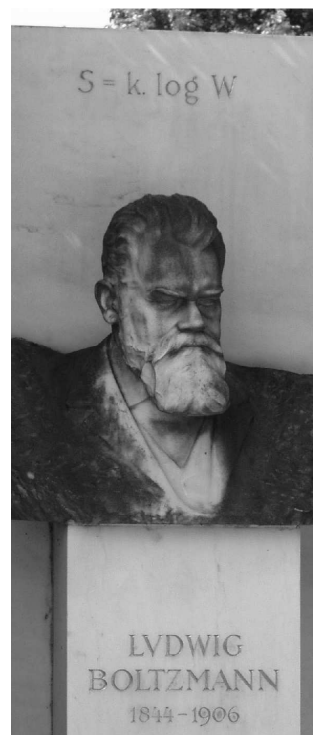}
  \end{center}
\vspace{-20pt}
  \caption{Boltzmann's gravestone.\label{boltz}}
\end{wrapfigure}
One of the basic postulates of statistical mechanics is that
for an isolated system, all its possible microstates are equivalent, i.e one
is not more important or more probable than another.
As a consequence, for an isolated system, macrostates that correspond to more microstates
are more probable than macrostates that correspond to fewer microstates.
This is closely connected to Boltzmann's famous definition of entropy as
a statistical quantity: given a set of macrostates $M$, the entropy
$S$ is proportional to the logarithm of $W$,
where $W$ is the number of microstates corresponding to $M$:
\begin{equation}\label{be} S = k_B\log W \end{equation} 
where $k_B$ is Boltzmann's constant. For continuous microscopic
configuration spaces, $W$ is the `volume' of the set of microstates corresponding
to the macrostate.
Saying that a system in equilibrium is characterized by a maximal entropy
is thus the same as saying that it is characterized by the macrostates
that correspond to the most microstates, i.e. the most probable macrostates.
One of the major achievements of statistical mechanics is that this entropy
coincides with its thermodynamic counterpart mentioned above. As a testament
to the significance of this formula (\ref{be}), it was engraved on the tombstone of Boltzmann.

\paragraph{Ensembles}
\index{ensemble}
The macrostate $M$ that a system is in at any moment depends on the microstate $x$
of the system: $M = M(x)$. Statistical mechanics explains the measured value
of $M$ from the viewpoint of the microstate $x$ the system can be in.
It does this by working with ensembles. An ensemble is a set
of (a large number of) imagined copies of a system: one for each microstate 
the system could be in. One then has to choose a probability distribution
 $\rho(x)$ on this set of possible microstates. The probability of observing
a certain macrostate $M$ is then the sum (or integral) of $\rho(x)$ over
all microstates that correspond to $M$. The statistical average of $M$
can be computed by
\[ \left<M\right> = \sum_x \rho(x)M(x) \]
where the sum can be an integral, depending on the system.
Within the theory of equilibrium ensembles the definition of entropy was
provided by Gibbs\index{entropy!Gibbs}:
\[ S = -k_B\sum_x \rho(x)\log \rho(x) \] 
which is equivalent to the Boltzmann entropy for the microcanonical ensemble (see below).

As macroscopic systems typically have $~10^{23}$ particles, and thus a large number of microstates,
the law of large numbers from probability theory dictates that
the measured value of $M$ is practically always equal to the computed average $\left<M\right>$.
Therefore, by choosing the right probability distribution $\rho$, one
can recover thermodynamics by considering the statistical averages of macrostates.
The most commonly used probability distributions are:
\begin{itemize}
 \item The microcanonical ensemble\index{ensemble!microcanonical} is used for isolated systems, i.e. systems
that have a fixed energy $E$. The set of possible microstates is thus restricted
to microstates that have an energy $E$. Apart from that an equal probability is assigned to
all those microstates.
 \item The canonical ensemble\index{ensemble!canonical} is used for systems that can exchange energy only in the form of heat
with an environment at a fixed temperature $T$. The canonical distribution is
 \[\rho(x) = \frac{1}{Z}e^{-\beta E(x)}  \]
where $Z$ is a normalization factor, called the canonical partition function, $\beta = \frac{1}{k_BT}$
and $E(x)$ is the energy of the system in microstate $x$.
 \item The grand canonical ensemble\index{ensemble!grand canonical} is used for systems that can exchange heat and particles
with a reservoir at a fixed temperature $T$ and chemical potential $\mu$:
\[ \rho(x) = \frac{1}{\mathcal{Z}}e^{\beta\mu N(x)-\beta E(x)} \]
where $\mathcal{Z}$ is a normalization factor, called the grand canonical partition function,
and $N(x)$ is the number of particles of the system in microstate $x$.
\end{itemize}
As an example of how the link to thermodynamics is made: we explicitly substitute the distribution for the canonical ensemble in the Gibbs entropy:
\[ S = k_B \log Z + \frac{\left<E\right>}{T}  \]
The average energy $\left<E\right>$ coincides by the law of large numbers to the measured energy $U$, so
\[ -\frac{1}{\beta}\log Z = U - TS = F \] 
where $F$ is the Helmholtz free energy of the system.

\section{Out of equilibrium}\label{intro-ooe}

In the end, the purpose of statistical mechanics is twofold: on the one hand to describe the macroscopic
world, and derive its physical behaviour, starting from the microscopic level. 
On the other hand, statistical
mechanics makes more predictions than thermodynamics, because it also describes
the deviations (fluctuations) from the average behaviour. 

Equilibrium statistical mechanics is in this sense very powerful, because 
it is seen that measurable quantities, using the law of large numbers, can be computed through averaging 
over equilibrium distributions
of microstates. These are themselves expressed solely in terms of external constraints (temperature, chemical potential),
and conserved quantities like energy and particle number.  Going beyond thermodynamics, for example
the fluctuation-dissipation theorem is a useful and well-known physical relation. 
(Part of this text discusses the generalization of this to 
nonequilibrium systems).

However, up to the moment that this text is written,
there is no general paradigmatic theoretical framework that 
describes systems \emph{out of} equilibrium, neither in thermodynamics
nor in statistical mechanics, not even for stationary systems.
This poses a problem, because many processes in real life can't be described as reversible processes, and many systems
are not in equilibrium.

\paragraph{How to be out of equilibrium}
In general we distinguish two ways for systems to be out of equilibrium.
First, systems that are in the process of relaxing to equilibrium:
when a parameter determined by the environment changes, like temperature or
volume, the system always needs some time to relax to that new equilibrium state.
For example, when a hot cup of coffee is placed in a room
at room temperature, heat will start to flow from the coffee 
to the air of the room (we ignore heat exchange by radiation with the walls of the room). 
After some time the coffee and the air will relax to a new equilibrium state:
the coffee cools down to room temperature. Before that time, however, the coffee is not in equilibrium
with the room,
and the process is not reversible. If it was reversible, then we would not be surprised to observe heat to 
flow spontaneously from the air to the coffee, heating it up again.
Sometimes the external parameters change continuously and fast enough such that the system
never has enough time to relax to equilibrium. Think for example of combustion engines.

Secondly there are systems that are driven from equilibrium by what we call thermodynamic forces.
Here we distinguish three subgroups, corresponding to thermal, diffusive and mechanical
exchanges of energy: 
\begin{itemize}
 \item Systems in contact with parts of the environment at different temperatures.
For example a wall of a house: on one side it is in contact with the warm air inside,
and on the other side it is in contact with cold air. In such systems there are constantly
 heat currents. A part of the environment that is in thermal contact with the system, is called a heat bath\index{heat bath}.
The thermodynamic force here is the temperature difference.
 \item Systems in contact with parts of the environment at different chemical potentials.
Think of a cell membrane with a bigger particle density on the inside than on the outside.
In such systems one sees particle currents. A part of the environment that is in diffusive contact with the system is called a 
particle reservoir\index{particle reservoir}.
The thermodynamic force here is the chemical potential difference.
 \item Systems under the influence of mechanical nonconservative forces. A nonconservative force is a force
which is not the derivative from a potential. Think of the pressure difference that makes water run out of a tap.
\end{itemize}
To simplify this second group of systems, one usually assumes that the parts of the environment
in contact with the system are and stay in equilibrium during the process. For this one should 
assume that there is no interaction between the different parts of the environment, and that 
the interaction between system and environment is weak,
i.e. changes in each part of the environment are small enough such that it relaxes fast enough to equilibrium for
our description of the system. One also assumes
that the heat baths and particle reservoirs are big, such that during the process their temperatures
and densities do not change measurably. 
Systems under such conditions can often relax to a stationary state\index{stationary state}. This is a state
in which the variables (macrostates) with which we describe the system do not change in time.
Such variables include heat or particle currents in contrast to equilibrium systems.
One of the goals of nonequilibrium thermodynamics is to describe such stationary regimes
and correctly predict the directions and sizes of their currents.

\paragraph{Dynamics}

An important property of nonequilibrium systems, in contrast to equilibrium, is that time
plays an important role. First of all because nonequilibrium processes are irreversible:
an arrow of time is introduced. The question of how irreversibility emerges when going
from the reversible microscopic world to the macroscopic world was already discussed by
Boltzmann himself. 

Apart from that, it seems clear that nonequilibrium systems
are by their very nature dynamical. This is because systems are either
out of equilibrium because they are driven from equilibrium and are in some stationary regime in which there are particle or
heat currents present, or because they are in the process of relaxing to equilibrium (or to a stationary state). 
This means that not only the microstates themselves are important,
but also the way in which they change in time. The dynamics of the process should enter our
(theoretical) description.

\paragraph{Statistical mechanics of stochastic processes: the mesoscopic level}

Ideally, one would therefore like to answer physical questions
about average behaviour and fluctuations, by starting from the microscopic Hamiltonian
dynamics. However, this is often just impossible.
This is not necessarily a disaster, because we expect (hope) that not every detailed
aspect of that dynamics is relevant for the macroscopic world.
Therefore most models used in nonequilibrium statistical mechanics
are already reduced descriptions, meaning that they do not contain all information
of the microscopic world. For example, one usually reduces the description
of the environment of the system to some parameters as temperature or chemical potential. 
On the other hand, to be able to describe fluctuations from the typical macroscopic behaviour,
a more detailed description than the macroscopic one is needed. 
We then say that we are working on the mesoscopic level\index{mesoscopic level}.

One should thus think of a mesoscopic model as describing a small system,
for which the description is not detailed enough to be Hamiltonian,
but is not big enough for the law of large numbers to apply.
Fluctuations around the statistical averages are important.
As a consequence a model describing a mesoscopic system can be a  stochastic process.
A stochastic process is therefore a very important tool in nonequilibrium
statistical processes. An important part of the present day research 
is therefore committed to finding `recipes' for defining stochastic processes
that are physically relevant. One way to do this is via the local detailed
balance assumption, which will be discussed in the next chapters, and is
used throughout this text. 

\paragraph{Applications}

As said before, many processes in everyday life are not reversible,
and many quantities of interest are some form of current, which
is not found in equilibrium systems.
On the macroscopic scale we find important examples in ecology and meteorology.
In meteorology, we see that the temperature at the poles is on average
much lower than at the equator. That is responsible for major energy currents
in the earth's weather system. On a smaller scale one is interested in predicting local
weather, like temperature changes and air flows (wind), 
given observational data concerning high and low pressure areas etc.
In ecology one is interested in the energetic and material (food) flows between
different species of organisms. 

On a much smaller scale we see that mesoscopic models are not only useful as a step towards macroscopic
theories, but also as a description of interesting but very small systems. 
In recent years the nonequilibrium world of small systems has become more
accessible experimentally, opening and expanding research areas as biophysics
and nanotechnology. In biophysics a lot of attention nowadays is going to
e.g. the dynamics of DNA and RNA, transport of ions through cell membranes,
molecular motors, etc. In nanotechnology the electrical current in extremely small
devices or parts of devices is central. Finally, in everyday life there are many finite networks,
like transport and communication networks, on which processes occur which are well described on the mesoscopic scale.

A lot of interesting and useful research has been done and is being done
in these areas, where explicit models (stochastic processes) are examined
to derive results for each specific area of interest. However, this
is not the goal of the research reported in this thesis.
Instead, we try to find general physical structures
of nonequilibrium statistical mechanics. Such a scheme
 should in the long run and together with many other contributions
lead to a better understanding of nonequilibrium physics. Such
an understanding should then give useful predictions applicable
to the specific areas of interest mentioned above.

\section{Outline and preview of the results}

\paragraph{Part I: Stochastic processes}
Because of the lack of a general theory, 
it may come as no surprise that the present day 
research into nonequilibrium systems
focuses on simple physical systems. It is then important to 
define a stochastic process
that models it correctly. The first part of this thesis 
therefore contains an introduction to the stochastic
models that are used in the rest of the text, focusing on the relevant aspects
for understanding the next two parts.

Throughout this text, we restrict ourselves to
classical systems, meaning that we do not consider relativistic
or quantum effects. (Quantum mechanics only enters in some specific
models in the discretization of configuration space).

\paragraph{Part II: Fluctuation-dissipation relations}

In this part of the thesis, we investigate how a system responds to a perturbation, 
namely a small change in its energy. 
The central object that summarizes this is the response function.
To be more precise, we denote the energy of the system in configuration $x$ by $U(x)$.
As a perturbation this energy is changed by the addition of a potential: $U\to U - h_t V$,
where the time-dependent function $h_t$ is the amplitude of the perturbation. 
We restrict our possible class of systems to those in which a small amplitude $h_t$
only has small effects (excluding for example the regime of phase transitions).
Then we can write the expectation of an observable $Q$
in the perturbed system as the expectation in the unperturbed system  plus a small correction:
\[ \left<Q(x_t)\right>^h \approx \left<Q(x_t)\right>^0 + \int_0^tdsh_sR_{QV}(t,s)  \]
This defines the response function $R_{QV}(t,s)$.
In equilibrium systems it is known that this response function can be written in terms of
a correlation function in the unperturbed system (a system in equilibrium at inverse temperature
$\beta$):
\[R_{QV}(t,s) = \beta\frac{\partial}{\partial s}\left<Q(x_t)V(x_s)\right>^0\]
This is called the fluctuation-dissipation theorem. Our contribution has been to 
discuss the form of the response function for systems out of equilibrium.
We found that the response function can generally be written as the sum of two
correlation functions. One correlation function is linked to heat dissipation
into the environment, and thus to entropy changes. The other correlation
function has to do with a quantity which we call traffic, which describes in a sense
the activity of the system. The main results of this part are summarized
by formulae (\ref{response1a}), (\ref{response2}) and (\ref{fdr}). These
results are then applied to several explicit examples for which simulations
have provided some visualization and verification.

The research discussed in this part was reported before in
\cite{bbmw09,bmw09a,bmw09b}.

\paragraph{Part III: Dynamical fluctuations}

In this part of the thesis we use the theory
of large deviations to examine certain fluctuations in systems
out of equilibrium. For equilibrium statistical mechanics,
large deviation theory provides a natural mathematical framework
especially for the thermodynamic potentials discussed above.
It is therefore useful to investigate that theory also for nonequilibrium systems.

In this theory we consider the empirical occupation density $p$ (which describes
the fraction of time the system spends in each configuration)
and the empirical current $J$ (describing the changes of configuration
the system makes) as observables. The probability is then considered that
these observables take on some given values $p\approx\mu$
and $J\approx j$. Such a probability often shows an exponential
decay, whenever $\mu$ and $j$ do not correspond to the typical
values of the observables:
\[ \textrm{Prob}(p\approx\mu,J\approx j) \approx e^{-TI(\mu,j)}\]
where the duration $T$ of the process is very large,
and $I(\mu,j)$ is called the rate function. 

The questions in this part of the thesis are then: how
to compute the rate function, and what are the physical
quantities governing its form? As in the previous part,
entropy and traffic are the main ingredients. Moreover,
the rate function $I(\mu,j)$ is explicitly computed for
the classes of models considered and expressed in terms
of entropy and traffic, see (\ref{imuj}) and (\ref{imujmj1}).
The rate function for the occupations can be expressed entirely in terms
of traffic. We also show that this traffic can be seen as a thermodynamic
potential for currents. Finally, for the 
close-to-equilibrium regime, known variational principles such as the 
minimum entropy production principle are recovered.

The research discussed in this part was published before in
\cite{mnw08b,mnw08a}.

\cleardoublepage
\part{Stochastic processes}

\vspace*{6cm}

``Observe what happens when sunbeams are admitted into a building 
and shed light on its shadowy places. You will see a multitude 
of tiny particles mingling in a multitude of ways... their dancing 
is an actual indication of underlying movements of matter that are 
hidden from our sight... It originates with the atoms which move of 
themselves. Then those small compound bodies that 
are least removed from the impetus of the atoms are set in motion by the 
impact of their invisible blows and in turn cannon against slightly larger 
bodies. So the movement mounts up from the atoms and gradually emerges to the 
level of our senses, so that those bodies are in motion that we see in sunbeams, 
moved by blows that remain invisible.''

\begin{flushright}
 Lucretius, \textit{De rerum natura}, ca. 60 BC.
\end{flushright}

\cleardoublepage
\chapter{Describing stochastic processes}
\label{chap-traj}
\textit{
Stochastic processes are a useful tool
to explore nonequilibrium statistical mechanics. The basic ingredient here is not the microstate but
the trajectory, i.e. the sequence of states the system visits during the process.
In this chapter we therefore discuss how to work with 
such trajectories in statistical physics. To lift the mathematical
models to a more physical level, the local detailed balance assumption is made
and explained. As we will see, some well-known general physical 
results are a direct consequence of this assumption.}

\section{Stochastic processes}
Mathematically\index{stochastic process} we describe a system by its configuration\index{configuration} or state\index{state}.
This state contains all information of the system that is relevant for our description. 
Let us denote it by $x$,
being an element of the configuration space\index{configuration!space} $\Omega$.
This $x$ can be for example the vector of positions and momenta of a gas, or the set of 
spin-configurations of a magnet, or maybe just the position of one particle submerged in a fluid.
Consequently, the set $\Omega$ can be continuous or discrete, unbounded, compact or even
finite.\\ 

In the course of time the configuration of the system will change, i.e. the configuration depends on time:
$x = x_t$. How it changes, is determined by the dynamics\index{dynamics} of the process. 
In classical mechanics, the dynamics at the microscopic level is Hamiltonian and therefore deterministic.
Indeed, when all positions and momenta of the particles of a gas in an isolated container are given
at a time $t$, one can in principle determine the positions and momenta at any later time.\\

Unfortunately, this is not always possible. Most of the times the system under consideration
is not isolated, but interacts with some environment of which the exact configuration is unknown. 
Moreover it is usually even impossible
to measure the exact configuration of the system itself. Because of such factors, uncertainty
enters, and must enter our mathematical description. This gives rise to stochastic processes,
and the configuration $x_t$ of the system at each time becomes a stochastic variable $X_t$.
A stochastic process is then the sequence of these stochastic variables:
\[ \{X_t|0\leq t\leq T\} \]
When the history of the system is given, i.e. $x_t$ for $t\in ]-\infty,T]$, the stochastic
dynamics of the process gives a probability to find the system in a configuration $y$ at a later time.
\\

Again one has to be more modest, as the complete history of a system is rarely known.
A widely used and usually well-working approximation is the Markov\index{Markov process} approximation.
This consists of the assumption that the probability of finding the system at a later 
time in some configuration, only depends on the configuration of the present, not the past.
Informally, for a sequence of times $t_1<t_2<\ldots<t_n<t$:
\begin{equation}\label{marass} 
 Prob(x_{t} = x| x_{t_n}=y_n,\ldots,x_{t_1}=y_1) = Prob(x_{t} = x| x_{t_n}=y_n)
\end{equation}
A Markov process 
is a stochastic process for which the Markov assumption (\ref{marass}) is valid.
Note that Hamiltonian dynamics, governing the evolution of the state $(x,v)$, although deterministic, 
also satisfy this assumption. In the rest of this text we
always assume our processes to be Markovian.
In many cases the dynamics of the Markov process have the property that
\[ Prob(x_{t} = x| x_{s}=y) = Prob(x_{t-s} = x| x_{0}=y) \ \ \ \ \ \forall s<t \]
In such cases we speak of time-independent dynamics\index{dynamics!time-independent} 
(or time-homogeneous dynamics)\index{dynamics!time-homogeneous}.
However, this is not always the case in this thesis.\\

For the reader who is unfamiliar with probability theory and stochastic processes, 
we refer to \cite{gs01} for a thorough introduction.

\paragraph{A particle in a box}

As an easy introductory example,
consider a box with volume $V$ filled with air. In this box we place one
charged test-particle. We also apply an electric field
to move the particle in our favourite directions.
If we knew all the positions and velocities of all
the particles in the box, in principle the trajectory of test-particle
could be computed with Hamilton's equations. Practically, this is not possible,
but suppose that in some way the position of the test-particle can be measured
at time intervals of 1 second, the first measurement being at time zero. 
The precision of the measurement is limited:
one can only determine the position up to a small volume $\Delta V$. Therefore
we divide the box into $N = V/\Delta V$ parts, labelled by $x=1,2,\ldots$.
Given the configuration (position) $x$ after $n$ seconds, denoted by $x_n$, the position $x_{n+1}$
is not determined exactly. There are usually several possible positions
that the particle can be in, each having a probability determined by the
temperature of the air, the electric and magnetic fields, etc. 
These probabilities can also depend on the previous positions of the particle.
However, let us assume that the interaction with the air molecules is sufficiently
strong compared to the influence of the electromagnetic fields and inertial effects, such
that the particle loses its memory quickly enough for the Markov
assumption to be valid.

Having this physical example in our mind, let us write
the probabilities of the successive positions (transition probabilities) as follows:
\[  P_n(x,y) = Prob(x_{n+1} = y| x_{n}=x) \]   
Mathematically, this defines what is called a Markov chain,
which is the simplest example of a Markov process.
In the next two chapters we discuss two other classes
of Markov processes: Markov jump processes and diffusions,
but in this chapter we use these simple Markov chains
to illustrate the introduced concepts.

The $N\times N$ matrix $P_n$ is called the transition matrix at time $n$,
and it determines the dynamics of the process.
Note that we should have that $\sum_yP_n(x,y) = 1$
for every $n$, where the sum is of course over all configurations.
When the electromagnetic fields are constant in time,
the dynamics are time-homogeneous: $P_n=P$. When the fields
are time-dependent, so are the dynamics. The time-dependence
of the fields is determined by some parameter $\lambda_t$ which
continuously changes in the time $t$, so $P_n = P_{\lambda_n}$. This $\lambda_t$ is controlled externally, and
we imagine it to be deterministic. It is called the protocol\index{protocol}.

\section{Trajectories}

A realization of a stochastic process is called a trajectory\index{trajectory} or path\index{path}. We denote it by
$\omega = (x_t)_{0\leq t \leq T}$. 
If the dynamics of the stochastic process are given, one can in principle compute the probability measure\index{path-probability!measure}
on such paths $\omega$. Suppose that the configuration of the system at time zero $x_0$ is given. 
We denote by $d\mathcal{P}_{x_0}(\omega)$ the path-probability measure\index{path-probability!measure} of $\omega$ given $x_0$.
With it we can compute the expectation values of observables: take any observable $f(\omega)$ that depends
on the trajectory. Then its expectation value\index{expectation value} is
\begin{equation}\label{expectation}
 \left<f(\omega)\right>_{x_0} = \int d\mathcal{P}_{x_0}(\omega)f(\omega) 
\end{equation}
where one integrates (or sums, depending on the model) over all possible trajectories that start from $x_0$ at time zero.
More generally, given a probability distribution (or density) $\mu_0$ of initial configurations, we denote the probability
measure of a path by $d\mathcal{P}_{\mu_0}(\omega) = \mu_0(x_0)d\mathcal{P}_{x_0}(\omega)$ and
the expectation value of $f(\omega)$ by
\[ \left<f(\omega)\right>_{\mu_0} =  \int d\mathcal{P}_{\mu_0}(\omega)f(\omega) \]
where integration/summation is now over all possible trajectories.\\

Basic probabilistic rules dictate that the path-probability measure should be normalized:
\begin{equation}\label{pathnorm}
 \int d\mathcal{P}_{\mu_0}(\omega) = 1 
\end{equation}
Moreover, if we split the trajectories into $\omega_1 = (x_t)_{0\leq t \leq s}$ and $\omega_2 = (x_t)_{s < t \leq T}$,
for any $s$ in the interval $[0,T]$, then the path-probability measure of paths $\omega_1$, which we
just denote by $d\mathcal{P}_{\mu_0}(\omega_1)$, is found by integrating over all $\omega_2$:
\begin{equation}\label{pathmarg}
  d\mathcal{P}_{\mu_0}(\omega_1) = \int_{\omega_2} d\mathcal{P}_{\mu_0}(\omega)
\end{equation}
This is an important property. For example, in many computations
in this thesis we take expectation values of state functions, evaluated
at some time $s$. This gives
\[ \left<f(x_s)\right>_{\mu_0} =  \int d\mathcal{P}_{\mu_0}(\omega)f(x_s) = \int d\mathcal{P}_{\mu_0}(\omega_1)f(x_s) \] 
i.e. we only need to take an average over all paths in the interval $[0,s]$.

\paragraph{Particle in a box} 
For Markov chains, a trajectory is determined by the successive configurations: $\omega = (x_0,x_1,\ldots,x_m)$,
where $m$ corresponds to the time $T$.
The probability of a trajectory $\omega$ is just $\mathcal{P}_{x_0}(\omega) = \prod_{n=0}^{m-1}P_n(x_{n},x_{n+1})$. 
If we take for example a function $f(x)$
evaluated at time $m$, then 
\[ \left<f(x_m)\right>_{x_0} = \sum_{x_1,\ldots,x_m}\left[\prod_{n=0}^{m-1}P_n(x_{n},x_{n+1})\right]f(x_m) \]

\section{Probability distribution of states}
The probability of finding the system\index{distribution} in a configuration $x$ at time $t$
is denoted by $\mu_t(x)$, the time-evolved distribution\index{distribution!time-evolved}
(actually in the case of a continuous configuration space, $\mu_t(x)$ is a probability density). 
It can easily be written in terms of expectation values: 
\begin{equation}\label{mudef}
  \mu_t(x) = \left<\delta(x_t -x)\right>_{\mu_0}
\end{equation}
using $\delta_{x_t,x}$ instead of $\delta(x_t -x)$ for discrete state spaces.
Equivalently, for any state function $f(x)$:
\begin{equation}\label{fexp}
 \left<f(x_t)\right>_{\mu_0} = \int dx\mu_t(x)f(x)  
\end{equation}
The integral becomes a sum for discrete state spaces. In systems
with time-independent dynamics, there
often exists a distribution which does not change under the time-evolution.
This is called the stationary distribution\index{distribution!stationary}. If the system is in the stationary distribution, 
it is said to be in the `steady state'\index{steady state}. We always denote the stationary distribution with $\rho$:
\[ \rho(x) = \left<\delta(x_t -x)\right>_{\rho}\ \ \ \ \ \ \forall t \]
Under certain conditions on the dynamics,
it can be proven that such a distribution exists, is unique, and that all distributions $\mu_t$ converge to it in the long time limit
(the system relaxes to the stationary distribution). Unless stated otherwise,
we always assume this to be true for time-independent dynamics. See \cite{gs01} for a rigorous treatment.

The existence of such a stationary measure implies ergodicity\index{ergodicity}, meaning that for trajectories $\omega$
and an arbitrary state function $g$:
\begin{equation}\label{ergodicity}
 \frac{1}{T}\int_0^T dt g(x_t) \to \int dx \rho(x)g(x) \ \ \ \ \ \textrm{ for }\ T\to\infty
\end{equation}
almost surely. More precisely: the probability that the system follows a trajectory that satisfies
(\ref{ergodicity}) is equal to one.

\paragraph{Particle in a box} 
For a Markov chain we get:
\[ \mu_m(x) = \sum_{x_0,\ldots,x_{m-1}}\mu_0(x_0)\left[\prod_{n=0}^{m-2}P_n(x_{n},x_{n+1})\right]P_{m-1}(x_{m-1},x) \]
Note that the distribution one time-step later can be written as $\mu_{m+1}(x) = \sum_y\mu_m(y)P_m(y,x)$,
and thus:
\[ \mu_{m+1}(x) - \mu_m(x) = \sum_y[\mu_m(y)P_m(y,x) - \mu_m(x)P_m(x,y) ] \]
In the case that $P_n=P$, if there is a stationary distribution it has to satisfy
\[ 0 = \sum_y[\rho(y)P(y,x) - \rho(x)P(x,y) ] \]

\section{The action}
Suppose now that one has two stochastic processes, each having the
same configuration space but a different dynamics. The action\index{action} gives a way of switching between expectation values
computed in those dynamics. This is essential to the rest of this text.\\

We denote the path-probability measures of the processes by $d\mathcal{P}$ and $d\mathcal{P}^*$ respectively.
Suppose that $d\mathcal{P}^*$ is absolutely continuous with respect to $d\mathcal{P}$. This means that $\int_Md\mathcal{P}^*(\omega)=0$
for any set of trajectories $M$ for which $\int_Md\mathcal{P}(\omega)=0$.
Then one has
\begin{equation}\label{rnexp}
  \left<f(\omega)\right>^*_{\mu_0} = 
\int d\mathcal{P}_{\mu_0}(\omega)\frac{d\mathcal{P}^*_{\mu_0}}{d\mathcal{P}_{\mu_0}}(\omega)f(\omega) = 
\left<f(\omega)\frac{d\mathcal{P}^*_{\mu_0}}{d\mathcal{P}_{\mu_0}}(\omega)\right>_{\mu_0}
\end{equation}
The quantity $\frac{d\mathcal{P}^*_{\mu_0}}{d\mathcal{P}_{\mu_0}}(\omega)$ is called the Radon-Nikodym
derivative\index{Radon-Nikodym derivative} between the two processes \cite{niko30}. In words: it is the probability of a path in one dynamics 
divided by the probability of the same path in another dynamics.
It is common to write this in the following way:
\begin{equation}\label{pathprob}
 \frac{d\mathcal{P}^*_{\mu_0}}{d\mathcal{P}_{\mu_0}}(\omega) = e^{-A(\omega)}
\end{equation}
where $A(\omega)$ is called the `action.'\index{action} For Markov processes it is independent of $\mu_0$.\\

Normalization of the path-probability measure (\ref{pathnorm}) tells us that
\begin{equation}\label{actionnorm}
 \left<e^{-A(\omega)}\right>_{\mu_0} = \left<1\right>^*_{\mu_0} = 1
\end{equation}
for any initial distribution $\mu_0$. 


\paragraph{Particle in a box}
For a Markov chain the action is easily computed to be
\[  A(\omega) = \sum_{n=0}^{m-1} \log \frac{P_n(x_{n},x_{n+1})}{P^*_n(x_{n},x_{n+1})}  \]

\section{Time-reversal}\label{traj-eat}
To be able to talk about time-reversibility
we first define an operator $\theta$ which acts as a time-reversal\index{time-reversal!operator}
on trajectories. This means that it 
reverses all trajectories, also taking care that velocities change sign under time-reversal: 
\[\omega = (x_t)_{0\leq t\leq T} \mapsto \theta\omega = (\pi x_{T-t})_{0\leq t\leq T}\]
where $\pi$ changes signs of velocities. Moreover, if the dynamics of the process is time-dependent, we also reverse this 
time-dependence. For example if the dynamics is governed by a force $F_t(x)$,
then time-reversal changes this to $F_{T-t}(x)$. This means that the path-probability
measure will also change. Therefore, whenever we deal with time-dependent dynamics, 
we denote the new path-probability measure by $d\mathcal{P}^R$,
(the $R$ stands for `reversed'). These reversals together make the time-reversal
that intuitively corresponds to playing the movie of a process backwards. 
\\

Let us consider the following measure of irreversibility\index{irreversibility!measure of}, defined in terms of a Radon-Nikodym derivative:
\begin{equation}\label{moi}
 S_{\mu}(\omega) = \log \frac{d\mathcal{P}_{\mu_0}}{d\mathcal{P}^{R}_{\mu_T}\theta}(\omega)
\end{equation}
where $\mu_0$ is an arbitrary probability distribution from which the initial state of the system is sampled.
For the time-reversed process\index{time-reversed process} the initial distribution is taken to be $\mu_T$, which is defined (given $\mu_0$) through (\ref{mudef}).
This quantity $S_{\mu}(\omega)$ is a measure of irreversibility of a certain trajectory $\omega$. It is the probability
of $\omega$ divided by the probability of its time-reversed twin in the time-reversed dynamics.
Let us rewrite (\ref{moi}) as follows:
\begin{equation}\label{sdef}
   S_{\mu}(\omega) = \log\left(\frac{\mu_0(x_0)}{\mu_T(x_T)}\frac{d\mathcal{P}_{x_0}}
{d\mathcal{P}^{R}_{x_T}\theta}(\omega)\right) =: \log \frac{\mu_0(x_0)}{\mu_T(x_T)} +S(\omega)
\end{equation}
which defines the quantity $S(\omega)$. Note that for Markov processes $S(\omega)$ does not depend on the initial distribution of the system.

\paragraph{Particle in a box}
For Markov chains, time-reversal of the path gives $\theta\omega = (x_m,\ldots,x_1,x_0)$,
as long as the configurations do not contain velocities, like in the example of the particle in a box.
The reversal of the dynamics gives transition probabilities $P_n \to P_{m-n-1}$. Together they give
\begin{equation} \label{pbent}
 S(\omega) = \sum_{n=0}^{m-1}\log\frac{P_n(x_{n},x_{n+1})}{P_n(x_{n+1},x_{n})}
\end{equation}

\section{Equilibrium}
The most general way of defining equilibrium\index{equilibrium} is that systems in equilibrium
are stationary and time-reversible\index{reversibility}. The first condition means that the probability
of finding the system in a state $x$ does not depend on time, i.e. it is given by
the stationary distribution $\rho(x)$ (the dynamics have to be time-independent of course). The second condition means that when we are shown 
a movie of an equilibrium process, we can not tell if the movie is played normally 
or backwards.
More mathematically: $S_{\rho}(\omega)$ as defined in (\ref{moi}) with $\mu_0=\mu_T=\rho$ is zero for any trajectory $\omega$:
\begin{equation}\label{equilibrium}
 S_{\rho}(\omega) = \log \frac{d\mathcal{P}_{\rho}}{d\mathcal{P}^{R}_{\rho}\theta}(\omega)=0
\end{equation}
This has as an immediate consequence (see (\ref{sdef})) that
\begin{equation}\label{eqdyn}
  S(\omega) = \log \rho(x_T) - \log \rho(x_0)
\end{equation}
We can use our knowledge of equilibrium systems (see Section \ref{intro-esm}) to see what this means 
physically. E.g. for a system in contact with a heat bath in equilibrium at inverse temperature $\beta$, 
$\rho$ is given by the canonical distribution. We get:
\begin{equation}\label{eqdyn1}
  S(\omega) = -\beta[U(x_T)-U(x_0)]
\end{equation}
with $U(x)$ the energy of the system in configuration $x$. The change of energy of the system
is equal to minus the change of energy of the environment.
It becomes more clear what this quantity $S$ is when we consider a system in contact with a particle reservoir at inverse temperature $\beta$
 and chemical potential $\mu$. For this $\rho$ is the grand canonical distribution, and
we get
\begin{equation}\label{eqdyn2}
  S(\omega) = -\beta[U(x_T)-U(x_0)] + \beta\mu[N(x_T)-N(x_0)]
\end{equation}
This is exactly the change of entropy in the environment, stated in terms of the configurations of the system
(up to a factor of $k_B$, Boltzmann's constant, which we set to one for notational simplicity).
This means that $S(\omega)$ can be interpreted as the entropy flux\index{entropy flux!in equilibrium} from the system to its environment
during the trajectory $\omega$. In the following section we generalize the connection between entropy flux and 
$S(\omega)$ to nonequilibrium processes.\\

Here we would like to point out that there is a fundamental difference between
a system in equilibrium and a system with an equilibrium dynamics\index{equilibrium!dynamics}. As we can see
from (\ref{sdef}), $S(\omega)$ is independent of the initial distribution $\mu_0$,
it only depends on the dynamics of the process. So if the system has an equilibrium dynamics, 
(\ref{eqdyn}) holds for any $\omega$, even if the system is not in the stationary (equilibrium) distribution.
This is also called a detailed balanced dynamics\index{detailed balance}.
Only if the system is also stationary, then 
(\ref{moi}) is zero for any $\omega$ and we say that the system is in equilibrium.\\   

\paragraph{Particle in a box}
For a Markov chain to be in equilibrium, (\ref{equilibrium}) must hold for any path. It should therefore 
hold at least for paths that last only one step: $\omega = (x_0,x_1)$. So a necessary condition for equilibrium is:
\[ 0 = \log \left(\frac{\rho(x_0)P(x_0,x_1)}{\rho(x_1)P(x_1,x_0)}\right) \]
for any $x_0,x_1$, which is equivalent to
\[ \rho(x_0)P(x_0,x_1) = \rho(x_1)P(x_1,x_0) \]
This is called the detailed balance condition. One can easily check that it is also a sufficient condition
for an equilibrium dynamics.

\section{Irreversibility and entropy: local detailed balance}\label{sec-ldb1}
\index{entropy flux} 
For a very large class of nonequilibrium
systems we will assume that \index{entropy flux!out of equilibrium}
\[S(\omega) = \textrm{ \textbf{ the entropy flux from the system into the environment.}}\]
This assumption is called `local detailed balance,'\index{local detailed balance} because it is based on the assumption 
that locally in time and in space the system has a dynamics that is detailed balanced.
This assumption is restricted to the case that the reservoirs only interact with the system, not with
each other. Moreover the coupling between system and reservoirs should be sufficiently
weak and the reservoirs sufficiently big, such that the reservoirs stay in equilibrium throughout the process.
The local detailed balance assumption can be seen as a restriction on the possible
mathematical models that we use to describe the system. More constructively,
it gives a partial recipe to write down models that correspond to the
physical world. Partial, because local detailed balance does not fully
specify the dynamics. The ideas behind this assumption
have been used already for some time, see e.g. \cite{leb59}, but it was first
called local detailed balance in \cite{kls83}. A more rigorous treatment of
this assumption was then later done in \cite{maes99,mrv00,mrv01} and mostly in \cite{mn03}.
Local detailed balance is central to the results discussed in this thesis,
and we always assume it to be true.\\

Let us give a heuristic motivation why such an assumption is reasonable.
As said in Section \ref{intro-ooe}, there are several ways in which 
a system can be driven away from equilibrium. Let us first consider the case of a system
in contact with two heat baths which  are themselves in equilibrium
 at different temperatures $\beta_1$ and $\beta_2$. The baths do not interact
with each other, only with the system.
Microscopically the system never interacts with the two baths at the
exact same time. Imagine therefore that we let the system run for only a very short time-interval
$[0,T]$. This interval is so short that the system interacts only with heat bath 1 within $[0,t]$, 
an only with heat bath 2 in $[t,T]$.  (A more elaborate discussion of such a process can be found in \cite{jar99,jar00}).
By its mathematical definition (\ref{sdef}), $S(\omega)$ is then just the sum of contributions
from $[0,t]$ and $[t,T]$. As the dynamics are detailed balanced 
in each time interval separately (there is interaction with only one heat bath at a fixed temperature), 
we can use the results of the last paragraph to see that
\[ S(\omega) = -\beta_1[U(x_{t})-U(x_{0})]   -\beta_2[U(x_{T})-U(x_{t})] = \beta_1 Q_1 + \beta_2 Q_2 \]
where $Q_1$ and $Q_2$ denote the heat fluxes into heat bath 1 and 2 respectively.
Therefore, $S(\omega)$ is the change in entropy of the environment due to the process of the system, 
i.e.  the entropy flux from the system into the environment.
Given an arbitrary time-interval, it is therefore reasonable to assume that we can split
this interval into many small intervals which are short enough so that we can assume the
dynamics to be detailed balanced in each interval separately: we assume detailed balance
`locally in time'. Adding all the contributions to $S(\omega)$ from these small intervals
gives us that
\[ S(\omega) = \beta_1 Q_1 + \beta_2 Q_2 \]
Using the same reasoning we can expand this formula to the case in which 
the system interacts with several heat baths and particle reservoirs.
The conclusion stays the same: the quantity $S(\omega)$ is 
the entropy flux from the system into the environment.\\

Finally, a system can be driven from equilibrium by some nonconservative
forcing. 
Let us illustrate this last class of systems with an example:

\paragraph{Particle in a box}
Consider the following example: the box in which the charged particle resides is 
surrounded by an environment at equilibrium at a single temperature (air or water).
The air inside the box is also at equilibrium at that temperature.  
Suppose the box  has the shape of a thin torus,
with a circumference $L$, which we divide into $N$ segments of length $d$
($d$ would be the error of our measurement).
The configuration space is thus a ring with $N$ sites, labelled $x=1,2,\ldots, N, N+1\equiv 1$.
We assume that the particle can maximally move one site to the left or right during one time step.
An electric field $E$ (constant in time) is applied on the box. As a consequence, the particle gains an energy
$E(x,y)\cdot d$ when going to the left or right ($y=x \pm 1$). Note that $E(x,y)=-E(y,x)$.

First of all, suppose that the electric field is of the form $E(x,y)\cdot d = U(y)-U(x)$,
meaning that $U(x)$ is the energy of the particle at site $x$. In this case the forcing
is conservative, and the dynamics of the system should satisfy the detailed balance condition:
\[ \frac{P(x,y)}{P(y,x)} = e^{-\beta[U(y)-U(x)]} = e^{\beta E(x,y)\cdot d} \]
If we restrict our attention to only two sites $x,y$, it is impossible so say
if the electric field is conservative or not.
For a nonconservative force we therefore assume that  detailed balance holds locally
(i.e. between each pair of states separately):
\[ \frac{P(x,y)}{P(y,x)} = e^{\beta E(x,y)\cdot d} \]
The quantity $S(\omega)$ is then given by (\ref{pbent}):
 \[S(\omega) = \sum_{n=0}^{m-1}\log\frac{P(x_{n},x_{n+1})}{P(x_{n+1},x_{n})} = \beta \sum_{n=0}^{m-1}E(x_{n},x_{n+1})\cdot d \]
which is the entropy production in the environment (the entropy flux into the environment).

In general, note that the word `conservative' is a global property. If one considers a
small enough region of the configuration space, one can always define
a local potential from which the force is derived. So restricting
our observation momentarily to this small region, we assume that
the dynamics of the system is detailed balanced (detailed balance locally in space), and still conclude that
$S(\omega)$ is the entropy flux into the environment. Returning to the
global picture, the sum of all these small regions gives then
the total entropy flux into the environment. The only difference
with a real detailed balanced dynamics is that this entropy flux
is no longer the difference of a state function, but depends on the whole path $\omega$.

\section{Average entropy}\label{traj-ae}
\index{entropy!average}
It turns out that we can also attach a physical meaning to $S_{\mu}(\omega)$, as defined in (\ref{moi}), 
when the average is taken. 
This means again an average over all possible trajectories $\omega = (x_t)_{0\leq t\leq T}$.
\[ \left<S_{\mu}\right>_{\mu_0} = \left<\log\frac{\mu_0(x_0)}{\mu_T(x_T)}\right>_{\mu_0} + \left<S(\omega)\right>_{\mu_0} \]
The second term on the right-hand side is the average entropy flux\index{entropy flux!average} into the environment. The first term can be
rewritten, using the fact that for any state function $f$, we have that $\left<f(x_t)\right>_{\mu_0}=\int dx \mu_t(x)f(x)$
(where the integral becomes a sum for Markov jump processes). We see that the average of $S_{\mu}$ becomes
\begin{equation}\label{avent}
 \left<S_{\mu}\right>_{\mu_0} = s(\mu_T)-s(\mu_0) + \left<S(\omega)\right>_{\mu_0}
\end{equation}
with $s(\mu)$ the Shannon entropy of the distribution $\mu$
\[ s(\mu) = -\int dx \mu(x)\log\mu(x) \]
(again up to a factor of $k_B$).
When $\mu$ is an equilibrium distribution, the Shannon entropy is equal to the Gibbs entropy, 
which is in equilibrium statistical mechanics the (physical) entropy of the system. 
Therefore we still call it here the entropy of the system, making
the average of $S_{\mu}$ equal to the average of the total entropy change
in the world due to the process of the system.\\

The average entropy change is positive. This is a direct consequence of its definition, 
\[ \left< \frac{d\mathcal{P}^{R}_{\mu_T}\theta}{d\mathcal{P}_{\mu_0}}(\omega)\right>_{\mu_0} = \int d\mathcal{P}_{\mu_0}(\omega)
\frac{d\mathcal{P}^{R}_{\mu_T}\theta}{d\mathcal{P}_{\mu_0}}(\omega) = 1 \]
and Jensen's inequality:
\[ 1 = \left<e^{-S_{\mu}}\right>_{\mu_0} \geq e^{-\left<S_{\mu}\right>_{\mu_0}} \]

\section{Entropy and traffic}\label{sec-eat}

In (\ref{pathprob}) we wrote the Radon-Nikodym derivative, which compares the path-probability
measures of two different dynamics. This defined an action $A(\omega)$.
Similarly, we can define an action for the time-reversed process:
\[
 \frac{d\mathcal{P}^{*,R}_{\mu_0}}{d\mathcal{P}^R_{\mu_0}}(\theta\omega) = e^{-A^R(\theta\omega)}
\]

As an immediate consequence of the local detailed balance assumption, we see that
\begin{equation}\label{exent}
  A^R(\theta\omega) - A(\omega) = S^*(\omega) - S(\omega) = S_{ex}(\omega)
\end{equation}
In words: the time-antisymmetric\index{action!time-antisymmetric part} part of the action $A$ is equal to
the difference in entropy fluxes in the two dynamics. We call it an
excess entropy flux\index{entropy flux!excess}. Here we clearly see the limits of local detailed balance:
it only specifies the time-antisymmetric part of the action. 
And, in contrast to equilibrium dynamics, the time-symmetric part
does play an important role in statistical physics out of equilibrium.
Therefore it has been proposed \cite{bmn07,dm07,mn07} to examine the following quantity
more closely\index{action!time-symmetric part}:
\[ \mathcal{T}_{ex}(\omega) = \mathcal{T}^*(\omega) - \mathcal{T}(\omega) = A^R(\theta\omega) + A(\omega) \] 
The quantity $\mathcal{T}(\omega)$ is called traffic\index{traffic} \cite{mn08,mnw08a,mnw08b},
and the time-symmetric part of the action is therefore an excess traffic\index{traffic!excess}.
The action $A$ can thus be written as
\[ A(\omega) = \frac{\mathcal{T}_{ex}(\omega)-S_{ex}(\omega)}{2} \]
and the Radon-Nikodym derivative:
 \begin{equation}\label{pathprob1}
 \frac{d\mathcal{P}^*_{\mu_0}}{d\mathcal{P}_{\mu_0}}(\omega) = e^{\frac{S_{ex}(\omega)-\mathcal{T}_{ex}(\omega)}{2}}
\end{equation}
The physical and operational meaning of traffic is up to now not completely clear. 
From the research discussed in this thesis, it seems to be of great importance in results of nonequilibrium statistical physics.

\section{General physical results}\label{sec:physres}

Using as an assumption only local detailed balance, a number of well-known
results can be very generally derived. We conclude
this chapter with these derivations.

\subsection{Fluctuation theorem}\label{sec:flucthm}
\index{fluctuation theorem}
A recent and celebrated result of out-of-equilibrium statistical physics is the fluctuation theorem for entropy fluxes \cite{ecm93,es94,gc95},
valid for systems with a time-independent dynamics:
\begin{equation}\label{fluctheorem}
\lim_{T\to\infty}\frac{1}{T}\log\left(\frac{P\left(S(\omega) = T\sigma\right)}{P\left(S(\omega) = -T\sigma\right)}\right)=\sigma
\end{equation}
where $P\left(S(\omega) = T\sigma\right)$ is the probability density that the entropy flux is equal to a value $T\sigma$. 
We use local detailed balance to relate entropy fluxes to the path-probability
measure. With this the derivation is not difficult.
 We first consider the probability density for $S_{\rho}(\omega)$ (started from the stationary distribution $\rho$), and we let the process
run for a finite time $T$.
\[ P\left(S_{\rho}(\omega) = T\sigma\right) = \int d\mathcal{P}_{\rho}(\omega) \delta\left(S_{\rho}(\omega) - T\sigma\right)   \]
where, as before, the integral is taken over all possible paths $\omega=(x_t)_{0\leq t\leq T}$. Of course, computing this distribution
is not possible in general, but we can rewrite it:
\begin{eqnarray*}
 P\left(S_{\rho}(\omega) = T\sigma\right) &=& \int d\mathcal{P}_{\rho}(\theta\omega)\frac{d\mathcal{P}_
{\rho}}{d\mathcal{P}_{\rho}\theta}(\omega) \delta\left(S_{\rho}(\omega) - T\sigma\right)\\
&=& e^{T\sigma}\int d\mathcal{P}_{\rho}(\theta\omega)\delta\left(S_{\rho}(\omega) - T\sigma\right)\\
\end{eqnarray*}
Where in the last line, we used the definition of $S_{\rho}(\omega)$ and the fact that the delta function
makes sure that $S_{\rho}(\omega) = T\sigma$. We now make a change of variables $\omega \to \theta\omega$,
which is an involution. Note that this changes $S_{\rho}$ to $-S_{\rho}$:
\begin{eqnarray*}
 P\left(S_{\rho}(\omega) = T\sigma\right)
&=& e^{T\sigma}\int d\mathcal{P}_{\rho}(\omega)\delta\left(S_{\rho}(\omega) + T\sigma\right)\\
&=& e^{T\sigma}P\left(S_{\rho}(\omega) = -T\sigma\right)
\end{eqnarray*}
This is what is called a fluctuation theorem for finite times for the quantity $S_{\rho}$.
However, the local detailed balance assumption only 
says that $S(\omega)$ is an entropy flux.
Without going into mathematical rigour, we do expect that, typically:
\[ \lim_{T\to\infty} \frac{S(\omega)}{T} = \lim_{T\to\infty} \frac{S_{\rho}(\omega)}{T} \]
because the difference between $S$ and $S_{\rho}$ is only in temporal boundary terms. 
With this we get the fluctuation theorem (\ref{fluctheorem}).
Note that it is possible that the boundary terms stay important even in the long time limit. In such
cases the fluctuation theorem is not valid, see \cite{rh08}.
\\

\subsection{Work relations}
\index{work relation}
Nonequilibrium work relations \cite{cro98,cro99,jar97} can be derived for a system connected to a single heat bath, with a time-dependent dynamics
parametrized by a parameter $\lambda_t$. This parameter is called the protocol\index{protocol}. Of course, because $\lambda_t$ changes in time,
the system is pulled out of equilibrium. Usually it is assumed that for each fixed value
$\lambda$ the dynamics is detailed balanced.  Actually, to have work relations, the dynamics only 
needs to be detailed balanced at the beginning and the end of the process.\\

We start at time zero with the system prepared in the equilibrium distribution
corresponding to the value $\lambda_0=A$. We then let the process run in the time-dependent dynamics until time $T$.
During the process there is local detailed balance, but at time $T$ there is again a detailed balanced dynamics
with $\lambda_T=B$.
Consider the following quantity:
\[ \frac{d\mathcal{P}_{\rho_A}}{d\mathcal{P}^{R}_{\rho_B}\theta}(\omega) = \frac{\rho_A(x_0)}{\rho_B(x_T)}e^{S(\omega)} \]
where $S(\omega)$ is by the local detailed balance assumption equal to $\beta$ times the heat flux $Q(\omega)$ into the environment .
Remember that with the time-reversal, also the protocol $\lambda_t$ changes to $\lambda_{T-t}$ (see Section \ref{traj-eat}),
hence the superscript $R$.
Substituting the explicit expression of the equilibrium distributions ($\rho_{A}(x)=\frac{1}{Z_{A}}e^{-\beta U(A,x)}$ and similarly
for $\lambda=B$), we get
\[ \frac{d\mathcal{P}_{\rho_A}}{d\mathcal{P}^{R}_{\rho_B}\theta}(\omega) = e^{\beta[F_A-F_B - U(A,x_0)+U(B,x_T)+Q(\omega)]}=
e^{\beta[-\Delta F  +W(\omega)]} \]
where $\Delta F = F_B-F_A = \frac{1}{\beta}[\log Z_A - \log Z_B]$ is the change of free energy between the equilibrium states corresponding to the values $A$ and $B$ of the parameter $\lambda$,
and $W$ is the work done on the system.\\

From this basic result a work relation\index{work relation!Crooks} can be derived, which was done for the first time by
Crooks \cite{cro98,cro99}: let $P(W=w)$ be the probability density that the work during the process is equal to $w$.
Then
\begin{eqnarray*}
 P(W = w) &=& \int d\mathcal{P}_{\rho_A}(\omega)\delta(W(\omega) - w)\\
&=& \int d\mathcal{P}^{R}_{\rho_B}(\theta\omega)e^{\beta[-\Delta F  +W(\omega)]}\delta(W(\omega) - w)\\
&=& e^{\beta[-\Delta F  +w]}\int d\mathcal{P}^{R}_{\rho_B}(\omega)\delta(W(\omega) +w)\\
&=& e^{\beta[-\Delta F  +w]}P^{R}(W=-w)
\end{eqnarray*}
where we defined $P^R(W=-w)$ as the probability density that the work during the reversed process 
is equal to $-w$, and we used that $W(\theta\omega) = -W(\omega)$. From this relation the following 
equality\index{work relation!Jarzynski} can easily
be derived (obtained by Jarzynski in \cite{jar97}):
\begin{eqnarray*}
\left<e^{-\beta W}\right>_{\rho_A} &=& \int dw P(W = w)e^{-\beta w}\\
 &=& \int dw e^{-\beta\Delta F}P^{R}(W=-w)\\
&=& e^{-\beta\Delta F}
\end{eqnarray*}
which relates nonequilibrium work-values to equilibrium free energies.
Again we see that these results are an immediate consequence of the local detailed balance assumption.

\subsection{McLennan formula}
\index{Mclennan formula}
The McLennan formula \cite{mcl59,mcl89} gives an approximation of the stationary measure for a 
dynamics that is close to detailed balance\index{detailed balance!close to}. For a system in contact with two
heat baths (particle reservoirs) close to detailed balance means that the difference in the temperatures 
(chemical potentials) is small, while for a system under the influence of a nonconservative
force this means that the force is small.
In general, we can parametrize the `distance' from detailed balance by a 
parameter $\epsilon$, (e.g. $\epsilon = T_2-T_1$ or the nonconservative force $F=\epsilon f$).
In the case that $\epsilon$ is small, one can expand the stationary distribution
around $\epsilon=0$. Up to first order in $\epsilon$ this gives the McLennan formula.\\

We derive the McLennan formula in the following way:
suppose that until time $t=0$ the system is detailed balanced ($\epsilon=0$), and at time $t=0$
the system has the corresponding equilibrium distribution $\mu_0=\rho^0$. At time zero, we drive the system
from equilibrium, parametrized by $\epsilon$. After a time $t$ the probability distribution
of being in a state $x$ is then given by
\[ \mu_t(x) = \left<\delta(x_t-x)\right>^{\epsilon}_{\rho^0} \]
where the average is an average over all possible paths in the interval $[0,t]$, computed in the dynamics with $\epsilon$ (hence the superscript).
To rewrite expectation values in the nonequilibrium system
into expectation values in the equilibrium system,
let us write down the Radon-Nikodym derivative (\ref{pathprob}):
\[ e^{-A(\omega)}=\frac{d\mathcal{P}^\epsilon_{\rho^0}}{d\mathcal{P}^0_{\rho^0}}(\omega) \]
so that the probability distribution becomes:
\[ \mu_t(x) = \left<\delta(x_t-x)e^{-A(\omega)}\right>^0_{\rho^0} \]
with the superscript $0$ denoting that the average is taken in the equilibrium process.
The expectation value of an observable $Q(\omega)$ in an equilibrium process, is the same as that of
the time-reversed observable $ Q(\theta\omega)$. This is a simple consequence of (\ref{equilibrium}).
This means that we have
\begin{eqnarray*}
  \mu_t(x) &=& \left<\delta(x_0-x)e^{-A(\theta\omega)]}\right>^0_{\rho^0}\\
&=& \left<\delta(x_0-x)e^{-S_{ex}(\omega)}e^{-A(\omega)}\right>^0_{\rho^0}\\
&=& \left<\delta(x_0-x)e^{-S_{ex}(\omega)}\right>^{\epsilon}_{\rho^0}
\end{eqnarray*}
where the excess entropy flux $S_{ex} = S_{\epsilon} - S_0$ is the difference in entropy fluxes in the nonequilibrium and the equilibrium system
(see (\ref{exent})).
One can see from the definition of $S_{ex}$ that it is zero when $\epsilon$ is zero. Moreover, in physical
systems (and certainly the ones we discuss in this thesis) , $S_{ex}$ is at least of first order in $\epsilon$:
$\lim_{\epsilon\to 0 } S_{ex}/\epsilon <\infty$. 
An expansion in $\epsilon$ gives
\[ \mu_t(x) = \rho^0(x)-\left<\delta(x_0-x)S_{ex}(\omega)\right>^0_{\rho^0} + o(\epsilon) = \rho^0(x)[1-\left<S_{ex}(\omega)\right>^0_{x}] + o(\epsilon) \]
where the last expectation is an average over all trajectories starting from the state $x$.
Up to first order in $\epsilon$ this formula gives the probability distribution of finding the system in state $x$
at an arbitrary time. Letting the time go to infinity, this will converge to the stationary state $\rho^\epsilon(x)$
corresponding to the nonequilibrium dynamics.

\subsection{Fluctuation-dissipation theorem in equilibrium}\label{sec-fdt}
\index{fluctuation-dissipation relation!in equilibrium}\index{fluctuation-dissipation theorem}
The equilibrium fluctuation-dissipation theorem \cite{cw51,kubo66} tells us how systems respond
to a small change in their Hamiltonian (energy), at least in the linear regime. The framework in which we work here
is almost the same as for the McLennan formula: up to time zero
the system enjoys an equilibrium dynamics with a Hamiltonian $H_0$, and at time zero it has relaxed to the corresponding
equilibrium distribution $\rho \propto \exp(-\beta H_0)$. At time zero an extra potential is added to the dynamics:
$H = H_0 - hV$, where $h$ is a small parameter. For finite times after this perturbation has been made, the system
is not yet in equilibrium. We want to compute the expectation values of observables in this perturbed (nonequilibrium)
system. We consider only observables that are state functions, denoting them by $Q(x)$. With a reasoning completely 
analogous to the one made for the McLennan formula (the only difference is that the observable $\delta(x_t-x)$ is replaced by $Q(x_t)$) one then finds that:
\[ \left<Q(x_t)\right>^h_{\rho^0} = \left<Q(x_0)e^{-S_{ex}(\omega)}\right>^h_{\rho^0} \]
where the superscript $h$ denotes that the averages are taken in the perturbed system, and $S_{ex}$
is the excess entropy flux, excess of the perturbed system versus the unperturbed system.
As the only difference between the two dynamics is the addition of the potential $hV$, this excess
must be equal to $\beta h[V(x_t)-V(x_0)]$.

In linear perturbation theory\index{linear perturbation theory} one is restricted to systems for which such
a small perturbation has only a small influence. One can therefore work up to linear order
in the small parameter $h$:
\begin{eqnarray*} 
\left<Q(x_t)\right>^h_{\rho^0} &=& \left<Q(x_0)\right>^0_{\rho^0} - \left<Q(x_0)S_{ex}(\omega)\right>^0_{\rho^0} + o(h)\\
&=&  \left<Q(x_0)\right>^0_{\rho^0} + \left<Q(x_t)S_{ex}(\omega)\right>^0_{\rho^0} + o(h)
\end{eqnarray*}
In the last step we have again used that in an equilibrium dynamics the expectations of observables 
and their time-reversals are the same. Using the explicit expression for the excess entropy flux, we get
\begin{equation}\label{fdt1}
  \left.\frac{\partial}{\partial h}\left<Q(x_t)\right>^h_{\rho^0}\right|_{h=0} = \beta\left<Q(x_t)[V(x_t)-V(x_0)]\right>^0_{\rho^0}
\end{equation}
which is the fluctuation-dissipation theorem in equilibrium. it is valid for any time $t$. 
The left hand side of this equation is called a response function\index{response function}.
It is the response of the expectation
value of an observable to an added potential and is thus equal to the correlation of the observable
with the extra entropy flux created by the potential.\\

The fluctuation-dissipation theorem is useful because it gives a relation between two quantities
in essentially different processes. One can e.g. determine the response of a system without actually perturbing it.
Part of the results in this thesis are about the generalization of this relation
to nonequilibrium systems.

\cleardoublepage
\chapter{Markov jump processes}
\index{Markov jump process}
\textit{Markov jump processes are Markov processes where the configuration changes discretely.
This means that in a finite time-interval, the configuration changes a finite number of times.
In this thesis we only work with discrete configuration spaces where Markov jump processes are concerned.
Such processes are widely used in physical modelling because they are relatively simple, but approximate
real physical systems quite accurately. The discrete configuration space can arise as a result of a discrete approximation
of a continuous space, or because of quantum mechanical principles
(e.g. spins in a magnetic field). We give a brief introduction to Markov jump processes
here, focusing on the properties relevant for the rest of the thesis. For a thorough
introduction, we refer to \cite{gs01}. As a motivation, we start with a very physical example.}

\section{An introductory example}

\begin{figure}[ht]
\begin{center}
\includegraphics[width=8cm]{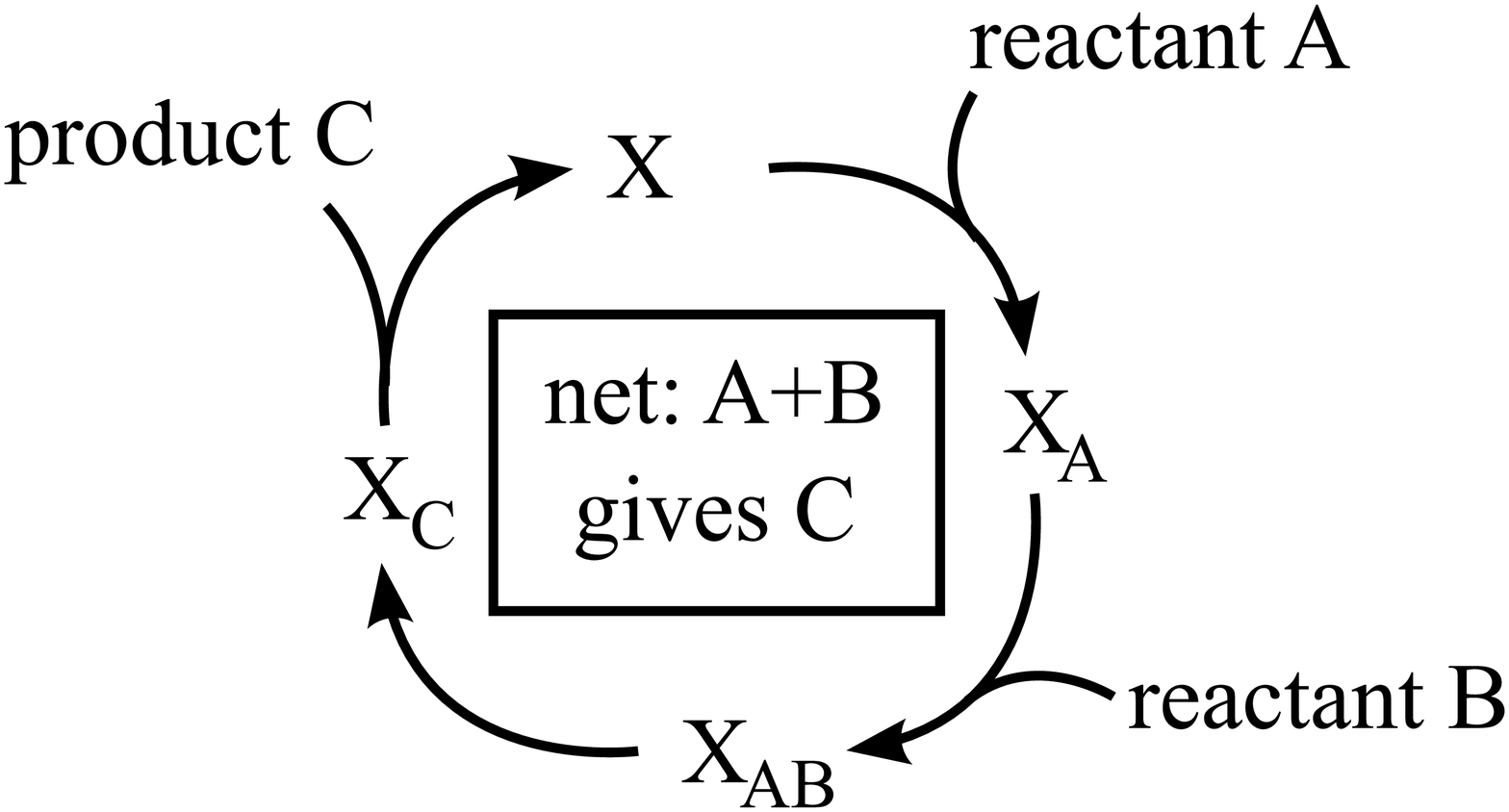}
\end{center}
\caption{A catalytic reaction cycle. \label{fig:catalyst}
}
\end{figure}

Imagine a chemical reaction where two reactants $A$ and $B$ react to form a product $C$.
The reaction is facilitated by the use of a catalyst $X$ as follows: first the molecule $A$
binds to the catalyst, forming a new molecule $X_A$. Then $B$ binds, forming $X_{AB}$. When bound to $X$, $A$ and $B$ react and form 
 $X_C$. Finally $C$ decouples from the catalyst, leaving the catalyst in its original state $X$, see Figure \ref{fig:catalyst}.

The details of the reaction $A+B\to C$ thus depend strongly on the reaction cycle of the catalyst $X$.
Apart from measuring the average speed or rate of such a cycle, we are also interested in more
detailed information. What are the fluctuations from this average speed? Does the inverse cycle also occur,
and with what probability? For such questions a more detailed model of the reaction is required.
As it is impossible to predict all the precise positions and velocities of all the molecules,
a Hamiltonian description is not wanted. Therefore we use a stochastic model. One can imagine
all the molecules involved in the reactions are `wandering' in some solution. A reaction can occur
when two molecules meet each other. The speed or rate
of each step in the reaction cycle then depends on parameters as concentrations of the molecules
in the solvent, temperature, and of course also on the probability that, given that two molecules
meet, they actually react. If the concentrations of the molecules in the solvent are not too big, 
it is a rather good assumption to think of this process as being Markovian.
Indeed, we assume that a reactant molecule makes many collisions with the solvent molecules before
meeting with another reactant, thus effectively losing its `memory' on a timescale that is much smaller
than we are interested in: the timescale of the successive reactions. Also, this allows us to treat
the molecules of the catalyst as independent particles.

To make a model of the reaction cycle, we therefore take one catalyst molecule as the system of interest.
This molecule can be in four different chemical states: $X,X_A,X_{AB}$ and $X_C$. At any time,
regardless of its history,
the molecule can change its state as a consequence of a reaction.
We thus arrive at a Markov process on a finite configuration space, but
in continuous time. This fits exactly in the framework of Markov jump processes.

This is only one example where Markov jump processes provide a good model. Other examples 
are traffic jams, transport of ions through nanotubes, the Ising model and other models
of spins in a magnetic field, population dynamics in ecological systems, etc.
Providing important and relevant models in physics, Markov jump processes are used
throughout this thesis. In this chapter we therefore introduce the aspects
relevant for the rest of this thesis. For a thorough
introduction, we refer to \cite{gs01}. At the end of this chapter we return to this model
and see how to describe it as a Markov jump process.

\section{Definition}
We work here with a discrete, even finite, set $\Omega$ of configurations $x,y,\ldots$.
The dynamics of a Markov jump process is defined as follows: let $P^{\Delta t}(y|x,t)$ denote
the probability that the system changes (jumps) from configuration $x$ to $y\neq x$ within $[t,t+\Delta t]$,
given that the system was in configuration $x$ at time $t$. Then 
\begin{equation}\label{jumpinf1}
 P^{\Delta t}(y|x,t) = k_t(x,y)\Delta t + o(\Delta t)
\end{equation}
The $k_t(x,y)$ are called transition rates\index{transition rate}. 
Obviously $k_t(x,y)\geq 0$. Also, as a convention, we take $k_t(x,x)=0$.
From (\ref{jumpinf1}) one also sees that the probability to jump
two times within a time span $\Delta t$ is of order $o(\Delta t)$.
Logically, the probability $P^{\Delta t}(x|x,t)$  that the system does not jump within $[t,t+\Delta t]$ is then
\begin{equation}\label{jumpinf2}
 P^{\Delta t}(x|x,t) = 1 - \lambda_t(x)\Delta t + o(\Delta t)\ \ \ \ \textrm{with}\ \ \ \ \lambda_t(x) = \sum_{y\neq x}k_t(x,y)
\end{equation}
This $\lambda_t(x)$ is called the escape rate\index{escape rate}, as it quantifies the probability that the system `escapes' from $x$.
We assume that
$\epsilon \leq \lambda_t(x) < \lambda$ for all $t,x$ and for some $\epsilon >0, \lambda<\infty$, to
make sure that the expected time for the system to jump is finite.\\

\begin{figure}[h]
\begin{center}
\includegraphics[width=5.2cm]{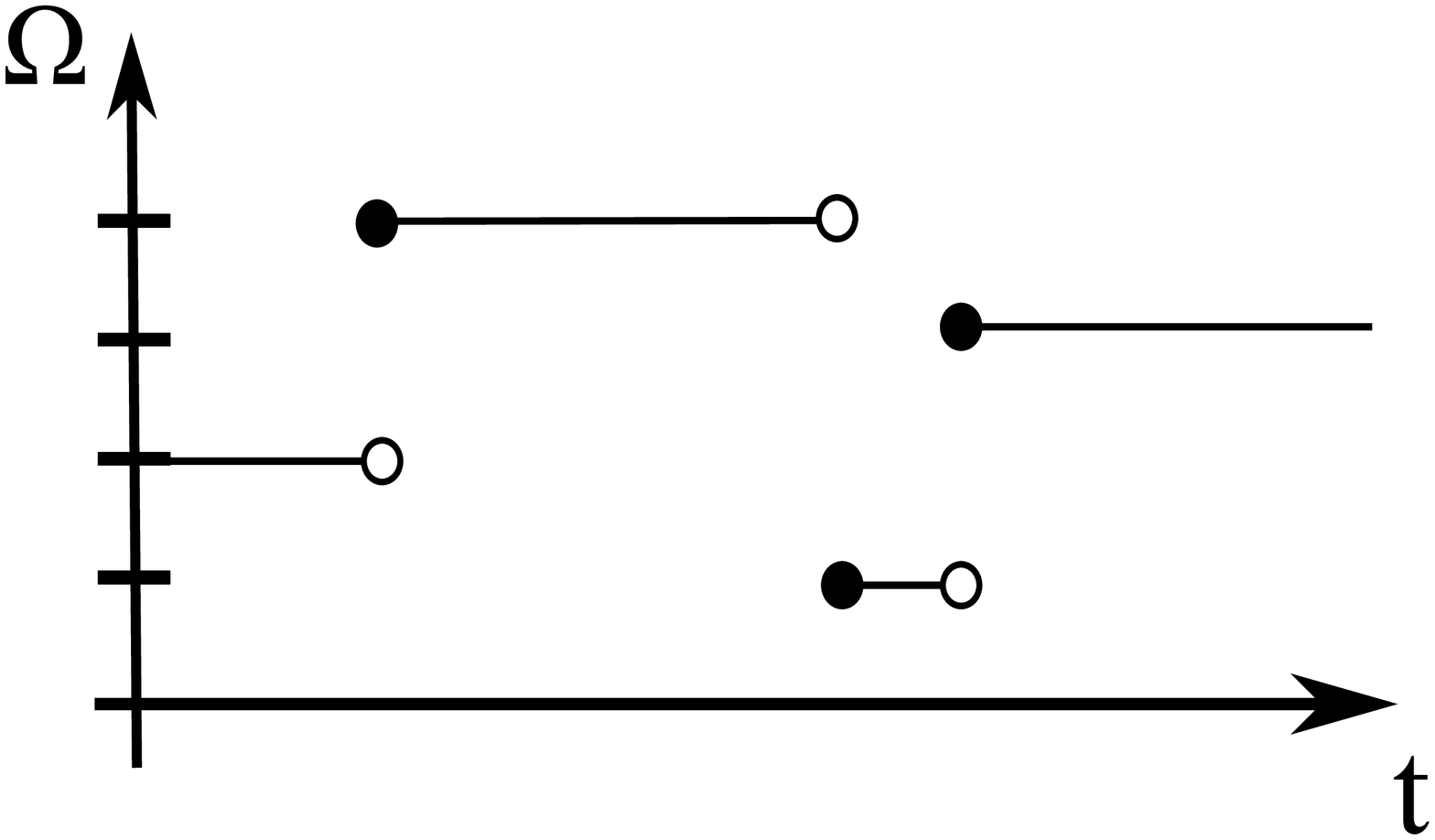}
\end{center}
\caption{A realization of a Markov jump process.\label{fig:markovjump}
}

\end{figure}

In Figure \ref{fig:markovjump}, an example
of a typical realization of a Markov jump process is shown: the system stays in a certain configuration
for an exponentially distributed time (we prove this later on in this chapter), and then jumps to the next state.
We adopt the convention that at the 
jump time, the system is in the state after the jump, i.e. in Figure \ref{fig:markovjump} the graph
is right-continuous.\\

\section{Probability distributions and the Master equation}

Suppose that we know the probability that the system is in state $x$ at time $t$: $\mu_t(x)$, for all $x$.
From this we can compute the probability that the system is in any state $x$ at time $t+\Delta t$. It is equal to the probability
that the system was in state $x$ at time $t$ and stayed there, plus the probability that it was in any state $y$
and jumped to $x$. With (\ref{jumpinf1}) and (\ref{jumpinf2}) it is clear that
\[ \mu_{t+\Delta t}(x) = \mu_t(x)[1-\lambda_t(x)\Delta t + o(\Delta t)] + \sum_{y\neq x}\mu_t(y)[k_t(y,x)\Delta t + o(\Delta t)] \]
This gives us a differential equation for $\mu_t$:
\begin{equation}\label{master}
  \frac{d\mu_t(x)}{dt} = \lim_{\Delta t \to 0}\frac{\mu_{t+\Delta t}(x) - \mu_t(x)}{\Delta t} = \sum_{y\neq x}[\mu_t(y)k_t(y,x)-\mu_t(x)k_t(x,y)]
\end{equation}
This equation is called the Master equation\index{Master equation}. Notice that it is a deterministic equation: all uncertainty is in the probability distribution
$\mu_t$, while the evolution of $\mu_t$ through time is deterministic.\\

The quantity $j_{\mu_t}(x,y) = \mu_t(x)k_t(x,y)-\mu_t(y)k_t(y,x)$ can be seen as the current between $x$ and $y$, or better even: 
a probability current\index{probability current}.
Indeed, the first term is the probability per unit of time that the system makes a jump from $x$ to $y$, while the second term
is the probability per unit of time that the system makes a jump from $y$ to $x$. With this definition of current we can rephrase the
Master equation as a continuity equation\index{continuity equation!for Markov jump process}:
\begin{equation}\label{master2}
  \frac{d\mu_t(x)}{dt} + \sum_{y\neq x}j_t(x,y) = 0
\end{equation}

\section{Stationarity and detailed balance}
\index{stationarity!for Markov jump process}
In the special case that the transition rates of the Markov process are independent of time ($k_t(x,y) = k(x,y)$),
we call the process time-homogeneous. As said before, for time-independent dynamics we assume that there 
exists a stationary distribution $\rho$, which is unique, and that all distributions $\mu_t$ converge to it in the long time limit.
This $\rho$ solves the Master Equation (\ref{master}) with the left-hand side zero. 
The conditions under which this assumption is valid are conditions of irreducibility: 
a jump process is called irreducible\index{irreducibility} if the probability of the system to go from 
any state $x$ to any other state $y$ in a finite time is non-zero. This is equivalent to saying there is a chain of states
$x=x_1,x_2,\ldots,x_n=y$ such that $k(x_i,x_{i+1})>0$ for all $i=1$ to $n$. The proof of this is beyond the scope of this text.\\

Finding the stationary distribution is difficult in general. There is however a special case in which computations simplify: if
there exists a function $f(x)$, ($f\neq 0$ and $f(x)\geq 0$), such that
\begin{equation}\label{detbal}
  f(x)k(x,y) = f(y)k(y,x)
\end{equation}
then $\rho(x)\propto f(x)$ solves the stationary Master Equation. If the transition rates satisfy this property,
then the process is said to satisfy detailed balance, and $\rho$ is called the equilibrium distribution. 
Once the system has relaxed to this distribution, we then also 
have that $j_{\rho}(x,y) = 0$, so that in equilibrium all currents between different states are zero.
This definition of equilibrium coincides with the one given in the previous chapter.



\section{Finite time probabilities}
To have a better idea of how the system evolves from configuration to configuration, we compute the following
probability: for any $t>s$, let $P(y,t|x,s)dt$ be the probability that the system remains in configuration $x$
until it jumps within $[t,t+dt]$ to state $y$, given that the system was in configuration $x$ at time $s$.
Then
\begin{equation}\label{jumpprob}
 P(y,t|x,s)dt = k_t(x,y)e^{-\int_s^t\lambda_u(x)du}dt
\end{equation}
We prove this as follows: divide the interval $[s,t]$ into $n$ pieces of length $\Delta t = \frac{t-s}{n}$.
Then from (\ref{jumpinf1}) and (\ref{jumpinf2}) we see that
\begin{eqnarray*}
 P(y,t|x,s)dt &=& k_t(x,y)dt\prod_{k=0}^{n-1}\left[1-\lambda_{s+k\Delta t}(x)\Delta t + o(\Delta t)\right]\\
&=&  \exp\left\{ \sum_{k=0}^{n-1}\ln \left[1-\lambda_{s+k\Delta t}(x)\Delta t + o(\Delta t)\right] \right\} k_t(x,y)dt\\
&=& \exp\left\{ \sum_{k=0}^{n-1}\left[-\lambda_{s+k\Delta t}(x)\Delta t + o(\Delta t)\right] \right\} k_t(x,y)dt
\end{eqnarray*}
If we now let $n\to\infty$, we get exactly (\ref{jumpprob}).

Let us examine this probability: the probability to jump to anywhere within\\ $[t,t+dt]$,
given that the system was in state $x$ at time $s$ is
\[ P(t|x,s)dt = \sum_yP(y,t|x,s)dt = \lambda_t(x)e^{-\int_s^t\lambda_u(x)du}dt \]
On the other hand, given that the system jumps at time $t$, the probability that it jumps to state $y$ is then
\[ p_t(x,y) = \frac{P(y,t|x,s)dt}{P(t|x,s)dt} = \frac{k_t(x,y)}{\lambda_t(x)} \]
The $p_t(x,y)$ are called the transition probabilities, and we see that $\sum_{y\neq x}p_t(x,y)=1$.
In conclusion, when
at time $s$ the system is in a state $x$, it stays there
for a time $t-s$ that is exponentially distributed, determined by the escape rate $\lambda_t(x)$. When it jumps, the probability
that it jumps to a state $y$ is $p_t(x,y)$.

\section{Trajectories}

A trajectory $\omega = (x_t)_{0\leq t\leq T}$
is completely described by giving the consecutive states $x_i$ (with $i=0,\ldots n$) the system visits plus the times $t_i$
at which it jumped:
\[ x_t = x_i \ \ \ \ \textrm{for }\ \  t_{i}\leq t < t_{i+1} \]
where $t_0=0$ and $t_{n}$ is the last jump time before time $T$.
Using (\ref{jumpprob}), we can compute the probability measure of such a path\index{path-probability!for Markov jump process}:
\[
 d\mathcal{P}_{\mu_0}(\omega) = \mu_0(x_0)\prod_{i=0}^{n-1}\left[k_{t_{i+1}}(x_{i},x_{i+1})e^{-\int_{t_{i}}^{t_{i+1}}
\lambda_u(x_{i})du}dt_{i+1}\right]e^{-\int_{t_{n}}^{T}\lambda_u(x_{n})du}
\]
The last factor on the right-hand side is the probability that the system does not jump in the interval $[t_n,T]$.
We rewrite this path-probability measure in a slightly more elegant form:
\newpage
\begin{eqnarray}
d\mathcal{P}_{\mu_0}(\omega) = \mu_0(x_0)\exp\left\{\sum_{i=0}^{n-1}\log k_{t_{i+1}}(x_{i},x_{i+1})-\int_{0}^{T}
\lambda_t(x_{t})dt\right\}dt_1\ldots dt_n\nonumber\\
\label{pmjp}
\end{eqnarray}
With this path-probability one can then compute expectation values, as defined in (\ref{expectation}),
and the time-evolved probability distribution of configurations $\mu_t$, as defined in (\ref{mudef}).
This definition of $\mu_t$ is completely consistent with the Master equation, of course.\\


\section{The Girsanov formula}


Suppose now that one has a different Markov jump process with rates $k^*(x,y)$ and one wants to calculate the expectation values of 
observables in this process. Then one can relate this to the expectation values in the original process by 
using the Radon-Nikodym derivative (see (\ref{rnexp}) and (\ref{pathprob})). The demand of absolute continuity
boils down to the demand that $k^*(x,y)\neq 0$ for any pair $x,y$ for which $k(x,y)\neq 0$.
It follows immediately from (\ref{pmjp}) that
\begin{equation}
 e^{-A(\omega)} = \exp\left\{ \sum_{i=0}^{n-1}
\ln\left(\frac{k^*_{t_{i+1}}(x_{i},x_{i+1})}{k_{t_{i+1}}(x_{i},x_{i+1})}\right) +\int_0^Tdt[\lambda_t(x_t)-\lambda_t^*(x_t)]\right\}
\end{equation}
with the action defined as in (\ref{pathprob})
This is often schematically written as\index{Girsanov formula!for Markov jump process}
\begin{equation}\label{girsanov}
 A(\omega) =  \int_0^Tdt[\lambda^*_t(x_t)-\lambda_t(x_t)] -\sum_{t\leq T}
\ln\left(\frac{k^*_{t}(x_{t^-},x_{t})}{k_{t}(x_{t^-},x_{t})}\right)
\end{equation}
where the sum is over jump times, and $t^-$ is the time just before the jump. This formula (\ref{girsanov}),
the Radon-Nikodym derivative for Markov jump processes,
is called the Girsanov formula, and is of much use throughout this text. More details can be found in Appendix
2 of \cite{kl99}.

\section{Local detailed balance}
\label{sec-ldb}
The Markov jump processes as described up to now are little more than mathematical models.
To make a connection with physics we make the local detailed balance assumption\index{local detailed balance!for Markov jump process}:
the quantity $S(\omega)$ that was defined in (\ref{sdef}), is equal to the entropy
flux into the environment during the trajectory $\omega$. For Markov jump processes
this becomes (see (\ref{pmjp})):
\[S(\omega)=\log\left(\frac{d\mathcal{P}_{x_0}}{d\mathcal{P}^{R}_{x_T}\theta}(\omega)\right) = \sum_{t\leq T}\log \frac{k_t(x_{t^-},x_t)}{k_t(x_t,x_{t^-})}\]
We see that there is only an entropy flux when there is a jump, and the total entropy
flux is just the sum of entropy fluxes for each jump:
 \[ \frac{k_t(x,y)}{k_t(y,x)} = e^{\sigma_t(x,y)}  \]
where $\sigma_t(x,y)$ is the entropy flux associated to one jump from a state $x$ to a state $y$.\\

\paragraph{Example 1:}
As a first example we imagine a system with time-homogeneous dynamics, immersed in a
 single heat bath at inverse temperature $\beta$. We denote the energy of the system,
(which depends on the configuration $x$) as $U(x)$. On top of that there is a
forcing,  meaning that for every jump from a state $x$ to a state $y$, there is an amount of work $W(x,y)$ needed.
Note that by definition $W(x,y)=-W(y,x)$.  
Using the first law of thermodynamics, we see that
\begin{equation}\label{ldbmj}
  \frac{k(x,y)}{k(y,x)} = e^{-\beta[U(y)-U(x)-W(x,y)]}
\end{equation}
Indeed, local detailed balance allows us to express a part of the transition rates
in terms of physical, measurable quantities.

If there exists a state function $V$ such that for all $x,y$ we have
$ W(x,y) = V(y) - V(x) $, then the forcing is conservative, and by (\ref{ldbmj})
the rates are detailed balanced with the corresponding equilibrium distribution
$\rho(x) = \frac{1}{Z}\exp\{-\beta[U(x)-V(x)]\}$.

\paragraph{Example 2:}

As a second example think of two big particle reservoirs with different particle densities.
Between these reservoirs there is a wall, with a narrow channel going through it
from one side to the other. The channel is so narrow that particles are effectively restricted to one dimension.
This is what happens for example in ionic transport through cell membranes. Because of the difference
in particle densities a current arises towards the side with the lower density.

\begin{figure}[h]
\begin{center}
\includegraphics[width=8.2cm]{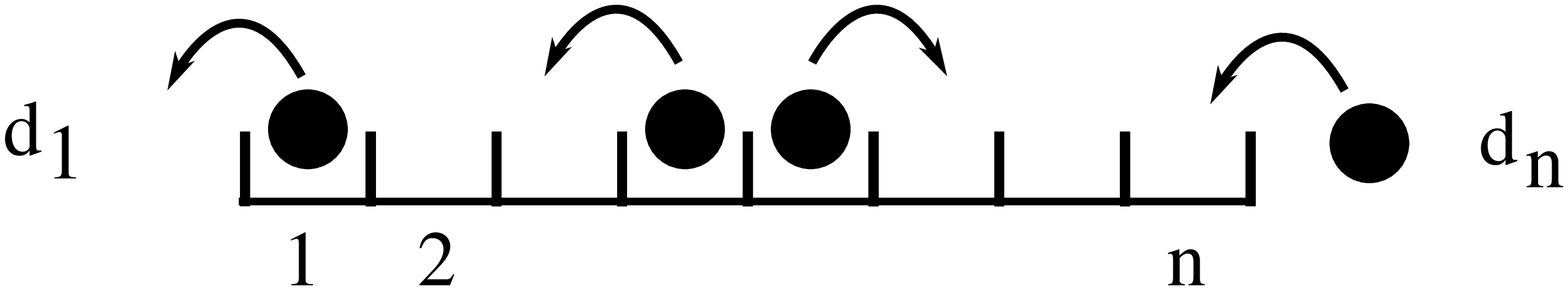}
\end{center}
\caption{A visualization of an exclusion process.\label{fig:exclusion}
}
\end{figure}

We model the channel as a set of sites labelled by $i=1,\ldots,N$. On each site there can either
be one or no particle. A configuration of the system is thus an array $x$, with empty sites $x_i=0$ and particles
$x_i=1$. Each particle can hop to the neighbouring site left or right if that site is unoccupied. Such a process
is called an exclusion process\index{exclusion process}. Hence the only transitions
allowed are $x\to x^{(ij)}$ where $j= i\pm 1$ and
\begin{equation}
x^{(ij)}_k = \left\{
\begin{array}{rll}
 x_k & \textrm{if }& i\neq k\neq j\\
 x_j & \textrm{if }& k = i\\
x_i & \textrm{if } & k=j
\end{array} \right.
\end{equation}
An exception occurs at the boundaries of the system $i=1,N$, where particles can enter and leave the system.
Hence transitions $x\to x^{(i)}$ are also allowed, with $i=1,N$ and
\begin{equation}
x^{(i)}_k = \left\{
\begin{array}{rll}
x_k & \textrm{if }& k\neq i\\
1-x_k & \textrm{if }& k=i\\
\end{array} \right.
\end{equation}
This means that at the left and right boundary the system is in contact with
particle reservoirs, which are characterized by a temperature and a chemical potential.
In Figure \ref{fig:exclusion} an example of such a process is visualized.
This is an example of Kawasaki dynamics\index{Kawasaki dynamics}.\\

Physically, we assume here that there is no interaction between the particles, except
for the exclusion. We take therefore the energy of the system zero. The particle number
of this system is not conserved however. If both reservoirs have the same inverse temperature
$\beta$ and chemical potential $\mu$, the system is detailed balanced,
and for any two states $x,y$:
\[ \frac{k(x,y)}{k(y,x)} = e^{\beta\mu[n(y)-n(x)]} \]
with $n(x)$ the number of particles in configuration $x$. We see from this that when a particle
hops within the system, the transition rate is symmetric: $k(x,x^{(ij)})=k(x^{(ij)},x)$. 
Detailed balance gives us as equilibrium distribution
\[ \rho(x) = \frac{e^{\beta\mu n(x)}}{(1+e^{\beta\mu})^N} \]
and the average particle density $d$
\[ d = \sum_x \rho(x)\frac{n(x)}{N} = \frac{e^{\beta\mu}}{1+e^{\beta\mu}} \]
This $d$ is the particle density of the particle reservoir, as it is equal to 
the density of the system when in equilibrium.

If the chemical potentials of the two reservoirs are not the same, we use local detailed balance.
We get our inspiration from the detailed balance case
to write
\begin{equation}\label{kk}
  \frac{k(x,x^{(i)})}{k(x^{(i)},x)} = e^{\beta\mu_i[n(x^{(i)})-n(x)]}
\end{equation}
for events where a particle enters or leaves the system ($i=1,N$),
and $k(x,x^{(ij)})=k(x^{(ij)},x)$ for events where a particle hops within the system.
Again inspired by the detailed balance case we define the particle densities of the reservoirs as
\[ d_i = \frac{e^{\beta\mu_i}}{1+e^{\beta\mu_i}} \]

Finding the stationary distribution in this case is possible, but certainly not trivial.
We refer to \cite{der07} for more details. 
We can say something about the current though. For this we use the fluctuation theorem
discussed in Section \ref{sec:flucthm}. For any specific $\omega$ we can write, using (\ref{kk})
\[ S(\omega) = \beta T\mu_1J_1(\omega) + \beta T\mu_NJ_N(\omega) \]
where the current $J_i(\omega)$ is the net number of particles per unit of time that has gone into the system through
site $i$. Note that the sum of these two currents gives the change of particle number
in the system per unit of time: $J_1(\omega) + J_N(\omega) = [n(x_T)-n(x_0)]/T$.
Therefore:
\[ \lim_{T\to\infty} [J_1(\omega) + J_N(\omega) ] = 0 \]
In the limit of large $T$ we therefore just use $J = J_1 = -J_N$.
The fluctuation theorem (\ref{fluctheorem}) can then be reformulated as
\[ 
\lim_{T\to\infty}\frac{1}{T}\log\left(\frac{P\left(J(\omega) = j\right)}{P\left(J(\omega) = -j\right)}\right)= \beta j (\mu_1 - \mu_N)
\]
The local detailed balance assumption on which the fluctuation theorem is based thus predicts what is the most probable
direction of the particle current. Not surprisingly the most probable direction of the current is towards the lower chemical potential.
The fluctuation theorem asserts that the reverse current is exponentially less probable. 

\paragraph{Example 3:}

Let us return to the introductory example from the beginning of this chapter. 
More precisely: $\Omega$ consists
of $4$ chemical states , labelled by $x=1,\ldots,4$ and lying on a ring, i.e. $5 \equiv 1$.
These states correspond to: $1\equiv X$, $2\equiv X_A$, $3\equiv X_{AB}$ and $4\equiv X_C$. 
We can see that $k(x,y)$ only differs from zero for $|y-x| = 1$.
Let us denote by $U(x)$ the energy of the molecule when it is in chemical state $x$,
and $\mu_i$ and $N_i$ are the chemical potentials and particle numbers of the reactant particles
in the solvent, where $i=A,B,C$. These reactant particles are seen as the reservoirs of
our catalyst molecule. As we imagine the reservoirs to be in equilibrium, their chemical
potentials are fixed. The entropy change in the environment for a transition $x\to y$ can thus be written
as \[ \sigma(x,y) = -\beta [U(y)-U(x)] -\beta\sum_i\mu_i\Delta N_i \]
where $\Delta N_i$ is the change in particles of species $i$. For example, if the catalyst molecule
makes a transition $X\to X_A$, we get as entropy flux to the environment:
\[ \log\frac{k(1,2)}{k(2,1)} = \sigma(1,2) = -\beta [U(2)-U(1)] + \beta\mu_A \]
It is the chemical potentials of the `reservoirs' that drive the system from equilibrium.
To show this, we take as trajectory one reaction cycle $X\to X_A \to X_{AB} \to X_C \to X$.  
We can compute the entropy flux into the environment during this cycle:
\[ \sum_{i=1}^4\sigma(i,i+1) = \beta[ \mu_A +\mu_B - \mu_C] \]
For an equilibrium dynamics this entropy flux should be equal to zero (see (\ref{eqdyn})).
Whenever $\mu_C \neq \mu_A + \mu_B$, the probability that the catalyst covers the reaction cycle
in one direction is not equal to the probability for the other direction. In fact,
when $\mu_C < \mu_A + \mu_B$ the reaction cycle will on average go in the direction as in Fig. \ref{fig:catalyst}.

\selectlanguage{english}
\cleardoublepage
\chapter{Diffusions}

\textit{Langevin dynamics (diffusions) are well-known and widely used as a combination of
Hamiltonian dynamics and stochastic noise. Mathematically they are more difficult to define than Markov
jump processes, but their physical meaning is much more clear, as the dynamics
are expressed in terms of measurable quantities as forces and positions, or electric fields
and currents. One
distinguishes in diffusions between the underdamped (inertial) regime and the overdamped regime.
In the latter accelerations are ignored due to high friction.  In this
chapter we briefly introduce diffusions, treating both the underdamped and the overdamped case.
For the interested reader, we suggest \cite{ckw04,kry95,mar04,mor69}.}

\section{The Langevin equation}\label{sec-lang}

For didactic purposes we start with the simplest case of one test-particle
in one dimension. The state space is here the phase space\index{phase space}, 
meaning the space of the possible positions and velocities of the particle.  Pure Hamiltonian dynamics would give two equations of motion:
\begin{eqnarray}
 dx_t &=& v_t dt\nonumber\\
 mdv_t &=& F_t(x_t)dt
\end{eqnarray}
which correspond to a particle that experiences a deterministic force $F_t$.
Note that $m$ is the mass of the particle, $x_t$ its position at time $t$ and $v_t$ its velocity.
Now we put the particle in a heat bath (imagine a glass of water) at inverse temperature $\beta$.
The particles of the heat bath will start colliding with our test-particle. Because we do not know
all the exact movements of these particles we model the effect of these collisions as a random force
on the test-particle. The modified equations then look like\index{diffusions!underdamped} this\index{Langevin equation!underdamped}:
\begin{eqnarray}\label{langevin}
 dx_t &=& v_t dt\nonumber\\
 mdv_t &=& F_t(x_t)dt -m\gamma v_tdt + \sqrt{2\mathcal{D}}dB_t
\end{eqnarray}
and are called the Langevin equations.
The term $-m\gamma v_t$ is the frictional force that the test-particle feels, with $\gamma$ the friction coefficient\index{friction coefficient}. 
Indeed, when the particle has a velocity,
then it has more head-on collisions than from behind, with a net effect equal to $-m\gamma v_t$.
This friction force is an average effect of the heat bath. That is why still another term is added, namely
$\sqrt{2\mathcal{D}}dB_t$, with $\mathcal{D}$ the diffusion coefficient\index{diffusion coefficient}. 
This should model the random kicks the particle receives. 
$B_t$ is what is called a Wiener
process\index{Wiener process}. It is a random variable with the property $B_t-B_s \sim \mathcal{N}(0,t-s)$, 
meaning that it has independent increments
with a Gaussian distribution with mean zero and variance $t-s$. This is the reason that $dB_t$ is usually called a standard Gaussian white 
noise\index{standard Gaussian white noise}. The random variable $\xi_t = \frac{dB_t}{dt}$ is often used here, but
mathematically ill-defined. It also has mean zero, but has the property that
 $\left<\xi_s\xi_t\right> = \delta(t-s)$. This means that we have approximated the physical process of collisions with
particles of the bath, as a succession of infinitely many, infinitely big kicks. As a consequence,
$dv_t$ is not really well-defined. Still the equations (\ref{langevin}) are mathematically well-defined 
if one integrates them with respect to time. This is not done here because the Langevin equations
do have a clear physical meaning in this form (\ref{langevin}).\\  

An important remark for things to follow is that we interpret $dv_t$ in the Langevin equation as $v_{t+dt}-v_t$, and $dx_t$ 
analogously. This is called the It\^o-interpretation\index{It\^o-interpretation}. It is also true that the inverse temperature $\beta$
of the heat bath has not yet entered the description. We come back to this at the end of this chapter,
when we discuss the local detailed balance assumption.

\section{The Fokker-Planck equation and stationarity}

We sample the initial positions and velocities (at time zero) from an initial probability density 
$\mu_0(x,v)$. One can then derive (see Appendix \ref{app-c}) from the Langevin equation  an evolution equation for the densities
at later times, which is called the Fokker-Planck equation\index{Fokker-Planck equation!underdamped}:
\begin{equation}\label{fokpla}
 \frac{\partial \mu_t}{\partial t} = -v\frac{\partial \mu_t}{\partial x} - 
\frac{\partial}{\partial v}\left[(\frac{F_t}{m}-\gamma v)\mu_t\right] + \frac{\mathcal{D}}{m^2}\frac{\partial^2 \mu_t}{\partial v^2}
\end{equation}
Just as in the case of Markov jump processes, this evolution equation is deterministic, and we can rewrite
it in the form of a continuity equation\index{continuity equation!for underdamped diffusions}:
 \begin{equation}\label{fokplacont}
 \frac{\partial \mu_t}{\partial t} + \nabla\cdot j_{\mu_t} = 0
\end{equation}
with a probability current\index{probability current!for underdamped diffusions} defined by
\begin{equation}\label{probcur}
j_{\mu_t} =  \left(v\mu_t,  
(\frac{F_t}{m}-\gamma v)\mu_t - \frac{\mathcal{D}}{m^2}\frac{\partial \mu_t}{\partial v}\right)
\end{equation}
When the force is time-independent, we assume as always that there is a stationary distribution 
$\rho$,\index{stationarity!for underdamped diffusions}
(which solves the Fokker-Planck equation with left-hand side zero). Unfortunately, it is often impossible to just solve the
Fokker-Planck equation.\\

In the special case however that we have $F(x) = - \frac{\partial U(x)}{\partial x}$,
then the solution is immediately found to be
\[ \rho(x,v) = \frac{1}{Z}e^{-\beta[U(x)+\frac{mv^2}{2}]} \]
where $\beta = \frac{m\gamma}{\mathcal{D}}$. Indeed, in this special case the particle is in equilibrium
with the surrounding fluid at inverse temperature $\beta$ and has the well-known Gibbs distribution, giving the probability of a 
state $(x,v)$ as the exponential of minus $\beta$
times the energy of the state. There is one difference with
the Markov jump case, namely that the probability current in equilibrium is not zero here:
\[ j_{\rho} =  \left(v\rho,  
-\frac{1}{m}\frac{\partial U}{\partial x}\rho\right) \]
On the other hand, this is equal to the current one would have in a purely Hamiltonian (deterministic) dynamics,
so one could say that in equilibrium, the diffusive part of the current is zero.

\section{Trajectories}\label{sec-trajud}

The position and velocity of the particle evolve according to the Langevin equation, and thus
follow a trajectory $\omega$ through phase-space: $\omega = (x_t,v_t)_{0\leq t\leq T}$.
One can compute the probability of such a path. We give a heuristic derivation of this,
for a mathematically rigorous treatment one should consult \cite{ls78}.\\

Given the initial position and velocity of the particle, the trajectory $\omega$ is completely
determined by the increments of the Wiener process. Let us consider the case that $x_t$ and $v_t$ are 
given, and compute the probability of having $(x_{t+dt},v_{t+dt})$ an infinitesimal time later. We know from the Langevin equation (\ref{langevin})
that 
\[dB_t = \frac{1}{\sqrt{2\mathcal{D}}}[mdv_t - F_t(x_t)dt+m\gamma v_tdt ]\]
And the fact that
$dB_t = B_{t+dt}-B_t\sim\mathcal{N}(0,dt)$ means that we have a probability density $P$\index{path-probability!for underdamped diffusions}
given by:
\begin{eqnarray}
  dP(v_{t+dt},x_{t+dt}|x_t,v_t) &=& P(v_{t+dt},x_{t+dt}|x_t,v_t)dx_tdv_t\nonumber\\
&=& \frac{1}{N}\delta(x_{t+dt}-x_t-v_t dt)e^{-L(x_t,v_t,dv_t)}dx_tdv_t \label{probdens}
\end{eqnarray}
where $L$ is given by
\begin{eqnarray*}
L(x_t,v_t,dv_t) &=& \frac{1}{4\mathcal{D}dt}[mdv_t - F_t(x_t)dt+m\gamma v_tdt ]^2\\
&=&  \frac{dt}{4\mathcal{D}}[m\frac{dv_t}{dt} - F_t(x_t)+m\gamma v_t ]^2
\end{eqnarray*}
Note that $N = \frac{\sqrt{4\pi \mathcal{D} dt}}{m}$ is a normalization constant. Furthermore we know that $x_{t+dt} = x_t + v_t dt$,
which is why a delta function is inserted. From now on, we restrict the space of all trajectories $\omega$ to
those trajectories which satisfy this constraint. This means that the delta function is no longer necessary.
Let us now divide the interval $[0,T]$ into $n$ intervals of length $\Delta t$ (so $n\Delta t = T$).
We should take $\Delta t$ small enough, so that (\ref{probdens}) can be used. (We take the limit of $\Delta t \to 0$). 
Define $t_i = i\Delta t$ for $i=0,\ldots ,n$, 
then the probability density of $\omega$ is just the product of the probability densities of all the steps $v_{t_i}\to v_{t_{i+1}}$,
because the increments of a Wiener process are mutually independent:
\begin{eqnarray*}
 \mathcal{P}_{\mu_0}(\omega) = \frac{\mu_0(x_0,v_0)}{N^n}\exp\left\{ -\frac{1}{4D}\sum_{i=0}^{n-1}\Delta t[m\frac{v_{t_{i+1}}-v_{t_i}}{\Delta t} - F_t(x_{t_i})+m\gamma v_{t_i} ]^2 \right\}\\
\end{eqnarray*}
which we denote in the limit of $\Delta t\to 0$ by
\begin{eqnarray*}
\mathcal{P}_{\mu_0}(\omega) =  \frac{\mu_0(x_0,v_0)}{\mathcal{N}}\exp\left\{ -\frac{1}{4\mathcal{D}}\int_0^Tdt[m\dot{v}_t - F_t(x_t)+m\gamma v_t ]^2 \right\}
\end{eqnarray*}
where $\mathcal{N}$ is a normalization constant, and $\mu_0(x_0,v_0)$ is the probability distribution
from which we sampled $x_0$ and $v_0$. The integral and $\dot{v}_t$ in the exponent are just notation, 
because they are not really well-defined (like `Riemann integrals' and `derivatives') 
as a consequence of the properties of the Wiener process.
Again, we do not go into the full mathematical details here (for that, see \cite{ls78}). 
A more important quantity for the purpose of this thesis
arises if we compare the path-probabilities of different dynamics to each other, by 
considering the action of the process with respect to a reference process. 
For this reference we define a new Langevin dynamics:
\begin{eqnarray}\label{langevinref}
 dx_t &=& v_t dt\nonumber\\
 mdv_t &=& -m\gamma v_tdt + \sqrt{2\mathcal{D}}dB_t
\end{eqnarray}
Note that it is the same as before, only we have taken the force $F$ to be zero. We denote the path-probability measure of this
process by $d\mathcal{P}^0$. Then we have that
\begin{equation}\label{actionlang}
 e^{-A(\omega)}=\frac{d\mathcal{P}_{\mu_0}}{d\mathcal{P}^0_{\mu_0}}(\omega) = \exp\left\{ \frac{1}{4\mathcal{D}}\int_0^Tdt[2m\dot{v}_t +2m\gamma v_t- F_t(x_t) ]F_t(x_t) \right\}
\end{equation}
This is the Girsanov-formula for our one-dimensional diffusion process\index{Girsanov formula! for underdamped diffusions}.
This quantity is well-defined, as the term
in the exponent with $\dot{v}_t$, actually a notation for
\begin{equation}\label{ito}
  \int_0^T dt \dot{v}_tF_t(x_t) = \lim_{n\to \infty}\sum_{i=0}^{n-1}(v_{t_{i+1}}-v_{t_i})F_{t_i}(x_{t_i})
\end{equation}
is mathematically well-defined. However, because of the properties of the Wiener process 
we can't treat this `integral' with the normal (Riemann) rules
of integration, and have entered the domain of stochastic integrals. The integral as defined in (\ref{ito}) is called
a `stochastic integral in the It\^o interpretation,' or just an `It\^o integral.'
We explain more on these stochastic integrals at the end of this section. We conclude for now by saying that, 
like in the previous chapters, we use this path-probability density to define expectation values for observables. 
If one defines the time-evolved probability density as $\mu_t(x,v) = \left<\delta(x_t-x)\delta(v_t-v)\right>_{\mu_0}$, then
one can prove that its evolution equation is exactly the Fokker-Planck equation,
see Appendix \ref{app-c}. 

\newpage
\section{Overdamped diffusions}
\index{diffusions!overdamped}
In many physical cases (e.g. molecular motors) the system is submerged in a highly
viscous environment. This means that the friction coefficient $\gamma$ is very big.
If on top of that the mass of the system is very small, one enters the overdamped
region, in which the system after each random kick relaxes quickly to the local expected
velocity, determined by the force $F$ and the friction. Because of this, one is not so much
interested in the changes of the velocity anymore, just in the positions.\\

Mathematically one takes the limit of $\gamma \to \infty$ and $m\to 0$, while keeping
$\chi = \frac{1}{m\gamma}$ constant and finite. This limit is not trivial \cite{ssd82,tit78,wil76}, but the result
is that one can ignore the term $mdv_t$ in the Langevin equation (\ref{langevin}).
If one then replaces in (\ref{langevin}) the term $m\gamma v_t dt$ by $m\gamma dx_t$,
one is left with the single equation
\begin{equation}\label{langover}
 dx_t = \chi F_t(x_t)dt + \sqrt{2 D}dB_t
\end{equation}
with $D = \chi^2\mathcal{D}$. 
We call this the overdamped Langevin equation\index{Langevin equation!overdamped}. 
To distinguish the previous Langevin equation from this one, we call from now on (\ref{langevin})
the underdamped Langevin equation.
The corresponding Fokker-Planck equation\index{Fokker-Planck equation!overdamped} for probability 
distributions is 
\begin{equation}
 \frac{\partial \mu_t}{\partial t} + \frac{\partial }{\partial x}j_{\mu_t}=0\ \ \ \ \textrm{with }\ \ \ \ \ j_{\mu_t} = 
\chi F_t\mu_t - D\frac{\partial \mu_t}{\partial x}
\end{equation}
which defines the probability current\index{probability current! for overdamped diffusions}.
This equation is in the literature often called the Smoluchowski equation\index{Smoluchowski!equation},
and the high-friction limit the Smoluchowski limit\index{Smoluchowski!limit}.
In the special case that $F(x) = - \frac{\partial U(x)}{\partial x}$, we can easily find the stationary distribution $\rho$:
\[ \rho(x) = \frac{1}{Z}e^{-\beta U(x)} \]
with $\beta = \frac{\chi}{D}$. Just as in the underdamped case, this corresponds to equilibrium. Moreover, in this case
the stationary probability current $j_{\rho}$ is actually zero.\\

Let us also provide a simple example of a nonequilibrium dynamics: suppose the particle moves on a circle
of length 1. We take the force $F(x) = f$ a constant. This can't be written as the derivative of a potential.
One can solve the stationary Fokker-Planck equation, though: the stationary distribution is just the uniform distribution $\rho(x) = 1$.
The stationary current is $j_{\rho} = \chi f$, which is a constant but is not zero.


\section{Overdamped diffusions in more dimensions}\label{sec-odiff}
\index{diffusions!overdamped}
To be more general, let us work in a $d$-dimensional space, and let $x_t$ be a $d$-dimensional vector,
representing the position of a single particle. Then the most general 
set of $d$ overdamped Langevin equations\index{Langevin equation!overdamped}, one for each component of $x_t$, is
\begin{equation}\label{langgencomponent}
 dx^i_t = \sum_j[\chi_{ij}(x_t) F^j_t(x_t)dt +\frac{\partial  D_{ij}}{\partial x^j}(x_t)dt+\sqrt{2D(x_t)}_{ij}dB^j_t]
\end{equation}
In this equation $F_t$ still represents a force, $D$ and $\chi$ have become $d\times d$ symmetric matrices
both depending on the position $x_t$,
and there are $d$ Wiener processes $dB^j_t$. The term with the derivatives of $D$ is a result of the fact that
a non-constant diffusion coefficient gives an extra net force on the particle \cite{vhbwb10}. We also stress that we make no
difference between indices that are placed as superscripts or as subscripts, i.e. $x^i=x_i$.
To reduce notation, we schematically write these Langevin equations in a matrix-form:
\begin{equation}\label{langgen}
 dx_t = \chi(x_t) F(x_t)dt +\nabla \cdot D(x_t)dt+\sqrt{2D(x_t)}dB_t
\end{equation}
In such a notation, $\chi F$ is to be interpreted as the matrix $\chi$ acting on the vector $F$.
The corresponding Fokker-Planck equation\index{Fokker-Planck equation!overdamped} is then
\begin{equation}\label{gsd1}
 \frac{\partial \mu_t}{\partial t} + \nabla\cdot j_{\mu_t}=0\ \ \ \ \textrm{with }\ \ \ \ \ j_{\mu_t} = 
\chi F_t\mu_t - D\nabla \mu_t
\end{equation}
The path-probability density\index{path-probability!for overdamped diffusions} is found in the same way as for underdamped diffusions:
\[ \mathcal{P}_{\mu_0}(\omega) = \frac{\mu_0(x_0)}{\mathcal{N}}\exp\left\{ -\frac{1}{4}\int_0^Tdt[\dot{x}_t - \chi F_t-\nabla \cdot D]
\cdot D^{-1}[\dot{x}_t - \chi F_t-\nabla \cdot D] \right\} \] 
where a dot $\cdot$ denotes a scalar product of vectors.
Again it is mathematically safer to look at the Radon-Nikodym derivative of
this process with respect to a reference. Let us take as a reference (\ref{langgen}) with
the force $F$ put to zero. We denote the corresponding path-probability by $\mathcal{P}^0$  and get
\begin{equation}\label{actiongen}
 e^{-A(\omega)} =\frac{d\mathcal{P}_{\mu_0}}{d\mathcal{P}^0_{\mu_0}}(\omega) = 
\exp\left\{ \frac{1}{4}\int_0^Tdt[2\dot{x}_t-2\nabla \cdot D - \chi F_t ]\cdot D^{-1}\chi F_t \right\}
\end{equation}

This formula is the Girsanov-formula for overdamped diffusions\index{Girsanov formula!for overdamped diffusions}
where, like in the underdamped case, we still have stochastic integrals in our formula,
meaning integrals which contain $\dot{x}$.

\section{Stochastic integrals}
\index{stochastic integral}
In this section we shortly review some technical aspects of stochastic integration. For more information,
see \cite{gar04,kry95}.

\paragraph{It\^o integral}\index{It\^o integral} \index{stochastic integral!It\^o}
Let us denote by $q$ the configuration of the system. For underdamped diffusions this is $(x,v)$,
and for overdamped diffusions this is $x$.
In (\ref{ito}) and (\ref{actiongen})  one encounters integrals of the form $\int_0^Tdt\dot{q}^k_t g_t(q_t)$,
usually written as $\int_0^Tdq^k_t g_t(q_t)$, with $g_t$ some function of the configuration and $q^k$ an
arbitrary component of $q$. This is actually
just notation for the following: divide the interval $[0,T]$ into $n$ intervals of length $\Delta t$ (so $n\Delta t = T$) and
define $t_i = i\Delta t$ for $i=0,\ldots n$, then for any function $g_t(q)$
\begin{equation}\label{itogen}
 \int_0^Tdq^k_t g_t(q_t) = \lim_{n\to \infty}\sum_{i=0}^{n-1}(q^k_{t_{i+1}}-q^k_{t_i})g_t(q_{t_i})
\end{equation}
Because of the special properties of Wiener processes, an integral defined as in (\ref{itogen}) can not
be treated as a Riemann integral. For example, for a Riemann integral, it does not matter where the function $g_t$
is evaluated in each time interval, for stochastic integrals like (\ref{itogen}) it does. 
From the way we defined the Langevin equation it is natural that $g_t$ is evaluated 
at the beginning of each interval. A stochastic integral with this property is called an It\^o integral.\\

To compute what the integration rules are for It\^o integrals, note the following:
the increments of the Wiener process $B_{t+\Delta t} - B_t$ have a Gaussian distribution with zero mean and variance
$\Delta t$. From this, one can prove that the quantity $(B_{t+\Delta t} - B_t)^2$ has a mean $\Delta t$ and a variance
of order $(\Delta t)^2$. Therefore, one can safely write 
$(B_{t+\Delta t} - B_t)^2 = \Delta t + o(\Delta t)$. 
Analogously: $(q^k_{t_{i+1}}-q^k_{t_i})(q^j_{t_{i+1}}-q^j_{t_i}) = 2D_{kj}\Delta t + o(\Delta t)$, 
for any $k,j=1,\ldots,d$ (Using $\mathcal{D}$ instead of $D$ for underdamped diffusions).\\

We compute here
what the actual rules for integration are for It\^o integrals. Take an arbitrary function $g(q)$ not explicitly dependent
on time.
A Taylor-expansion gives:
\begin{eqnarray*}
  g(q_{t_{i+1}}) &=& g(q_{t_i}) + \sum_j(q^j_{t_{i+1}}-q^j_{t_i})\cdot \frac{\partial }{\partial q^j} g(q_{t_i})\\
&& + 
\frac{1}{2}\sum_{j,k}(q^j_{t_{i+1}}-q^j_{t_i})(q^k_{t_{i+1}}-q^k_{t_i})\frac{\partial^2 }{\partial q^j\partial q^k}g(q_{t_i}) + o(\Delta t)\\
&\approx& g(q_{t_i}) + (q_{t_{i+1}}-q_{t_i})\cdot \nabla g(q_{t_i}) + (D\nabla)\cdot \nabla g(q_{t_i})\Delta t  + o(\Delta t)
\end{eqnarray*}
As a consequence
\begin{eqnarray*}
 \int_0^T dq_t\cdot \nabla g(q_t) &=& \lim_{n\to \infty}\sum_{i=0}^{n-1}(q_{t_{i+1}}-q_{t_i})\cdot\nabla g_t(q_{t_i})\\
&=& \lim_{n\to \infty}\sum_{i=0}^{n-1}[g(q_{t_{i+1}})-g(q_{t_i}) - (D\nabla)\cdot \nabla g(q_{t_i})\Delta t + o(\Delta t)] \\
&=& g(q_T)-g(q_0) - \int_0^T dt (D\nabla)\cdot \nabla g(q_{t})
\end{eqnarray*}
The integral over time on the right-hand side is a normal Riemann integral.
Similarly one proves that for explicitly time-dependent functions $g_t(q)$ one gets
\begin{equation}
 \int_0^T [dq_t\cdot \nabla g_t(q_t)+dt\frac{\partial g}{\partial t}(q_t) ]
= g_t(q_T)-g_t(q_0) - \int_0^T dt (D\nabla)\cdot \nabla g_t(q_{t})
\end{equation}

\paragraph{Stratonovitch integral}\index{Stratonovitch integral}\index{stochastic integral!Stratonovitch}
Another stochastic integral that is useful is the Stratonovitch integral, denoted and defined by
\begin{equation}\label{strat}
 \int_0^Tdq^k_t\circ g_t(q_t) = \lim_{n\to \infty}\sum_{i=0}^{n-1}(q^k_{t_{i+1}}-q^k_{t_i})g_t(\frac{q_{t_i}+q_{t_{i+1}}}{2})
\end{equation}
A nice property of the Stratonovitch integral is that it is antisymmetric with respect to
time reversal, if $q$ only contains positions, not velocities ($q=x$). 
When velocities are involved one also has to reverse the signs of the velocities,
and the Stratonovitch integral may not be time-antisymmetric in this case. 
One can easily prove that for any function $g$
\begin{equation}\label{itostrat}
 \int_0^Tdq^k_t\circ g_t(q_t) = \int_0^T dq^k_t g_t(q_t) + \int_0^T dt (D\nabla)^k  g_t(q_{t})
\end{equation}
so that the Stratonovitch integral actually has the normal integration rules:
\begin{equation}\label{stratrules}
 \int_0^T [dq_t\circ \nabla g_t(q_t)+dt\frac{\partial g_t}{\partial t}(q_t) ]
= g_t(q_T)-g_t(q_0)
\end{equation}

\paragraph{Work}
From mechanics we know that work should be equal to the integral
over a path of force times displacement. But in diffusion systems
this definition does not completely specify work, as one has to choose
which stochastic integral to use. However, we also know that
the work should be antisymmetric under time-reversal.
Because of this, it can be seen that the work performed by a force $f_t(x)$
during $\omega$ is defined with a Stratonovitch integral:
\begin{equation}\label{work}
 \int_0^Tdx_t\circ f_t(x_t) = \sum_k \int_0^Tdx^k_t\circ f^k_t(x_t)
\end{equation}
where $x_t$ denotes a position, not a velocity.

\section{Local detailed balance}\label{sec-ldb2}

Let us see what restrictions the local detailed balance assumption (Section \ref{sec-ldb1})
puts on our models.

\paragraph{Overdamped diffusions}
\index{local detailed balance!for overdamped diffusions}
Let us consider general overdamped diffusions (\ref{langgen}).
One can then compute the quantity $S(\omega)$ (as defined in (\ref{sdef}))
in the following way. The reference process, which is defined through (\ref{langgen}) but with
$F$ put to zero, is a pure diffusion, which is an equilibrium process with the uniform
distribution as equilibrium distribution. Actually in unbounded state spaces this distribution is not normalizable, 
so technically we do not have an equilibrium process. In any case we still have the important property that
\[ \frac{d\mathcal{P}^0_{x_0}}{d\mathcal{P}^0_{x_T}\theta}(\omega) = 1 \]
For the original process we can thus write
\[ S(\omega) = \log\frac{d\mathcal{P}_{x_0}}{d\mathcal{P}^0_{x_0}}(\omega) - \log\frac{d\mathcal{P}_{x_T}\theta}{d\mathcal{P}^0_{x_T}\theta}(\omega)
= A(\theta\omega) - A(\omega) \]
which after a short computation, using (\ref{actiongen}) and the definition of the Stratonovitch integral, becomes
\[ S(\omega) = \int_0^T dx_t \circ D^{-1}(x_t)\chi(x_t) F_t(x_t) \]
By the local detailed balance assumption this should be equal 
to the entropy flux into the environment. Let us consider the case
where the environment consists of one heat bath at inverse temperature $\beta$.
As an addition we write the force $F_t$  as the sum of a conservative part $-\nabla U$ with
$U$ representing the energy of the system, and a nonconservative part $f_t$.
Then we see that local detailed balance dictates that $\chi = \beta D$, because only in this case
we have
\begin{equation}\label{ldbdiff}
  S(\omega) = -\beta[U(x_T)-U(x_0) -\int_0^T dx_t \circ f_t(x_t)]
\end{equation}
where in the brackets we have change of energy minus work done on the system (see (\ref{work})).
This relation\index{Einstein relation}
between mobility and the diffusion coefficient is called the 
Einstein relation. It is commonly used in Langevin dynamics, but here we see it
as a consequence of the local detailed balance assumption.

\paragraph{Underdamped diffusions}
\index{local detailed balance!for underdamped diffusions}
Let us return to the one-dimensional underdamped case (\ref{langevin}). To compute $S(\omega)$
we should take into account that the velocities change sign under time-reversal:
\[ S(\omega) = \frac{m\gamma}{\mathcal{D}}\int_0^T dt v_t F_t(x_t) \]
We can apply the same argument as for overdamped diffusions, and we see that
local detailed balance dictates that $\mathcal{D} = \frac{m\gamma}{\beta}$, which is again called the Einstein relation\index{Einstein relation}.

\cleardoublepage

\cleardoublepage

\cleardoublepage
\part{Fluctuations and Response}

\vspace*{6cm}

``Recent developments in non equilibrium statistical physics have convinced us
that times are ripe for a review of the vast subject concerning the fluctuations
of systems described by statistical mechanics. This issue is important even
beyond the `traditional' applications of statistical mechanics, e.g. in a wide
range of disciplines ranging from the study of small biological systems to turbulence,
from climate studies to granular media etc. Moreover, the improved
resolution in real experiments and the computational capability reached in
numerical simulations has led to an increased ability to unveil the detailed
nature of fluctuations, posing new questions and challenges to the theorists.''

\begin{flushright}
 Umberto Marini Bettolo Marconi, Andrea Puglisi,
Lamberto Rondoni and Angelo Vulpiani, in their introduction to \textit{Fluctuation-Dissipation: Response Theory in
Statistical Physics} (2008), \cite{mprv08}
\end{flushright}

\cleardoublepage
\chapter{Response}\label{chap-response}

\textit{ In this part of the thesis, we investigate how a system responds
to a perturbation, namely a small change in its energy. The central object that summarizes this
is the response function. The goal is to relate this response function
to correlation functions in the unperturbed process. In equilibrium
systems this has already been done and the resulting relation is called the fluctuation-dissipation
theorem. Out of equilibrium one needs to modify this relation. Extra terms appear that have
to do with traffic, as defined in previous chapters. In this chapter we
describe the general framework; in the next chapters more explicit models are examined.
The research in this chapter was summarized earlier in \cite{bmw09a}.}

\section{Introduction}

The fluctuation-dissipation theorem\index{fluctuation-dissipation relation!in equilibrium}
\index{fluctuation-dissipation theorem} is a standard chapter in statistical 
mechanics \cite{kubo66,mprv08,set06}. In Chapter \ref{sec-fdt} we already
explained that a small change in an equilibrium system that changes the Hamiltonian  $H_0 \rightarrow H = H_0 - h \,V$ from time zero on, 
changes expectation values of observables $Q(x_t)$ at any time $t$. This change, up to linear order in $h$, is called the `response'
and is given by 
\begin{equation}\label{fdt2}
\left.\frac{\partial}{\partial h}\left<Q(x_t)\right>^h_{\rho^0}\right|_{h=0} = \beta\left<Q(x_t)[V(x_t)-V(x_0)]\right>^0_{\rho}
\end{equation}
where the superscript $h$ means that the average is taken in the perturbed system, and the superscript $0$ stands for
the unperturbed system, which is here an equilibrium system.
This equality (\ref{fdt2}) is called the fluctuation-dissipation theorem. Actually it is usually stated in a different but equivalent
form. In this different form the parameter $h$ is time-dependent: $h=h_t$, where $h_t=0$ for $t<0$. 
The time-dependence of $h_t$ is something that is provided `externally,' i.e.
it is deterministic, it is controlled by the experimentalist. 
As an example, think of an Ising spin system, where the perturbation
is a time-dependent magnetic field. The response is
now the (functional) derivative of the expectation value of $Q(x_t)$ with respect to $h_s$ with $s<t$.
The fluctuation-dissipation theorem for equilibrium systems then states that
\begin{equation}\label{fdt3}
\left.\frac{\delta}{\delta h_s}\left<Q(x_t)\right>^h_{\rho^0}\right|_{h=0} = \beta\frac{\partial}{\partial s}\left<Q(x_t)V(x_s)\right>^0_{\rho^0}
\end{equation}
We will derive this form of the fluctuation-dissipation theorem as a special case of the more general formula later on.

An early example of this theorem is present in Einstein's
treatment of Brownian motion, where the diffusion constant,
expressed as a velocity auto-correlation function, is found
to be proportional to the mobility \cite{ein1905}. We will come back to this in the following chapters. 
Other famous examples include the
Johnson-Nyquist formula for electronic white noise \cite{john28,nyq28} and the Onsager
reciprocity for linear response coefficients \cite{ons31a,ons31b}.
The fluctuation-dissipation theorem is useful because it gives a relation between two quantities
in essentially different processes. One can e.g. determine the response of a system without actually perturbing it.

In this part of the thesis, we discuss and explain the results obtained in \cite{bbmw09,bmw09a,bmw09b}.
This research investigates linear response in systems out of equilibrium. 
So far, approaches deriving
a fluctuation-dissipation relation (FDR) for 
systems out of equilibrium have
not found a physical unification and do not appear as
textbook material.  One reason may be that
previous work has not been seen to identify a sufficiently general
structure with a clear corresponding statistical thermodynamic
interpretation.  

Aiming to provide a simple and general approach to FDR's, in \cite{bmw09a}
 we have put forward a FDR for nonequilibrium regimes in a
framework that may represent a unifying scheme for previous
formulations. This is discussed in the remainder of this chapter.
The application of this theory to Markov jump processes and
overdamped diffusions was investigated in \cite{bmw09b},
and the application to underdamped diffusions in \cite{bbmw09},
and is explained in the next chapters. Additional work on
fluctuation-dissipation relations for Markov jump processes was
also reported in \cite{mw09}, but we do not discuss this here.

\section{Linear Response}
\index{linear response}\index{response}
The unperturbed systems we consider are driven from equilibrium, but with a time-independent
(time-homogeneous) dynamics. This dynamics can be far from equilibrium.
We denote the energy of the system in a configuration $x$ by $U(x)$.
As a perturbation \index{perturbation} the energy is changed by the addition of a potential: $U\to U - h_t V$,
where the time-dependent function $h_t$ is the amplitude of the perturbation. We assume
that it is bounded: $|h_t| < h <\infty$, with $h$ a small number, and $h_t = 0 $ for $t<0$.
We also assume that it is a continuous differentiable function of time. 
Furthermore, if the configuration of the system includes velocities, we always assume that
energies and potentials that we use are symmetric under the sign-reversal of those velocities.
With the notation of Section (\ref{traj-eat}): e.g. $V(x) = V(\pi x)$.

We want to compute the influence of this perturbation on expectation values of observables $Q(x)$.
We restrict ourselves to a regime where a small perturbation only has small consequences,
which excludes for example phase-transitions. This regime allows us to make an expansion
of expectation values
in orders of the parameter $h$, which we do up to the first, linear, order:
\begin{equation}\label{resdef}
  \left<Q(x_t)\right>_{\mu_0}^h = \left<Q(x_t)\right>_{\mu_0}^0 + \int_0^tdsh_sR_{QV}(t,s) + o(h)
\end{equation}
which defines the response function $R_{QV}$, \index{response function} which actually also depends on the initial distribution
$\mu_0$:
\begin{equation}\label{resfun}
  R_{QV}(t,s) = \left.\frac{\delta}{\delta h_s}\left<Q(x_t)\right>^h_{\mu_0}\right|_{h=0}
\end{equation}
Applying the framework explained in
Chapter \ref{chap-traj}: we use the action to rewrite expectation
values of observables in the perturbed dynamics into expectation values in the unperturbed
dynamics. Henceforth we denote by $\mathcal{P}^h$ the path-probability density of the 
perturbed process and by $\mathcal{P}^0$ that of the unperturbed one:
\[ \left<Q(x_t)\right>_{\mu_0}^h = \left<\frac{\mathcal{P}^h_{\mu_0}}{\mathcal{P}^0_{\mu_0}}(\omega)Q(x_t)\right>_{\mu_0}^0 =
\left<e^{-A(\omega)}Q(x_t)\right>_{\mu_0}^0  \]
The averages are over all paths in the interval $[0,t]$.
Because $h_t<h$ is small, the action is also small, and we can expand the exponential up to linear
order in $h$:
\[ \left<Q(x_t)\right>_{\mu_0}^h = \left<Q(x_t)\right>_{\mu_0}^0 - \left<A(\omega)Q(x_t)\right>_{\mu_0}^0 + o(h)\]
Up to now this is only mathematics. Physics comes in when
we split the action in its time-antisymmetric and time-symmetric parts (see Section \ref{sec-eat}):
\begin{eqnarray*}
 S_{ex}(\omega) &=& A(\theta\omega) - A(\omega)\\ 
\mathcal{T}_{ex}(\omega) &=& A(\theta\omega) + A(\omega)
\end{eqnarray*}
where $S_{ex}(\omega)$ can, by the local\index{entropy flux!excess} detailed balance assumption, be interpreted
as the excess entropy flux into the environment during the process $\omega$. Excess
is meant as excess of the perturbed process with respect to the unperturbed process.
$\mathcal{T}_{ex}(\omega)$ is called the (excess) traffic\index{traffic!excess}. With this in mind we get
\begin{equation}\label{response1}
  \left<Q(x_t)\right>_{\mu_0}^h = \left<Q(x_t)\right>_{\mu_0}^0 + \frac{1}{2}\left<S_{ex}(\omega)Q(x_t)\right>_{\mu_0}^0 - 
\frac{1}{2}\left<\mathcal{T}_{ex}(\omega)Q(x_t)\right>_{\mu_0}^0 + o(h)
\end{equation}
We have thus written the change in the expectation value of $Q$ as the sum of the correlation functions
of $Q$ with entropy and traffic. This is equivalent to writing that the response function (\ref{resfun}) equals
\begin{equation}\label{response1a}
  R_{QV}(t,s) = \frac{1}{2}\left.\frac{\delta}{\delta h_s}\left<S_{ex}(\omega)Q(x_t)\right>_{\mu_0}^0\right|_{h=0} - 
\frac{1}{2}\left.\frac{\delta}{\delta h_s}\left<\mathcal{T}_{ex}(\omega)Q(x_t)\right>_{\mu_0}^0\right|_{h=0}
\end{equation}
This constitutes our most general but also our most vague result: the response function is the sum
of two terms. One of the terms is expressed using the entropy flux into the environment, which is in principle known and measurable.
The other term is up to here only mathematically defined and needs more investigation.
Before we do that, we first discuss some general properties of the response function.



\section{General properties of the response function}

\subsection{Relaxation}

We assume that the unperturbed dynamics is time-homogeneous. Then, as we assume in this thesis, 
the probability distribution of states relaxes to the stationary distribution, meaning that
\[ \left<Q(x_t)\right>^0_{\mu_0} \to \left<Q(x_t)\right>^0_{\rho}  \ \ \ \ \ \textrm{for}\ \  t\to\infty \]
Suppose now that the perturbation we added to the system, is removed again after a 
time $t_1$, i.e. $h_t = 0$ for $t>t_1$. Then logically the system relaxes again to the stationary
regime:
\[ \left<Q(x_t)\right>^h_{\mu_0} \to \left<Q(x_t)\right>^0_{\rho}  \ \ \ \ \ \textrm{for}\ \  t\to\infty\]
Using (\ref{resdef}), this gives
\[ \int_0^{t_1}dsh_sR_{QV}(t,s) \to 0 \ \ \ \ \ \textrm{for}\ \  t\to\infty \]
As this is true for any $h_t$, we can safely say that $R_{QV}(t,s) \to 0$ for $t\to\infty$. However,
throughout this text, we make a stronger assumption, namely that
\begin{equation}\label{finite}
 \int_s^{\infty} dt |R(t,s)| < \infty 
\end{equation}
for any $s$. The reason for this assumption is that the excess heat dissipation
remains finite, see Section \ref{sec-diss}.

\subsection{Causality}
\index{causality}
Physically, the response function should be zero whenever $s>t$. This is a consequence of causality,
because the expectation value of an observable at one time can not be influenced by a perturbation at
a later time. What we want to show here, is that causality is already embedded in our framework,
and not something we have to put in afterwards by hand. In the discussion after (\ref{pathmarg}) we argued
that to take the expectation value of $Q(x_t)$, we only need to average over paths in the interval $[0,t]$.
The perturbed dynamics that govern the probability of this path, and thus also the action $A(\omega)$,
can logically only depend on $h_s$ with $0\leq s\leq t$. So the perturbed expectation value of $Q(x_t)$
is independent of $h_s$ with $s>t$. This is indeed causality.
In (\ref{response1}) causality gives us that for $s>t$:
\begin{equation}\label{causality}
 \left.\frac{\delta}{\delta h_s}\left<S_{ex}(\omega)Q(x_t)\right>_{\mu_0}^0\right|_{h=0} = 
\left.\frac{\delta}{\delta h_s}\left<\mathcal{T}_{ex}(\omega)Q(x_t)\right>_{\mu_0}^0\right|_{h=0}
\end{equation}

\subsection{A constant perturbation}\label{sec-constant}
\index{perturbation!constant} \index{integrated response} \index{generalized susceptibility|see{integrated response}}
For experiments and simulations, it is often more convenient to work with a constant perturbation.
This means that $h_s=h$ for $s\geq 0$ and is equal to zero for $s<0$. 
In this case one will not directly measure the response function, but rather the time-integral of it.
This is because (\ref{resdef}) reduces in this case to:
\[ \left<Q(x_t)\right>_{\mu_0}^h = \left<Q(x_t)\right>_{\mu_0}^0 + h\int_0^tdsR_{QV}(t,s) + o(h) \]
and the response is then rather defined as
\[ \left.\frac{\partial}{\partial h}\left<Q(x_t)\right>_{\mu_0}^h\right|_{h=0} = \int_0^tdsR_{QV}(t,s) \]
This integrated response is sometimes called a generalized susceptibility, as a generalization
of the susceptibility defined in thermodynamics.

\subsection{Response in the frequency domain}\label{sec-frequency}

Apart form a constant perturbation, a periodic perturbation is also often convenient in experimental settings,
and response functions are often measured in the frequency domain (using Fourier transforms).
Let us consider therefore a perturbation with an amplitude $h_t = h_0\cos(kt)$.
We also restrict ourselves to the case that the unperturbed system before the perturbation
was stationary:
\begin{equation}\label{frequency}
  \left<Q(x_t)\right>_{\rho}^h = \left<Q(x)\right>_{\rho}^0 + h_0\int_0^tds\cos(ks)R_{QV}(t-s) + o(h_0)
\end{equation}
Up to now we have always taken $t=0$ as the moment that the perturbation was turned on. If we change this time to $t=-T$,
then it only changes the lower limit of the integral in (\ref{frequency}) from $0$ to $-T$.
We suppose that the perturbation was taken a long time ago, so we can effectively take $-T\to -\infty$.
Our assumption (\ref{finite}) ensures that this makes sense. The upper limit of the integral (\ref{frequency})
can also be changed to $+\infty$, as $R_{QV}(t-s) =0$ for $s>t$ due to causality. This naturally
leads us to rewriting (\ref{frequency}) in terms of Fourier transforms:
\begin{eqnarray*}
 \int_{-\infty}^{+\infty}ds\cos(ks)R_{QV}(t-s) &=& \int_{-\infty}^{+\infty}ds\frac{e^{iks}+e^{-iks}}{2}R_{QV}(t-s) \nonumber\\
&=& \frac{1}{2}\left[ \tilde{R}_{QV}(k)e^{-ikt} + \tilde{R}_{QV}(-k)e^{ikt} \right]\label{frequency2}
\end{eqnarray*}
where the Fourier transform of the response function is defined as $\tilde{R}_{QV}(k) = \int_{-\infty}^{+\infty}dtR_{QV}(t)e^{ikt}$.
Using properties of Fourier transforms, and the fact that $R_{QV}(t)$ is a real-valued function, we arrive at
\[ \left<Q(x_t)\right>_{\rho}^h = \left<Q(x)\right>_{\rho}^0 + h_0\textrm{Re}(\tilde{R}_{QV}(k)e^{-ikt}) + o(h_0) \]
where $\textrm{Re}(z)$ is the real part of a complex number $z$. So for these kinds of perturbations,
what is actually measured is the Fourier transform of the response function.

Using complex analysis, one can prove \cite{mar04} that $\tilde{R}_{QV}(k)$ is a continuous bounded function,
as a consequence of the assumption (\ref{finite}).

\section{The case of one heat bath}\label{sec-oneheat}

The result (\ref{response1}) is a very general, but also a very vague result. 
To get more physical results we can of course investigate
explicit examples, which we do in the next chapters. Before doing this however,
we write down a more explicit
form for the response function by utilizing the fact that entropy is a known physical quantity.
To do this we restrict ourselves to systems in contact with an environment at a single
temperature. This means the system can be driven away from equilibrium by a nonconservative force
or by particle reservoirs at different chemical potentials. Throughout the rest of this chapter
and the next, we keep this assumption.
In the example of underdamped diffusions
we will come back to the case of heat baths at different temperatures.

\paragraph{Entropy, work and heat:}
In the case that the environment of the system is described by a single temperature, we can
write the entropy flux into the environment as
\[ S(\omega) = \beta Q(\omega) = \beta[-\Delta E + W(\omega)] \]
with $Q$ the heat flow into the environment, $\Delta E$ the change of energy of the system,
and $W$ the work done on the system. The excess entropy flux is 
the extra entropy flux created by the addition of the potential $-h_tV$, and is thus equal to
\begin{equation}\label{hwork}
  S_{ex}(\omega) = \beta Q_{ex}(\omega) = \beta[ h_tV(x_t)- h_0V(x_0) - \int_0^t ds \frac{\partial h_s}{\partial s}V(x_s) ]
\end{equation}
where the last term is the work due to the added potential $-h_tV$. We give a short argument
to see why this is work:\index{heat!excess} \index{work!excess}
divide the time interval $[0,t]$ into $n$ segments of length $\Delta t$, and define $t_i = i\Delta t$
for $i=1,\ldots , n$. We then have that
\begin{eqnarray}
 h_tV(x_t)-h_0V(x_0) &=& \sum_{i=1}^n[h_{t_i}V(x_{t_i})-h_{t_{i-1}}V(x_{t_i})]\nonumber\\
&&+\sum_{i=1}^n[h_{t_{i-1}}V(x_{t_i})-h_{t_{i-1}}V(x_{t_{i-1}})]\label{heatwork} 
\end{eqnarray}
The left-hand side of this equation is a change of energy, and can thus be split into work and heat.
In our discussion in Section \ref{intro-esm} we defined heat as the work due to (stochastic) forces we do not detect,
and can not be externally controlled.
If we apply this definition here, we can see that the second term on the right-hand side exists due
to changes in the configuration, which are stochastic and can't be controlled.
The first term is due to changes in the parameter $h_t$, which is controlled externally.
Therefore the first term in (\ref{heatwork}) is interpreted as work and the second as heat
(see also \cite{cro98}).
Taking the limit of $n\to \infty$ for the work-term, we get exactly the last term in (\ref{hwork}).
Note that we can't write the heat in terms of a partial derivative of $V$ with respect to $x$,
because that derivative usually does not exist, or even makes sense (e.g. for Markov jump processes).

\paragraph{Response function}
With this knowledge we can rewrite the correlation of the entropy flux with the observable as follows:
\begin{eqnarray}
 \left<S_{ex}(\omega)Q(x_t)\right>_{\mu_0}^0  &=& \beta[ h_t\left<V(x_t)Q(x_t)\right>_{\mu_0}^0- h_0\left<V(x_0)Q(x_t)\right>_{\mu_0}^0\nonumber\\  
&&-\int_0^t ds \frac{\partial h_s}{\partial s}\left<V(x_s)Q(x_t)\right>_{\mu_0}^0]\nonumber\\
&=& \beta\int_0^t ds h_s \frac{\partial}{\partial s}\left<V(x_s)Q(x_t)\right>_{\mu_0}^0\label{entcorr}
\end{eqnarray}
To write down the response function (\ref{resfun}), it is convenient to define the functional derivative of the traffic:
\begin{equation}\label{frenesy}
  \tau(\omega,s) = \left.\frac{\partial}{\partial h_s}\mathcal{T}_{ex}(\omega)\right|_{h=0}
\end{equation}
With this notation and using (\ref{entcorr}) the response function (\ref{resfun}) becomes\index{fluctuation-dissipation relation!one heat bath}:
\begin{equation}\label{response2}
 R_{QV}(t,s) = \frac{\beta}{2}\frac{\partial}{\partial s}\left<V(x_s)Q(x_t)\right>_{\mu_0}^0 -\frac{1}{2}\left<\tau(\omega,s)Q(x_t)\right>_{\mu_0}^0
\end{equation}
This is the most general formula for Markovian systems out of equilibrium, but in contact with only one heat bath at inverse temperature $\beta$.
We see that the first term on the right-hand side coincides formally with one half of the response function for equilibrium systems.
It is of course the second term that is different for nonequilibrium systems with respect to equilibrium.
For a constant perturbation $h_t=h$, we get
\begin{eqnarray*}
  \int_0^tdsR_{QV}(t,s) = \frac{\beta}{2}\left<[V(x_t)-V(x_0)]Q(x_t)\right>_{\mu_0}^0 -
\frac{1}{2}\int_0^Tds\left<\tau(\omega,s)Q(x_t)\right>_{\mu_0}^0\\
\end{eqnarray*}

\subsection{Causality and Stationarity}

For the case of one heat bath, we can combine (\ref{response2}) with (\ref{causality}) to find for $t>s$:
\begin{equation}\label{entropytraffic}
 \beta\frac{\partial}{\partial t}\left<V(x_t)Q(x_s)\right>_{\mu_0}^0 = \left<\tau(\omega,t)Q(x_s)\right>_{\mu_0}^0
\end{equation}
We can use this causality relation to rewrite (\ref{response2}) for $t>s$:
\begin{eqnarray}
 R_{QV}(t,s) &=& \frac{\beta}{2}\left[\frac{\partial}{\partial s}\left<V(x_s)Q(x_t)\right>_{\mu_0}^0 - 
\frac{\partial}{\partial t}\left<V(x_t)Q(x_s)\right>_{\mu_0}^0\right]\nonumber\\
&& + \frac{1}{2}\left[\left<\tau(\omega,t)Q(x_s)\right>_{\mu_0}^0 -\left<\tau(\omega,s)Q(x_t)\right>_{\mu_0}^0\right]  \label{response3}
\end{eqnarray}
Consider now the case that the system was in the stationary state when the perturbation was added, meaning $\mu_0 = \rho$
is the stationary distribution. Stationarity of the process means that it is time-translation invariant. As a
consequence, the response function can be written as $R_{QV}(t,s) = R_{QV}(t-s)$.
Moreover, correlation functions computed in the stationary regime satisfy
\[ \left<V(x_s)Q(x_t)\right>_{\rho} =  \left<V(x_0)Q(x_{t-s})\right>_{\rho} \]
This means that in this case (\ref{response3}) simplifies to 
\begin{eqnarray}
 R_{QV}(t,s) &=& \frac{\beta}{2}\frac{\partial}{\partial s}\left[\left<V(x_s)Q(x_t)\right>_{\rho}^0 + 
\left<V(x_t)Q(x_s)\right>_{\rho}^0\right]\nonumber\\
&& + \frac{1}{2}\left[\left<\tau(\omega,t)Q(x_s)\right>_{\rho}^0 -\left<\tau(\omega,s)Q(x_t)\right>_{\rho}^0\right]  \label{response4}
\end{eqnarray}
valid for $t>s$.

\subsection{Equilibrium}

We prove here the fluctuation-dissipation relation for equilibrium systems\index{fluctuation-dissipation relation!in equilibrium}. 
A system in equilibrium
is necessarily in contact with only one heat bath, so (\ref{response2}) applies here.
A system in equilibrium is also time-reversible. The time-reversal operator is defined as $\theta\omega = (\pi x_{t-s})_{0\leq s\leq t}$,
where $\pi$ reverses the sign of velocities.
Time-reversibility means that the expectation values of observables are equal to the expectation
values of their time-reversed twins, e.g:
\[ \left<V(x_s)Q(x_t)\right>_{\rho}^0 = \left<V(x_t)Q(\pi x_s)\right>_{\rho}^0 \]
where we used the assumption that $V(x) = V(\pi x)$.
From the definition of $\tau(\omega,s)$ (see (\ref{frenesy})), it also follows that $\tau(\theta\omega,s)= \tau(\omega,t-s)$,
so that for $t>s$:
\[ \left<\tau(\omega,s)Q(x_t)\right>_{\rho}^0 = \left<\tau(\omega,t-s)Q(\pi x_{0})\right>_{\rho}^0 = \left<\tau(\omega,t)Q(\pi x_s)\right>_{\rho}^0 \]
which by the causality relation (\ref{entropytraffic}) equals
\[ \left<\tau(\omega,t)Q(\pi x_s)\right>_{\rho}^0 =  \beta\frac{\partial}{\partial t}\left<V(x_t)Q(\pi x_s)\right>_{\rho}^0
= -\beta\frac{\partial}{\partial s}\left<V(x_0)Q(x_{t-s})\right>_{\rho}^0  \]
Substituting this equality in (\ref{response3}) we arrive at
\[ R_{QV}(t,s) = \beta\frac{\partial}{\partial s}\left<V(x_s)Q(x_t)\right>_{\rho}^0  \ \ \ \ \ \ \textrm{for }\ \ \  t>s     \]
which is exactly the fluctuation-dissipation theorem in equilibrium.
Note that for a constant perturbation $h_t=h$ the response becomes
\[ \int_0^tdsR_{QV}(t,s) = \beta\left<[V(x_t)-V(x_0)]Q(x_t)\right>_{\rho}^0 \]
as in (\ref{fdt2}).

\subsection{Dissipation}\label{sec-diss}

\index{dissipation}
In equilibrium it is known that the response function is
closely related to the energy dissipation of the system into the
environment \cite{mar04}. This is already seen in the fact that the response function
is then expressed through the fluctuation-dissipation theorem as a
correlation functional between the observable and entropy flux.
In systems driven out of equilibrium things get more
complicated: even when the system is not perturbed there is
already a heat dissipation. This is called the
`housekeeping' heat that is needed to maintain the (unperturbed)
nonequilibrium stationary state. Perturbing the system then gives additional
heat. We see here that the prediction in \cite{op98} that the
usual equilibrium relation between response and dissipation is
preserved when taking into account only the excess heating
and ignoring the
housekeeping heat, is indeed true in the following sense.\\

For a system in contact with only one heat bath, the entropy flux
into the environment is equal to $\beta$ times the heat dissipated
into the environment. For each trajectory $\omega$, this heat can be split
in the heat of the unperturbed process and the excess heat of the perturbed process with respect
to the unperturbed process. In (\ref{hwork}) we already wrote
down explicitly the excess heat. We assume a perturbation of the form
$h_t=h_0\cos(kt)$, like we did in Section (\ref{sec-frequency}), and compute
the excess heat over one period of the perturbation $[0,T=\frac{2\pi}{k}]$
in the perturbed system:
\begin{eqnarray*}
  \left<Q_{ex}(\omega)\right>^h_{\rho} &=& h_T\left<V(x_T)\right>^h_{\rho} - h_0\left<V(x_0)\right>^h_{\rho} - 
\int_0^T dt\frac{\partial h_t}{\partial t}\left<V(x_t)\right>^h_{\rho}\\
&=& \int_0^T dt h_t \frac{\partial}{\partial t}\left<V(x_t)\right>^h_{\rho}
\end{eqnarray*}
We can write the expectation of $V$ using the response function as in (\ref{frequency2}):
\[ \left<V(x_t)\right>^h_{\rho} = \left<V(x)\right>^0_{\rho} + \frac{h_0}{2}\left[ \tilde{R}_{VV}(k)e^{-ikt} + \tilde{R}_{VV}(-k)e^{ikt} \right] \]
Substituting this, and the explicit expression of $h_t$ into the expression of the heat, gives us
\begin{eqnarray*}
  \left<Q_{ex}(\omega)\right>^h_{\rho}
&=& \frac{ikh_0^2}{2}\int_{0}^Tdt  \cos(kt)\left[-\tilde{R}_{VV}(k)e^{-ikt} +\tilde{R}_{VV}(-k)e^{ikt} \right] +o(h_0^2)\\
&=& \frac{ikh_0^2T}{4}\left[ \tilde{R}_{VV}(-k) - \tilde{R}_{VV}(k) \right]+o(h_0^2)\\
&=& \pi h_0^2\textrm{Im}[\tilde{R}_{VV}(k)]+o(h_0^2)
\end{eqnarray*}
We see that the excess dissipated heat can be expressed through the imaginary part of the 
Fourier transform of the response function. For systems that are (unperturbed) in equilibrium,
there is no housekeeping heat, so the excess heat is then total dissipated heat.
Although the excess dissipated heat only specifies the imaginary part, the Kramers-Kronig relations 
in complex analysis connect the real and imaginary parts of complex functions
to each other \cite{mar04}. This means that when one of them is known, the other can be computed from it.

Let us conclude here by noting that the imaginary part of the 
Fourier transform is equal to the Fourier transform of the 
time-antisymmetric part of the response function (\ref{response2}):
\begin{eqnarray*}
 r(t-s) &=& \frac{1}{2}[R_{VV}(t,s)-R_{VV}(s,t)]\\
 &=& \beta\frac{\partial}{\partial s}\left<V(x_s)V(x_t)\right>_{\rho}^0\nonumber
 + \frac{1}{2}\left[\left<\tau(\omega,t)V(x_s)\right>_{\rho}^0 -\left<\tau(\omega,s)V(x_t)\right>_{\rho}^0\right]
\end{eqnarray*}
which is the same as (\ref{response4}) for $Q=V$, but (\ref{response4}) is only valid for $t>s$.
For $t>s$, we have that $r(t-s) = R_{VV}(t-s)$, while for $t<s$ the response function is zero.
The antisymmetric part of the response function therefore contains all information of the whole
response function.

\cleardoublepage
\chapter{Markov jump processes and overdamped diffusions}\label{chap-mjpod}

\textit{In this chapter the general framework outlined in the
previous chapter is applied to the cases of Markov jump processes and
overdamped diffusions. These two classes of models share a common property,
simplifying the fluctuation-dissipation relation, as we shall see in (\ref{fdr}).
For a few explicit examples, simulations have been done to visualize
the terms in  the fluctuation-dissipation relations. The work 
explained here was written down in \cite{bmw09b}.}

\section{An explicit formula}

In this chapter we only consider systems in contact with one heat bath,
so that we can use formulae like (\ref{response2}) and (\ref{response3}).
The advantage of considering specific models is that we can get more explicit results.
To do this, we need to compute the traffic, i.e. the time-symmetric part of the action
for our models. More precisely, we need to compute the functional derivative
of traffic, as defined in (\ref{frenesy})

\subsection{Overdamped diffusions}
Consider the case of general overdamped diffusions in $d$ dimensions (\ref{langgen}).
Adding the potential $-h_tV(x_t)$ in Langevin equations is straightforward. The perturbed Langevin equation becomes
(we are still working in the It\^o convention):
\[ dx_t = \chi(x_t) [F(x_t)+h_t\nabla V(x_t)]dt +\nabla \cdot D(x_t)dt +\sqrt{2D(x_t)}dB_t \]
Remember that the local detailed balance assumption here implies that $\chi = \beta D$.
In the same way as we derived (\ref{actiongen}), we can here derive the action that describes
the relative probability of paths of the perturbed process with respect to the unperturbed process:
\begin{eqnarray*}
  A(\omega) &=& -\log \frac{d\mathcal{P}^h}{d\mathcal{P}^0}(\omega)\\ &=&  -\frac{\beta}{2}\int_0^t dx_s h_s \nabla V
+ \frac{\beta}{2}\int_0^tdsh_t\nabla V \cdot [\nabla \cdot D + \chi F + \frac{1}{2}h_t\chi \nabla V]
\end{eqnarray*}
The time-antisymmetric part of the action is given by a Stratonovitch integral:
\[ S_{ex}(\omega) = A(\theta\omega)-A(\omega) =  \beta\int_0^t dx_s\circ h_s \nabla V \]
as a consistency check, let us rewrite this using (\ref{stratrules}):
\[ \int_0^t dx_s\circ h_s \nabla V  = h_tV(x_t)-h_0V(x_0) - \int_0^tds\frac{\partial h_s}{\partial s}V(x_s)  \]
which is exactly (\ref{hwork}). More importantly, we can compute explicitly the 
time-symmetric part of the action:
\[ \mathcal{T}_{ex}(\omega) = A(\theta\omega)+A(\omega) = \beta\int_0^tds h_s[\nabla \cdot (D\nabla ) + \chi F\nabla  + \frac{1}{2}h_s(\nabla  V)\chi \nabla]V \]
so that the functional derivative with respect to $h_s$ becomes:
\begin{equation}\label{trafod}
 \tau(\omega,s) = \beta\left.\frac{\partial}{\partial h_s}\mathcal{T}_{ex}(\omega)\right|_{h=0} = 
[\nabla \cdot (D(x_s)\nabla ) + \chi(x_s) F(x_s)\nabla]V(x_s)
\end{equation}
and the fluctuation-dissipation relation\index{fluctuation-dissipation relation! overdamped diffusions}:
\begin{eqnarray}
 R_{QV}(t,s) &=& \frac{\beta}{2}\frac{\partial}{\partial s}\left<V(x_s)Q(x_t)\right>_{\mu_0}^0\label{fdrod}\\
 &&-\frac{\beta}{2}\left<[\nabla \cdot (D(x_s)\nabla V(x_s) ) + \chi(x_s) F(x_s)\nabla V(x_s)]Q(x_t)\right>_{\mu_0}^0\nonumber
\end{eqnarray}

\subsection{Markov jump processes}

For Markov jump processes, it is more ambiguous to define what `adding a potential' means.
Starting from a general Markov jump process with transition rates $k(x,y)$, the local
detailed balance assumption asserts that (see (\ref{ldbmj})):
\[ \frac{k(x,y)}{k(y,x)} = e^{-\beta[U(y)-U(x)-W(x,y)]} \]
where $U$ is the energy of the system, and $W(x,y)$ is the work needed for a jump $x\to y$.
Adding a potential $-h_tV$ changes the transition rates to new time-dependent rates $k_t(x,y)$ which,
again by the local detailed balance assumption, have to satisfy
\[ \frac{k_t(x,y)}{k_t(y,x)} = e^{-\beta[U(y)-U(x)-W(x,y)-h_tV(y)+h_tV(x)]} \]
Still, this does not completely specify the perturbation. There are many possibilities. The
most common in literature however is the following one, which we use throughout this
chapter:
\[ k_t(x,y) = k(x,y)e^{\frac{\beta h_t}{2}[V(y)-V(x)]} \]
For the treatment of fluctuation-dissipation relations for more general perturbations, see \cite{die05,mw09,rit03}.
From (\ref{pmjp}) we arrive straightforwardly at
\begin{eqnarray*}
  A(\omega) &=& -\log \frac{d\mathcal{P}^h}{d\mathcal{P}^0}(\omega)\\ &=&  
-\frac{\beta}{2}\sum_{s\leq t} h_s[V(x_s)-V(x_{s^-})]\\ && + \int_0^t ds \sum_yk(x_s,y)[ e^{\frac{\beta h_t}{2}[V(x_s)-V(x_{s^-})]}-1 ]
\end{eqnarray*}
The time-antisymmetric part of the action is here
\[ S_{ex}(\omega) =  \beta\sum_{s\leq t} h_s[V(x_s)-V(x_{s^-})] \]
one can check that this is consistent with (\ref{hwork}). The traffic is the time-symmetric part of the action:
\[ \mathcal{T}_{ex}(\omega) = A(\theta\omega)+A(\omega) = 2 \int_0^t ds \sum_yk(x_s,y)[ e^{\frac{\beta h_t}{2}[V(y)-V(x_{s})]}-1 ] \]
and its functional derivative:
\begin{equation}\label{trafmj}
 \tau(\omega,s) = \left.\frac{\partial}{\partial h_s}\mathcal{T}_{ex}(\omega)\right|_{h=0} = \beta\sum_yk(x_s,y)[V(y)-V(x_{s})]
\end{equation}
The fluctuation-dissipation relation then becomes\index{fluctuation-dissipation relation! Markov jump processes}
\begin{eqnarray}
 R_{QV}(t,s) &=& \frac{\beta}{2}\frac{\partial}{\partial s}\left<V(x_s)Q(x_t)\right>_{\mu_0}^0\nonumber\\
 &&-\frac{\beta}{2}\left<\sum_yk(x_s,y)[V(y)-V(x_{s})]Q(x_t)\right>_{\mu_0}^0\label{fdrmj}
\end{eqnarray}

\subsection{A general formula}

An important property of both Markov jump processes and overdamped diffusions, is that 
the functional derivative of traffic is a state function, see (\ref{trafod}) and (\ref{trafmj}), i.e.
$\tau(\omega,s) = \tau(x_s)$. With this in mind we use the equality (\ref{entropytraffic}):
\begin{equation}\label{tussen}
  \beta\frac{\partial}{\partial s}\left<V(x_s)Q(x_t)\right>_{\mu_0}^0 = \left<\tau(x_s)Q(x_t)\right>_{\mu_0}^0
\end{equation}
which is true for any $s>t$, $Q$ and $\mu_0$.  We use the (backward) generator
of the unperturbed Markov process, which is defined (see Appendix \ref{chap-gen}) as follows: 
there exists an operator $L$ such that for any state function $f$ and any initial distribution $\mu_0$:
\[ \frac{d}{d t}\left<f(x_t)\right>^0_{\mu_0} = \left<Lf(x_t)\right>^0_{\mu_0} \]
This $L$ is called the backward generator of the (unperturbed) Markov process.
If we use this generator and take $Q=1$ in (\ref{tussen}) then we get:
\[ \left<\tau(x_s)\right>_{\mu_0}^0 = \beta\left<LV(x_s)\right>^0_{\mu_0} \]
and because this is true for any $s$ and any $\mu_0$, we actually have that 
\begin{equation}\label{trafgen}
 \tau(x_s) = \beta LV(x_s)
\end{equation}
Comparing (\ref{trafod}) and (\ref{trafmj}) to (\ref{bgmj}) and (\ref{bgod}) we can check that this is indeed the case.
This expression also gives an interpretation of $\tau(x)$. Using the definition of the generator $L$, we see that
\[ \tau(x) = \beta LV(x) = \beta\frac{d}{d t}\left.\left<V(x_t)\right>^0_{x}\right|_{t=0} \] 
where the superscript $x$ means that the initial configuration is fixed at $x$. Thus $\tau(x)$ is the instantaneous expected
change in the potential $V$ when started at the state $x$.

The fluctuation-dissipation relation becomes\index{fluctuation-dissipation relation! overdamped diffusions}
\index{fluctuation-dissipation relation! Markov jump processes}:
\begin{equation}\label{fdr}
 R_{QV}(t,s) = \frac{\beta}{2}\frac{\partial}{\partial s}\left<V(x_s)Q(x_t)\right>_{\mu_0}^0 -\frac{\beta}{2}\left<LV(x_s)Q(x_t)\right>_{\mu_0}^0
\end{equation}
which is the most general formulation when restricted to Markov jump processes and overdamped diffusions.
These are large and useful classes of models. Moreover, notice that the result is true for any observable $Q(x_t)$, any potential
$V$ and any initial distribution $\mu_0$. The result is thus true in three nonequilibrium situations:
\begin{enumerate}
 \item The case in which there are nonconservative forces.
 \item The case in which there are different particle reservoirs at the same temperature
but with different chemical potentials.
 \item The case in which the system has an equilibrium dynamics, but has not yet relaxed
 to equilibrium.
\end{enumerate}
Of course combinations of these nonequilibrium conditions are also possible. In this sense
the result (\ref{fdr}) is both very general, and explicit in terms of the generator of the process.
The downside is that experimentally, this generator is not always known. On the other hand,
in the next section we show some examples where this formula leads to
(in principle) measurable results.

\section{Examples}\label{exs}
In this section we give three physical examples to clarify the structure of the fluctuation-dissipation
relations. For two of these examples, simulations have been made by Marco Baiesi, to provide some visualization
and a check of the results \cite{bmw09b}. For simulations it is often more convenient to use a constant
perturbation: $h_s = h, s \ge 0$. As discussed in Section \ref{sec-constant}, this gives an integrated version of the 
fluctuation-dissipation relation. For this section, it is convenient to introduce the following notation:
\begin{itemize}
 \item The generalized susceptibility\index{generalized susceptibility}:
\[ \chi(t) = \lim_{h\to 0}\frac{1}{h}\left[ \left<Q(x_t)\right>_{\mu_0}^h - \left<Q(x_t)\right>_{\mu_0}^0 \right]  \]
If extended to $t\uparrow \infty$, $\chi(t)$ gives the change 
in nonequilibrium stationary expectation when adding a small potential.
\item The correlation function, originally coming from the entropic term in (\ref{fdr}):
\[
C(t) = \left<Q(x_t)V(x_t)\right>_{\mu_0}^0 -  \left<Q(x_t)V(x_0)\right>_{\mu_0}^0
\]
\item The term coming from the traffic term in (\ref{fdr}), (extra with respect to equilibrium):
\[
K(t) = - \int_0^t ds \left<LV(x_s)Q(x_t)\right>_{\mu_0}^0
\]
representing an integrated correlation function. 
\item The average of $C$ and $K$:
\[  C_{NE}(t) = \frac{1}{2}[ C(t) + K(t) ]  \]
\end{itemize}
The fluctuation-dissipation relation
is in this notation expressed by
\begin{equation}\label{chit}
\chi(t) =  \beta C_{NE}(t)
\end{equation}

\subsection{Driven Kawasaki dynamics}\label{ex1}
\index{Kawasaki dynamics}
As a first example we consider an exclusion process as a model of ionic transport through a narrow channel.
The model we use is very similar to the exclusion process described as example 2 in Section \ref{sec-ldb},
differing only by the addition of an interaction potential.
We repeat quickly: the model is described
by a collection of $n$ sites, labelled by $i=1,\ldots,n$, each holding
either one particle ($x^i=1$) or none ($x^i=0$).
A configuration $x$ is thus an array of ones and zeros. 
In the bulk of this system no particles are
created or annihilated, only jumping to neighbouring sites is
allowed (Kawasaki dynamics). 
At the edges $i=1,n$ particles can move in or out from reservoirs
with density $d_1$ and $d_n$, respectively. 
To this a nearest neighbour
interaction is added, with energy $U(x) = -\sum_{i=1}^nx^i x^{i+1}$. 
Moreover, we can add an ``electric'' field $E$ promoting particle jumps to the right.

We have to construct transition rates
for particles hopping to neighbouring sites and rates for creation and
annihilation at the edges. For example, consider the case that a particle enters into site $i=1$ from the reservoir.
Denote the configuration before that jump $x$, and after the jump $y$, where $y=x$ except that $x^1=0$ while $y^1=1$.
Using the local detailed balance assumption, we know that
\[ \frac{k(x,y)}{k(y,x)} = e^{\beta\mu_1}e^{ -\frac{\beta}{2} [U(y) - U(x)]} = \frac{d_1}{1-d_1}e^{ -\frac{\beta}{2} [U(y) - U(x)]} \]
This does not specify the rates fully of course.
For simulational purposes we therefore define the rate for a particle entering at $i=1$ as
\[
k(x,y) = d_1 \psi(x,y)\,\exp\left\{ -\frac{\beta}{2} [U(y) - U(x)]\right\},\quad
\psi(x,y)=\psi(y,x)
\]
where for the moment $\psi(x,y)$ is an arbitrary function.
Then of course, a particle leaving gives
\[ k(y,x) =(1 - d_1) \psi(x,y)\,\exp\left\{ \frac{\beta}{2} [U(y) - U(x)]\right\} \]
and similar for particles entering and leaving at site $n$.
In the bulk of the system particles can hop, e.g. to the right. Let $x$ and $y$ denote configurations
which differ by the fact that a particle at site $i$ has hopped to the right:  $x^i=1, x^{i+1}=0$ and $y^i=0, y^{i+1}=1$.
Then local detailed balance gives
\[
\frac{k(x,y)}{k(x,y)} = \exp\left\{ -\beta [U(y) - U(x) - E]\right\}
\]
Where $E$ is an electric field promoting hops to the right.
We therefore take the hopping rates to be:
\[
k(x,y) = \psi(x,y)\,\exp\left\{ -\frac{\beta}{2} [U(y) - U(x) - E]\right\},\quad
\psi(x,y)=\psi(y,x)
\]
and similarly for jumps to the left. For simulational purposes we have specified
$\psi(x,y)$ by an additional condition, namely that $k(x,y)+k(y,x)=1$ for all $x,y$.

The system is driven from equilibrium by:
\begin{itemize}
\item[i)] setting different reservoir densities $d_1 \ne d_n$, or
\item[ii)] setting a nonzero electric field $E>0$ in the bulk.
\end{itemize}
We choose the total number of particles ${\cal N}(t)=\sum_{i=1}^n x^i$ as observable. We also introduce
a perturbation $V(s)$ equal to ${\cal N}(s)$. This particular perturbation is equivalent to
 changing  the chemical potential of both reservoirs
with a common shift: $\mu_i \to \mu_i + h$:
\[ k_h(x,y) = k(x,y)e^{-\frac{\beta h}{2}[{\cal N}(y)-{\cal N}(x)]} \]
Transition rates for the perturbed process are thus multiplied by
a factor $e^{\beta h/2}$ if a particle enters the system, and by $e^{-\beta h/2}$ when a
particle leaves; transitions in the bulk are left unchanged.

In this case the traffic term is
\[L{\cal N}(x) = \sum_yk(x,y)[{\cal N}(y) - {\cal N}(x)] = {\cal J}(x)\]
which is a current. This current represents the expected change of ${\cal N}$ per unit time,
i.e., it is the rate of change in the number of particles from the two possible
transitions at the boundary sites.
Thus, for $x\to y$ the transition modifying $x^1$, and for $x\to z$ the transition
modifying $x^n$, we have
\[
{\cal J}(x)\! =\!
[{\cal N}(y) - {\cal N}(x)] k(x,y)+
[{\cal N}(z) - {\cal N}(x)] k(x,z)
\]
We have numerically verified
that $\chi,C,K$ as defined earlier, are all equal to each other under equilibrium conditions. While $C = K$ to excellent
precision, the shape of $\chi$ depends weakly on $h$, and is found to
converge to $C$ only for $h$ sufficiently small. In fact, one can pretend
exact matching only in the limit $h\to 0$, but $h=0.01$ turns out to be
sufficiently small to achieve a good convergence.

\begin{figure}[!bt]
\begin{center}
\includegraphics[angle=0,width=10cm]{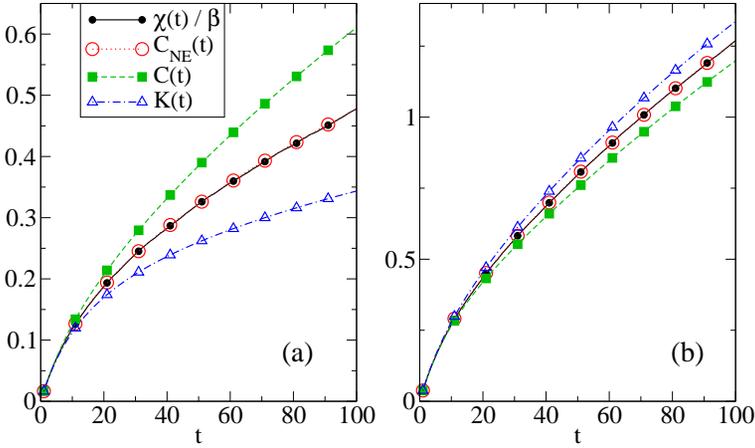}
\end{center}
\caption{Plot of the quantities involved in
Eq.~(\ref{chit}), for (a) case i) ($E=0$) with $n=10$, $\beta=1$, 
$h=-0.01$, and reservoir density unbalance $d_1=0.9$, $d_n=0.1$, 
and (b)  for case ii) ($d_1=d_n=0.5$) with $n=10$,
$\beta=1$, $h=-0.01$, $E=3$.
\label{fig:1}}
\end{figure}

A representative example for the nonequilibrium case i) is shown in
Fig.~\ref{fig:1}(a). One can see that each of the functions $C(t)$  and $K(t)$
is a poor approximation of the response, while
the agreement between $C_{NE}(t)$ and $\chi(t) / \beta$ is excellent.

\begin{figure}[!bt]
\begin{center}
\includegraphics[angle=0,width=10cm]{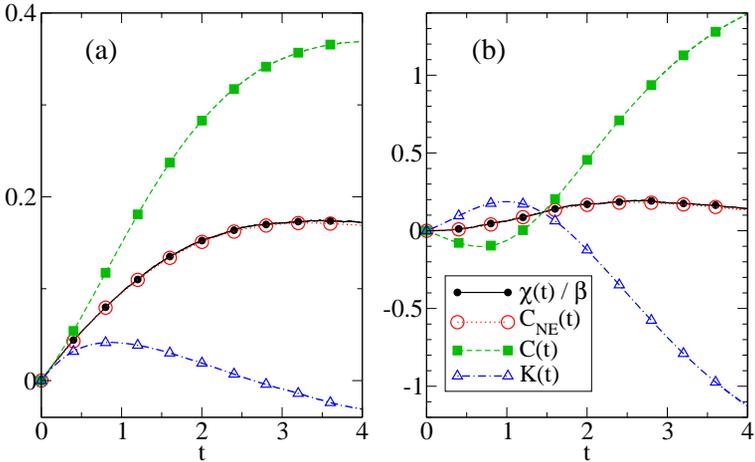}
\end{center}
\caption{Response and fluctuations of the overdamped particle
in a tilted periodic potential, as discussed in the text,
with inverse temperature $\beta=0.2$, 
mobility $\nu=1$  and perturbation $h=0.02$. 
(a) Steady state regime, with initial distribution equal to the
stationary one, and force $f=0.9$.
(b) Transient regime, with initial position $x_0 = 0$ and force $f=0$.
}
\label{fig:overd}
\end{figure}

In Fig.~\ref{fig:1}(b) we show an example of nonequilibrium condition ii).
Again, only $C_{NE}(t)$ matches $\chi(t) / \beta$.
Curiously, a comparison of this example with the previous one reveals that
$C$ can be either larger or smaller than $K$, even for two nonequilibrium conditions
that look pretty similar, in the sense that they yield a current in the
same direction for a relatively simple system.

\subsection{Diffusion on the circle}\label{ex2}

Consider the overdamped Langevin equation (\ref{langover}) for a particle position $x_t \in S^1$ (on a circle)
with a force $F(x) = f-U'(x)$. Here the prime denotes differentiation with respect to space. The constant force $f$
drives the particle around the circle and out of equilibrium. To this equation a perturbation is added:
\begin{equation}
 d x_t = \nu[f-U'(x_t)+h_tV'(x_t)]d t + \sqrt{2D}d B_t
\end{equation}
we renamed the mobility $\nu$ to avoid confusion with the integrated response $\chi(t)$.
 The diffusion constant $D$
and the mobility $\nu$ are related by the Einstein relation $\nu =
\beta D$,  with $\beta$ the inverse temperature (local detailed balance).

This example has been recently experimentally realized as reported in \cite{gpccg09},
for testing the results of \cite{cfg08,cg09}, see also \cite{bslsb07,ss06}.
Moreover, its nonequilibrium stationary distribution $\rho$ is known analytically, see
\cite{mnw08a}.

 Here we  show the result of a simple simulation of the overdamped
particle studied in the experiment,
with a potential $V(x) = \sin(x)$, and also $U(x)=Q(x)=V(x)$, and
a constant force $f$.
The functional derivative of the traffic $\tau(x) = \beta L V(x)$ of a particle at position $x$
 can be computed by applying the generator
of the overdamped dynamics
\[
L = (f-U')\frac{d}{d x} + \frac{1}{\beta}\frac{d^2}{d x^2}
\]
to the potential $V(x)$, which gives $\tau(x) =
\beta[f-\cos(x)]\cos(x) - \sin(x)$. We have measured the
correlations $C(t)$ and $K(t)$ as well as responses
$\chi(t)$ for small $h$.

Fig.~\ref{fig:overd}(a) shows that for strong stationary nonequilibrium
$f\gg 0$ there is a large difference between the integrated
correlation function of the traffic and that of the entropy
production. However, their average $C_{NE}(t)$ agrees very well
with the response $\chi(t)$, as it should.
It is also important to remember that our approach works for
non-stationary regimes as well. In Fig.~\ref{fig:overd}(b) we show
an example of a particle  starting at time $t=0$ from
a non-stationary initial distribution $\mu(x) = \delta_{x,0}$
(i.e. its position is $x_0=0$), but for the rest
not forced outside equilibrium, $f=0$, 
to emphasize the transient character of this situation.
Again, we can see that the response is well estimated by $C_{NE}(t)$.

In the next chapter we treat the inertial version
of this model.

\subsection{A generalized Einstein relation}\label{ex3}
\index{Einstein relation!generalized}
Consider the overdamped Langevin equation in $\mathbb{R}^d$ (\ref{langgen}) but with constant $\nu = \beta D$:
\begin{equation}\label{ger}
 d x^i_t = \nu^{ij}[F^j(x_t) + h\delta^{jk}]d t +\sqrt{2D}^{ij}d B^j_t
\end{equation}
with repeated indices $j=1,\ldots d$ summed over.
The $F$ includes all forces and $h$ is the small constant perturbation.
Note that this means that the perturbation potential is $V(x) = x^k$.
With such a potential the fluctuation-dissipation relation connects
mobility and diffusion which are actually defined as follows.
The true mobility $M$ (in contrast to $\nu$) is defined as the response
\begin{equation}\label{mmm}
 M^{ik} = \lim_{t\to\infty}\left.\frac{d}{dh}\left<
 \dot{x}^i_t \right>^h_{\rho}\right|_{h=0} = \lim_{t\to\infty}\frac{1}{t}\int_0^t\left.\frac{d}{dh}\left<
 \dot{x}^i_t \right>^h_{\rho}\right|_{h=0}
\end{equation}
while the real diffusion matrix (in contrast to $D$) is given by
\[
 \mathcal{D}^{ik} = \lim_{t\to\infty}\frac{1}{2t}\int_0^td s\int_0^td r\
\left<(\dot{x}^i_{s}-\left<\dot{x}^i\right>_{\rho})(\dot{x}^k_{r}-\left<\dot{x}^k\right>_{\rho}) \right>^0_{\rho}
\]
As (\ref{mmm}) is a response function, we can easily compute it
within our framework. Note that $LV(x) = \nu^{kj}F^j(x)$:
\[ \left.\frac{d}{dh}\left<
 \dot{x}^i_t \right>^h_{\rho}\right|_{h=0} = \frac{\beta}{2}\left<[x^k_t-x^k_0]\dot{x}^i_t\right>^0_{\rho} - 
\frac{\beta \nu^{kj}}{2}\int_0^tds\left< F^j(x_{s})\dot{x}^i_{t} \right>^0_{\rho} \]
We rewrite this response function by using the following equality, derived from
taking the average of the Langevin equation ((\ref{ger}) with $h=0$):
\[ \left<\dot{x}^i\right>^0_{\rho} = \left<\nu^{ij}F^j(x)\right>^0_{\rho} \]
because the average of $dB_t$ gives zero. 
Using this equality we get
\begin{eqnarray*}
  \left.\frac{d}{dh}\left<
 \dot{x}^i_t \right>^h_{\rho}\right|_{h=0} &=& \frac{\beta}{2}\int_0^tds[\left<\dot{x}^k_s\dot{x}^i_t\right>^0_{\rho}- \left<\dot{x}^k_s\right>^0_{\rho}\left<\dot{x}^i_t\right>^0_{\rho} ]\\
&& - \frac{\beta \nu^{kj}}{2}\int_0^tds\left< F^j(x_{s})(\dot{x}^i_{r}-\left<\dot{x}^i\right>^0_{\rho}) \right>^0_{\rho}
\end{eqnarray*}
Now all that is left is integrating this relation over $t$, dividing by $t$ and taking the limit for $t\to\infty$
to arrive at
\begin{equation}\label{er}
 M^{ik} = \beta\mathcal{D}^{ik}
 -\frac{\beta\nu^{kj}}{2}\lim_{t\to\infty}\frac{1}{t}\int_0^td
 s\int_0^td
 r\left< F^j(x_{s})(\dot{x}^i_{r}-\left<\dot{x}^i\right>^0_{\rho}) \right>^0_{\rho}
\end{equation}
Observe that in equilibrium ($F$ derives from a potential), the second term in (\ref{er})
vanishes because the observable $\dot{x}_t$ is anti-symmetric, and $\left<\dot{x}\right>^{\rm{eq}}_{\rho}=0$. Then,
the equilibrium fluctuation-dissipation relation holds with
$M=\beta {\cal D}$. Moreover, when $F=0$, (pure diffusion) these quantities can be explicitly calculated, and are found to be equal to
$M^{ik} = \nu^{ik}, \mathcal{D}^{ik} = D^{ik}$. We see that for pure diffusion the fluctuation-dissipation theorem
is equivalent to the Einstein relation. But even for equilibrium systems, when $F\neq 0$ the fluctuation-dissipation theorem
is not equivalent with the Einstein relation, although there is still a simple relation between mobility and diffusion.

Out of equilibrium, this simple relation is violated, as there is a correction, namely the second term in (\ref{er})).
Note that this is completely
compatible with the condition $\nu=\beta D$, because that is actually a relation for the reservoir, 
which is in thermal equilibrium at inverse temperature $\beta$. Similar results for a one-dimensional
overdamped diffusion were reported before in \cite{bslsb07,ss06}.

\section{Connection with effective temperature}\label{sec-efftemp}
\index{effective temperature}
A reason, not mentioned before, that the fluctuation-dissipation theorem in equilibrium
is so useful and powerful is that it specifies the temperature as a
universal parameter: regardless of what potential we use and what observable,
the relation between response and correlation function is always governed by
that same single parameter. In this sense it is not strange to wonder
what remains of this parameter out of equilibrium.

Therefore, even though the fluctuation-dissipation theorem is violated out of equilibrium, one
attempts to restore it by the
introduction of an effective temperature $T^{\textrm{eff}}$, in the
sense
\begin{equation}\label{ef}
R_{QV}^\mu(t,s)= \frac 1{k_BT^{\textrm{eff}}}\frac{d }{d s}
\langle V(s) Q(t)\rangle^0_{\mu}
\end{equation}
Many studies have been devoted to the study of this prefactor, in what sense it perhaps
resembles a thermodynamic temperature-like quantity for some classes of
observables and over some timescales; we refer to \cite{cg05,cr03,hp09,kur05} for an
entry into the extensive literature.  Clearly, whatever the
purpose of the discussion, an exact expression of the response should
help, especially when entirely in terms of explicit correlation functions.  The first calculations
in this sense are in \cite{cha03} and they have been referred to as the ``no field-method''
\cite{ric03}. In particular, for purposes of simulation or numerical verification of (\ref{ef}) 
we
do no longer need to perform the perturbation by hand.
In fact, now we can write the ratio $T/T^{\textrm{eff}}= X$ entirely
in terms of correlation
 functions
 \begin{equation}\label{x} X= X_{QV}(\mu;t,s) = \frac 1{2}\big[ 1 - \frac{\langle
LV(s)\,Q(t)\rangle^0_\mu}{\frac{d}{d s} \langle V(s)
Q(t)\rangle^0_\mu}\big]
\end{equation}
with numerator and denominator in (\ref{x}) each having a specific
physical meaning in terms of entropy and traffic. An effective
temperature is obtained as the ratio between the traffic and the
entropic term: if for some observables $(V,Q)$ and over
some time-scales,
\[
Y\;\frac{d}{d s}\langle
V(s)\,Q(t)\rangle^0_\mu = \langle LV(s)
\,Q(t)\rangle^0_\mu
\]
for some $Y$, then $X=(1-Y)/2$.  Equilibrium has   $X=1=-Y$.  
In the case where $LV\approx 0$ as for a conserved quantity, then
$Y=0$ and $T^{\textrm{eff}}=2T$. Finally,  $X$ and the effective temperature
$T^{\textrm{eff}}$ get negative when the traffic term overwhelms the
entropic contribution.

One should understand that
(\ref{ef}) represents a rather optimistic scenario. Formula (\ref{ef}) mimics
(\ref{fdt3}) by replacing a function depending a priori on the observable, the potential, the initial distribution and on other parameters
as temperature and time, by just one parameter. Why should
there be also out-of-equilibrium a single parameter and a useful notion of temperature in
its usual thermodynamic
 understanding? Moreover, how would it depend on
  the observables $V$ and $Q$? (See \cite{mbd09} for a very recent discussion.)
    Answers to these questions have
  been partially given but are often restricted within a context of mean
   field systems or for small fluctuations, effectively dealing with
calculations as in Section \ref{sec-meanfield}, similar to calculations for scalar fields as in \cite{ckp94}
and in \cite{cg05}. In fact, the optimism in (\ref{ef})
 is a sort of conservatism as it tries to return to equilibrium-like formulae.  
We take a different attitude: the violation of
the equilibrium fluctuation-dissipation relation (FDR) is an opportunity to discover new 
connections between response and the relevant newly emerging
physical quantities as traffic.

\subsection{Explicit calculations for pure diffusions}\label{sec-pure}
In simple overdamped diffusion equations, one can explicitly calculate the correlation functions,
and see when the concept of effective temperature could make sense. 
Similar calculations have been done in \cite{cg05,ckp94}.
The simplest example is Brownian motion in one dimension:
\[ dx_t = h_tdt + \sqrt{2D}dB_t \]
where we have taken the potential $V(x)$ equal to the position $x$.
The generator of the unperturbed dynamics is (see Appendix \ref{chap-gen})
$L=D\Delta$, so that $LV=0$ and $T^{\textrm{eff}}=2T$ cf.
Virasoro's example in \cite{ckp94}. In that last reference, what
are called ``flat directions'' can be associated to perturbations with $LV=0$.\\

\subsection{Explicit calculations for linear diffusions}\label{sec-meanfield}
Next, consider the following model for diffusions with a linear force (harmonic potential).
Fix parameters $\alpha, B \in \mathbb{R}, D>0$ and look at the linear
Langevin dynamics for a global order parameter $M \in \mathbb{R}$,
\[
dM(t) = -\alpha\,M(t)dt + h_t\,Bdt + \sqrt{2D}\,dB(t)
\]
The $h_t, t>0,$ is a small time-dependent field. The
generator of such a dynamics dynamics (on observables $f$) is (see Appendix \ref{chap-gen})
$L^hf(M) = [-\alpha \,M + h_t B]\, f'(M) + D\,f''(M)$, using a prime to denote
differentiation with respect to $M$.  Such a dynamics can arise as 
a Gaussian approximation to a relaxational dynamics
of the scalar magnetization $M$ (no conservation laws and no spatial
structure) valid in high enough dimensions (above $d=4$ for the standard Ising model). Then,
 in a way, $\alpha = 0$  corresponds to the
critical (massless) dynamics and $\alpha >0$ is a paramagnetic dynamics (high temperature), see \cite{cg05}.
By taking $D\downarrow 0$ we exclude the diffusive aspects and we can think then of gradient
relaxation in the low temperature regime.

The equilibrium (reversible stationary density on $\mathbb{R}$ for perturbation $B=0$) is
\[
\rho(M) =\frac 1{Z}\exp \left\{-\alpha\frac{M^2}{2D}\right\}
\]
with zero mean and variance $\langle M^2\rangle = D/\alpha$.
We now start from an initial distribution $\mu_0$. If $\mu_0\neq \rho$ then the system is not
in equilibrium. We can compute the response function by calculating the expectation value of $M$ at time $t$ in the
perturbed dynamics. We do that by deriving the following differential equation
\[ \frac{d}{dt}\langle M(t)\rangle^h_{\mu_0} = \langle L^hM(t)\rangle^h_{\mu_0} = -\alpha\langle M(t)\rangle^h_{\mu_0} + h_t B  \]
the solution of which is
\[\langle M(t)\rangle^h_{\mu_0} = \left<M(0)\right>^0_{\mu_0}\,e^{-\alpha t} + B\int_0^t d s\,h_s\,e^{-\alpha(t-s)}\]
or
\begin{equation}\label{res}
\frac{\delta}{\delta h_s}\left.\langle M(t) \rangle^h_{\mu_0}\right|_{h=0} = B\,e^{-\alpha (t-s)}
\end{equation}
which does in fact not depend on $\left<M(0)\right>^0_{\mu_0}$ (and thus also equals the equilibrium result).
In the same way, we use for $s<t$: 
\[\frac{d}{dt}\langle M(t)M(s)\rangle^0_{\mu_0} = \langle M(s)L^0M(t)\rangle^0_{\mu_0} 
= -\alpha\langle M(t)M(s)\rangle^0_{\mu_0}\]
to arrive at
\[ \langle M(t)M(s)\rangle^0_{\mu_0} = \langle M^2(s)\rangle^0_{\mu_0}e^{-\alpha t} \]
Using same strategy we can calculate the expectation of $M^2(s)$.
Denoting  $\overline{M_0^2} = \left<M^2(0)\right>^0_{\mu_0}$, the correlation function for $0<s<t$ is
\[\langle M(s)\,M(t)\rangle^0_{\mu_0}= \overline{M_0^2}\,e^{-\alpha(t+s)} + \frac{D}{\alpha}
\big[ e^{-\alpha(t-s)} - e^{-\alpha(t+s)}\big]\]
and hence
\[\frac{d}{d s}\langle M(s)\,M(t)\rangle^0_{\mu_0} = -\alpha \overline{M_0^2}\,e^{-\alpha(t+s)} + D
\big[ e^{-\alpha(t-s)} + e^{-\alpha(t+s)}\big]\]
If $\mu_0 = \rho$ in the last expression (thus replacing $\overline{M_0^2}$ by
$D/\alpha$) we find
\[\frac{d}{d s}\langle M(s)\,M(t)\rangle^0_{\rho} =  D\, e^{-\alpha(t-s)}\]
which, in comparison with (\ref{res}) specifies the equilibrium inverse temperature to be equal to
$\beta=B/D$.

The traffic term is  obtained from $LM= -\alpha M$, and thus
\[
\langle LM(s)\,M(t)\rangle^0_{\mu_0} =
-\alpha\,\overline{M_0^2}\,e^{-\alpha(t+s)} - D
\big[ e^{-\alpha(t-s)} - e^{-\alpha(t+s)}\big]
\]
Clearly,
\[
\frac{\delta}{\delta h_s}\langle M(t) \rangle^h_{\mu_0}(h=0) 
=\frac{B}{2D}\{\frac{d}{d
s}\langle M(s)\,M(t)\rangle^0_{\mu_0}-\langle LM(s)\,M(t)\rangle^0_{\mu_0}\}
\]
as it should.

For the issue of effective temperature we compute the ratio $Y$
as
\[
Y = Y(M_0;s,t) =
\frac{-\alpha \langle M(s) M(t)\rangle^0_{\mu_0}}{\frac{d}{d s}\langle
M(s)M(t)\rangle^0_{\mu_0}} =
\frac{D-\alpha \overline{M_0^2}  - De^{2\alpha s}}{D-\alpha \overline{M_0^2} + De^{2\alpha s}}
\]
Remark that $Y$ is independent of $t$, as long as $t>s$.
In that notation, the effective inverse temperature is $T^{\mbox{eff}} = 2T/(1-Y)$. In
equilibrium $Y=-1$ while $Y=1$ for $D=0$ and $M_0\neq 0$. If $\alpha=0$, then $Y=0$ as
 in the pure diffusion case. For $\alpha>0$, if we let $s\to\infty$ while keeping $t-s=u$
fixed, we get
\begin{eqnarray}
\lim_{s\uparrow+\infty} Y(M_0;s,s+u) = -1
\end{eqnarray}
so that $T^{\mbox{eff}} = T$. This case is referred to as the paramagnetic case, 
while $T^{\mbox{eff}} = 2T$ for $\alpha=0$ is the critical 
quench, cf. \cite{cg05}.

\subsection{Effective traffic}
\index{effective traffic}
Let us try to see how the notion of effective
temperature could be seen as a one-parameter reduction of a general
equilibrium-like FDR that is valid also outside equilibrium but with an effective dynamics.
The starting point is comparing the equilibrium formula (\ref{fdt3}) with the more general 
non-equilibrium one (\ref{fdr}). By this we see that in equilibrium (\ref{fdt3}) is equivalent with
\begin{equation}\label{efo}
  R_{QV}(t,s) \!=\!
   -\beta\langle (LV)(s)\,Q(t)\rangle^0_\rho
   \end{equation}
In other words,
the fluctuation-dissipation theorem in equilibrium can also be
called a fluctuation-traffic theorem; the two terms on the
right-hand side of (\ref{fdr}) are simply the same in equilibrium.  Therefore,
for the purpose of getting closer to equilibrium response formulae
one really has the choice to mimic either (\ref{fdt3}) or rather
(\ref{efo}).  The first leads to the ambition of effective
temperature (\ref{ef}), the latter to the new notion  of effective
traffic.  But the latter is also much richer.  In fact, a combination
of (\ref{fdr}) with (\ref{corrder})
shows that the exact nonequilibrium response formula (\ref{fdr})
can indeed be written in the equilibrium form (\ref{efo}):
\begin{equation}\label{inst}
 R_{QV}^\mu(t,s) =
  -\left<G_{\mu_s}V(s)Q(t)\right>^0_{\mu}
\end{equation}
with a new effective traffic
\begin{eqnarray}\label{Gs}
 G_{\mu}V &=& \frac{\beta}{2\mu}[L^{\dag}(\mu V)-VL^{\dag}\mu+\mu LV]
\end{eqnarray}
Here $L^{\dag}$ is the forward generator,
defined by (see Appendix \ref{chap-gen}, (\ref{backfor})):
\[ \int dx g(x)Lf(x) = \int dx f(x)L^{\dag}g(x) \ \ \ \ \ \forall f,g \]
where the integration has to be replaced by a sum for Markov jump processes. 
The operator $G_{\mu}$ acting on $V$ in
(\ref{Gs}) has the following exact property: it is itself a
generator but of a new dynamics for which $\mu$ is an
equilibrium distribution (i.e. $\langle f\,(G_{\mu}g)\rangle^0_\mu =
\langle g\,(G_{\mu}f) \rangle^0_\mu$ for all $f,g$, see (\ref{eqadj})).  Thus, in
(\ref{inst}) the generator $G_{\mu_s}$ is the instantaneous
equilibrium generator
 with respect to the time-evolved distribution $\mu_s$.

We can rewrite $G_{\mu}$ in terms of the adjoint generator 
(see (\ref{adjoint})), which is defined using the stationary distribution
of the process:
\[\int dx \rho(x)g(x)L^*f(x) = \int dx \rho(x)f(x)Lg(x)\ \ \ \ \ \forall f,g\]
With this we get
\begin{eqnarray}\label{Gsa}
 G_{\mu}V &=& \frac{\beta\rho}{2\mu}[L^*(\frac{\mu}{\rho} V)-VL^*(\frac{\mu}{\rho})+\frac{\mu}{\rho} LV]
\end{eqnarray}
In the stationary nonequilibrium case, we have $\mu_s=\rho$ and
(\ref{Gs}) is
\begin{equation}\label{sa} 
G_\rho = \frac{\beta}{2} (L + L^*)
\end{equation}
replacing $L$ in (\ref{efo}).
For equilibrium dynamics $L = L^*$ (see Appendix \ref{chap-gen}),
and (\ref{inst}) reduces to (\ref{efo}). In stationary nonequilibrium,  if the perturbation $V$ is
`time-direction independent' in the precise sense that $LV=L^*V$,
 then the nonequilibrium response
  reduces to the equilibrium formula (\ref{efo}) and $X=1$.
See  \cite{ns08} for very related conjectures and observations.\\

\cleardoublepage
\chapter{Underdamped diffusions}\label{chap-under}

\textit{In this chapter we treat the case of underdamped diffusions, for which 
the formula (\ref{fdr}) is no longer correct.
We can, however, still calculate the response function and express it in terms
of correlation functions of observable quantities. For several explicit examples
the response is calculated and simulated. This chapter describes work reported in
\cite{bbmw09}.}

\section{Model}

In this chapter we consider an extension of the one-dimensional underdamped diffusions
explained in Section \ref{sec-lang}. There we considered the dynamics of just one
particle. Now we consider $k$ particles in $d$ dimensions.
This means that the state of the system is now given by
 $(x,v) = (x_1,x_2,\ldots,x_n;v_1,v_2,\ldots,v_n)\in \mathbb{R}^{2n}$ of positions
and momenta, with $n=kd$. These particles are subject to mechanical forces:
a potential $U(x)$ taking care of the coupling and
pinning of the positions (the pinning is also thought to confine
the positions to some finite volume), and a nonconservative forcing $f_i(x)$.
Each particle is connected to its own heat bath at an inverse temperature $\beta_i$.
This gives the following set of $2n$ Langevin equations\index{diffusions!underdamped}:
\begin{eqnarray}
d x^i_t &=& v^i_t\, d t \nonumber\\
m_idv^i_t &=&
 [f_i(x)-\frac{\partial U}{\partial x_i}(x_t) -m_i\gamma_i v^i_t] d t 
 + \sqrt{2D_i}\,d B_i(t)
\label{ud}
\end{eqnarray}
Again, we make no distinction between upper or lower indices: $x_i = x^i$.
To avoid confusion, we always use $i,j,k$ to denote the components of the vectors,
and $s,t$ to denote times.
For the simple underdamped diffusions in (\ref{sec-lang}), we found by the
local detailed balance assumption the Einstein relation
between diffusion and friction coefficient. Let us investigate what it gives here.
For this we write down the path-probability density for paths $\omega = (x_t,v_t)_{0\leq t\leq T}$.
In the same way as explained in Section \ref{sec-trajud} we find, formally:
\begin{eqnarray}\label{ppud}
 \mathcal{P}_{x_0,v_0}(\omega) = \frac{1}{\mathcal{N}}\exp\left\{ -\sum_i\frac{1}{4D_i}\int_0^Tdt[m_i\dot{v}^i_{t} - 
F_i(x_t)+m_i\gamma_i v^i_{t} ]^2 \right\}
\end{eqnarray}
with $F_i = f_i-\frac{\partial U}{\partial x_i}$, and $\mathcal{N}$ a normalization factor.
The entropy flux into the environment is then
\begin{equation}
 S(\omega) = \frac{d\mathcal{P}_{x_0,v_0}}
{d\mathcal{P}_{x_T,v_T}\theta}(\omega) = -\sum_i\frac{m_i\gamma_i}{D_i}\int_0^Tdt v_i[m_i\dot{v}^i_{t} - 
F_i(x_t)] 
\end{equation}
Note that the quantity
\[ m_i\dot{v}_{i} - F_i = -m_i\gamma_i v_i + \sqrt{2D_i}\frac{d B_i}{dt} \]
is the (stochastic) force on particle $i$ originating from the $i$-th heat bath.
The integral over time of this force times the velocity of the particle thus gives the heat flux
from the $i$-th reservoir to the system. As a consequence $S(\omega)$ is the entropy flux
into the environment if and only if $D_i = \frac{m_i\gamma_i}{\beta_i}$ for all $i$.
Note that the system is driven out of equilibrium by the nonconservative force,
but also by the difference of the temperatures of the heat baths $\beta_i$.

At time zero, the probability density of the system being in state $(x,v)$ is denoted as always by
$\mu_0(x,v)$. The Fokker-Planck equation determining the time-evolution of this probability in this case is
 \begin{equation}\label{fk}
 \frac{d }{d t}\mu_t + \nabla \cdot J_{\mu_t} =0
 \end{equation}
for $\nabla = (\nabla_x,\nabla_v)$ and for the probability current $J_\mu = (J_\mu^x,J_\mu^v)$ with
 \begin{equation}\label{curre}
 (J_\mu^x)_i= m_iv_i\mu,\qquad (J_\mu^v)_i = (f_i-\frac{\partial U}{\partial x_i})\mu - 
\gamma_i \,v_i \,\mu - D_i\,\frac{\partial\mu}{\partial v_i}
 \end{equation}

\section{Perturbation}

The perturbation is a potential $V(x)$ added to
the unperturbed Hamiltonian\\ $H_o= \sum_i \frac{m_iv_i^2}{2} + U(x)
\rightarrow H_o - h_s\,V(x)$ with small
 time-dependent amplitude $h_s, s\geq 0$.
We compare the path-probabilities of the perturbed
versus the unperturbed process, using the action. By (\ref{ppud}) we find
\[ \frac{d\mathcal{P}^h_{x_0,v_0}}{d\mathcal{P}^0_{x_0,v_0}}(\omega) = e^{-A(\omega)} = \exp\left\{\frac{S_{ex}(\omega)-\mathcal{T}_{ex}(\omega)}{2}\right\} \]
where the action is split in its time-antisymmetric and symmetric parts. The time-antisymmetric part is:
\begin{equation}\label{eq:S}
S_{ex}(\omega)= \sum_{i=1}^n \beta_i \int_0^t \,h_s \frac{\partial
V}{\partial x_i}(x_s) \,v_s^i\,d s
\end{equation}
Note the following: if all temperatures are the same, $\beta_i=\beta$, then
$S_{ex}(\omega) = \beta \int_0^t \,h_s \frac{\partial
V}{\partial s}(x_s) d s$, which is consistent with (\ref{hwork}).  
The time-symmetric part of the action gives: 
 \begin{eqnarray}
\mathcal{T}(\omega) &=& \sum_i\frac{1}{D_i} \int_0^t \,h_s
\frac{\partial V}{\partial x_i}(x_s)\Bigl[\{f_i(x_s) -  \frac{\partial
U}{\partial x_i}(x_s)\}d s - m_id v^i_s\Bigr] + o(h) \nonumber\\
&&  \label{eq:tau}
\end{eqnarray}
The last stochastic integral can be interpreted in the It\^o or Stratonovitch sense,
this does not matter in this case, see (\ref{itostrat}). However, the stochastic integral
is the reason that we can't write the functional derivative of traffic as a state function.
Indeed, $dv_i$ contains information about two consecutive states, not one.

\section{Result}\label{sec-result}

The model that we use in this chapter is in contact with more than one heat bath, possibly at
different temperatures. Therefore we can't use the formulae derived in Section \ref{sec-oneheat}.
We therefore fall back on the more general formula (\ref{response1a}).
We consider general observables $Q(x_t,v_t)$ which we simply denote by $Q(t)$. The immediate
application of (\ref{eq:S}) and (\ref{eq:tau}) to (\ref{itostrat}) gives
\begin{eqnarray*}
 &&\frac{\delta}{\delta h_s} \left.\langle Q(t)\rangle^h_\mu\right|_{h=0}\ \ \ = \ \ \
\sum_{i=1}^n \frac{\beta_i}{2} \langle \frac{\partial V}{\partial
x_i}(x_s) \,v_s^i\,Q(t)\rangle_{\mu_0} \\ &&\ \ \ \ \ \ \ \ \ \ \ \ -\frac 1{2}\sum_{i=1}^n\frac{1}{D_i}\langle
\frac{\partial V}{\partial x_i}(x_s) \,[f_i(x_s) -  \frac{\partial
U}{\partial x_i}(x_s) - m_i\dot{v}_s^i]\,Q(t)\rangle_{\mu_0}
\end{eqnarray*}
However, because of the singular nature of the white noise, $\dot{v}$ is not a good
observable. We therefore rewrite that part of the correlation function. We take the stochastic integral in the It\^o sense,
which is easiest here. This means that we have to interpret the correlation function containing $\dot{v}_i$
as follows:
\[ \left<\frac{\partial V}{\partial x_i}(x_s) \,\dot{v}_s^iQ(t)\right>_{\mu_0}
= \lim_{\epsilon \to 0}\frac{1}{\epsilon} \left\{
\left<\frac{\partial V}{\partial x_i}(x_s) [v^i_{s+\epsilon}-v^i_s]Q(t)\right>_{\mu_0}\right\} \]
We rewrite this inspired by the product rule for derivatives to
\begin{eqnarray*}
 \lim_{\epsilon \to 0}\frac{1}{\epsilon} \left\{
\left<[\frac{\partial V}{\partial x_i}(x_{s+\epsilon}) v^i_{s+\epsilon}-\frac{\partial V}{\partial x_i}(x_s)v^i_s]Q(t)\right>_{\mu_0}\right\}\\
 - \lim_{\epsilon \to 0}\frac{1}{\epsilon} \left\{
\left<[\frac{\partial V}{\partial x_i}(x_{s+\epsilon}) -\frac{\partial V}{\partial x_i}(x_s)]v^i_{s+\epsilon}Q(t)\right>_{\mu_0}\right\}
\end{eqnarray*}
Finally taking the limit, we obtain a useful form:
\begin{eqnarray}\label{trafcorr}
\left<\frac{\partial V}{\partial x_i}(x_s) \,\dot{v}^i_sQ(t)\right>_{\mu_0}
&=& \frac{d}{d s}\langle \frac{\partial V}{\partial
x_i}(x(s)) \, v^i_s\,Q(t)\rangle_\mu \\ 
&& - \sum_j\langle \frac{\partial^2 V}{\partial x_j\partial
x_i}(x_s) \, v^j_s\,v^i_s\,Q(t)\rangle_\mu\nonumber 
\end{eqnarray}
With this the response function becomes
\begin{eqnarray}\label{ttt}
\frac{\delta}{\delta h_s} \left.\langle Q(t)\rangle^h_{\mu_0}\right|_{h=0} &=&
\frac{1}{2}\sum_{i=1}^n \beta_i \langle \frac{\partial V}{\partial
x_i}(x_s) \,v^i_s\,Q(t)\rangle_\mu\\
&& - \frac{1}{2} 
\sum_i\frac{1}{D_i}\langle
\frac{\partial V}{\partial x_i}(x_s)\,[f_i(x_s) -
\frac{\partial U}{\partial x_i}(x_s)]\,Q(t) \rangle_{\mu_0}\nonumber\\
&& +\frac{1}{2} \frac{d}{d s} \sum_i \frac{m_i}{D_i}\langle \frac{\partial V}{\partial x_i}(x_s)
\, v^i_s\,Q(t)\rangle_{\mu_0}\nonumber\\ 
&& - \frac{1}{2}\sum_{i,j} \frac{m_i}{D_i}\langle \frac{\partial^2 V}{\partial x_j\partial x_i}(x_s) 
\,v^j_s\,v^i_s\,Q(t)\rangle_{\mu_0}\nonumber
\end{eqnarray}
This may be a complicated formula, but we now have expressed the response function in terms of correlation functions of observables
which are in principle measurable.

\section{Examples}
The result (\ref{ttt}) is a complicated formula. This
is why we have investigated it in several physically interesting examples.
For two of these examples, simulations have been provided by Marco Baiesi \cite{bbmw09}.

We remind that a regime out of equilibrium can be created in (\ref{ud}) by 
letting the inverse temperatures $\beta_i$ differ from one another.
Another way of going out of equilibrium is to introduce nonconservative forces
$\{f_i\}$, like external fields that are rotational.
A final possibility that we can consider is to start from an initial condition 
that is not stationary, and thus to observe the response in a transient regime.
All of these possibilities are covered in the following examples.
As in the previous chapter, numerical results are better presented with integrated responses.
We use slightly different definitions than in Section \ref{exs}:
\begin{itemize}
 \item The integrated response, also called generalized susceptibility\index{generalized susceptibility}:
\[ \chi(t) = \int_0^t d s\, R_{QV}(t,s) = \lim_{h\to 0}\frac{\left<Q(t)\right>_{\mu_0}^h-\left<Q(t)\right>_{\mu_0}^0}{h}\]
 \item The correlation with entropy:
\[ C(t) = \frac{1}{h}\left<S(\omega) Q(t)\right>_{\mu_0} \]
 \item The correlation with traffic:
\[ K(t) = \left<\frac{\partial }{\partial h}\left.\mathcal{T}(\omega)\right|_{h=0} Q(t)\right>_{\mu_0} \]
 \item The average
\[ C_{\textrm{ne}}(t)= \frac{C(t) + K(t)}{2} \]
\end{itemize}
Contrary to the examples in the last chapter, here we have embedded $\beta_i$'s in the definitions of
the correlation functions. The integrated response relation is thus $\chi(t) = C_{\textrm{ne}}(t)$.
We choose for simplicity particles with mass equal to
one in all the examples.

\subsection{Langevin particle in a periodic potential}
Recently there have been experiments testing the response of an
overdamped particle (high viscosity limit) in a periodic
potential \cite{gpccg09}. In the previous chapter, section \ref{ex2}, we have 
discussed simulations of that system. Here we look for the changes in an
underdamped set-up, allowing e.g. for the particle to have a
considerable mass and to obey a noisy Hamiltonian dynamics.

We consider here a system consisting of one particle in one dimension:
 its position is $x_t \in S^1$ (on a circle) and the velocity is $v_t\in \mathbb{R}$.
 Then, the equations (\ref{ud}) simplify to
\begin{eqnarray}\label{lang}
  d x_t &=& v_t d t\nonumber\\
  d v_t &=& F(t) d t - \gamma v_t d t \,{ - h_t\, g(t)} d t + \sqrt{2 D} \,d B_t
\end{eqnarray}
where we abbreviate
\begin{eqnarray}
F(t)  &=& f - \frac{d U}{d x}(x_t) \qquad\textrm{(deterministic force)}\nonumber\\
{ g(t)} &=& { - \frac{d V}{d x}}(x_t) \qquad\textrm{(perturbing force)}
\nonumber
\end{eqnarray}
The nonconservative force $f$ is the driving and is taken constant over the circle, thus
effectively tilting the conservative potential. This drives the system out of equilibrium.
As there is only one particle, there is only one heat bath.

At time $s=0$ the unperturbed system is taken to be in the stationary nonequilibrium $\rho$
corresponding to (\ref{fk}), and then for $s>0$ a constant small $h$ is turned on.
Hence,  the integrated correlations are
\begin{eqnarray}
  C(t) &=&   =- \beta \int_0^t d s \,\left<v_s g(s) Q(t)\right>^0_\rho = \beta \,[ \left<V(t)Q(t)\right>^0_{\mu_0} - \left<V(0)Q(t)\right>^0_{\mu_0} ]
 \nonumber\\
  K(t) &=&   = \frac{1}{D} \left\{ \int_0^t ds \left< F(s) g(s) Q(t) \right>^0_\rho -
 \int_{0}^{t} \left<d v(s)  g(s) Q(t) \right>^0_\rho\right\}\nonumber\\
\end{eqnarray}
Note that we left $dv_s$ in the definition of $K(t)$. This can of course be rewritten
as explained in Section \ref{sec-result}. However, in simulations one discretizes
time, and $dv_s$ is known at every time-step. It can therefore be used as an observable here. 
 We
take $V(x) = U(x) = \cos (2\pi x)$ like in the previous chapter, and again also $Q = U$.

\begin{figure}[!bt]
\begin{center}
\includegraphics[angle=0,width=12.0cm]{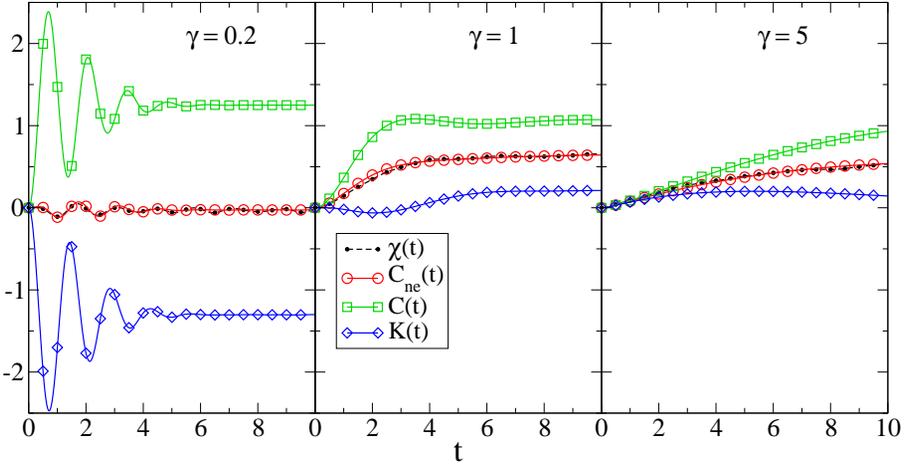}
\end{center}
\caption{Integrated correlation functions of the entropic term 
$C(t)$, of the traffic term $K(t)$, and their average $C_{\textrm{ne}}(t)$ giving the response in nonequilibrium,
and the integrated response $\chi(t)$ calculated directly with $h=0.01$.
Panels are for simulations with various friction coefficients: 
$\gamma=0.2$ (left), $\gamma=1$ (center), and $\gamma=5$ (right).
Other parameters: $T=1/\beta=0.2$, $f=0.9$.
\label{fig:Lang}}
\end{figure}

\begin{figure}[!bt]
\begin{center}
\includegraphics[angle=0,width=12.0cm]{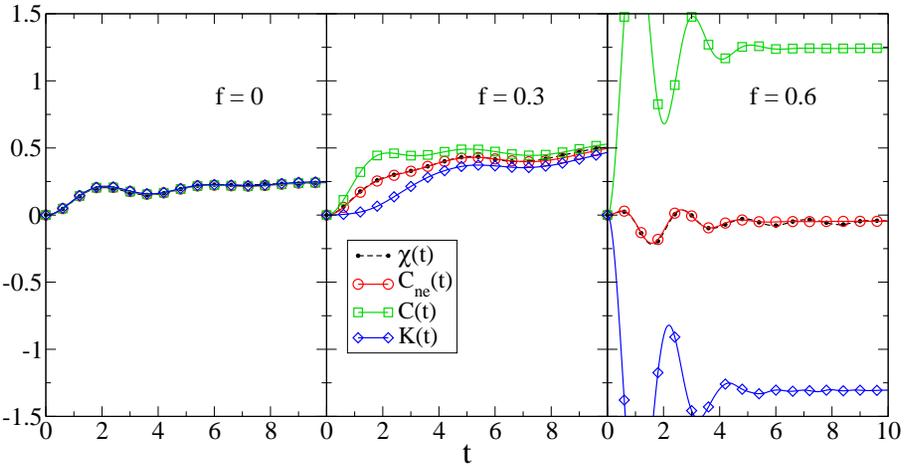}
\end{center}
\caption{As in figure \ref{fig:Lang}, but with fixed $\gamma=0.2$ and
varying force $f$, from left to right: $f=0$ (equilibrium), $f=0.3$, and $f=0.6$
(case $f=0.9$ is in the previous figure).
\label{fig:Lang2}}
\end{figure}

In Fig.~\ref{fig:Lang} we visualize the various terms of the fluctuation-dissipation relation,
for three scenarios with different viscosity, increasing from left to right.
The response is well reproduced by $C_{\textrm{ne}}(t)$, even if we
perform a numerical integration with $dt=10^{-3}$.
Oscillations in the response are visible for small viscosity; at higher friction there is
 a monotonous drift towards a new stationary state 
(right panel in Fig.~\ref{fig:Lang}).
The inertial regime is less sensitive to the perturbation,
as $\chi(t)$ displays only a small wiggling: a high entropy
production is almost compensated by a high traffic. 
For the high viscosity regime, with this setting ($f\lesssim 1$) the traffic term 
is close to zero compared with the entropic term,
and their combination yields $C_{\textrm{ne}}\approx C/2$.

In Fig.~\ref{fig:Lang2} we can follow the response as a function of the driving $f$, with 
$f=0$ for equilibrium.  For $f\neq 0$ the entropic term $C(t)$ can be quite different from
$C_{\textrm{ne}}(t)$. In equilibrium all the terms coincide, as expected.

\subsection{A generalized Einstein relation}
\index{Einstein relation!generalized}
Let us consider the underdamped version of the example in the last chapter, Section \ref{ex3}.
 For this we take the general Markov dynamics (\ref{ud}),
with the following simplifications: we take $V(x) = x_j$ for an
arbitrary $j$, and for the observable we take $Q =
v_k$ for a $k$ which can be different from $j$. We also
take the field $h_t=h$ to be constant. In this case the response function is related to the mobility of the system, i.e. the way
in which the average velocity changes under a constant added force. We fix a large time $u$ and define the time-averaged mobility 
by:
\[ \mathcal{X}_{jk} =  \frac{1}{u}\int_0^udt\left.\frac{\partial}{\partial h}\left<v^k_t\right>^h_{\mu_0}\right|_{h=0} \]
starting from $\mu_0$ at time zero.
We wish to connect this to the velocity fluctuations in the unperturbed (but driven) system:
\[ \mathcal{D}_{jk} = \frac{1}{2u}\left<[x^j_u-
x^j_0][x^k_u-x^k_0]\right>^0_{\mu_0} = \frac{1}{2u}\int_0^udt\int_0^uds\left<v^j_s v^k_t\right>^0_{\mu_0} \]
We try to be more general here than in the last chapter: we do not take the limit $u\uparrow +\infty$. 
In equilibrium, when $\mu_0$ is the Maxwell-distribution and
with $f$ and $U$ equal to zero and all the temperatures are equal $(\beta_i=\beta)$,
one can compute that $\mathcal{X}_{jk} = \frac{1}{\gamma_j}\delta_{j,k}$ and $\mathcal{D}_{jk} = \frac{1}{\beta\gamma_j}\delta_{j,k}$.
Indeed, also in the underdamped case, the Einstein relation 
coincides with the fluctuation-dissipation relation for pure diffusion. For general equilibrium systems
we still have $\mathcal{X}_{jk} = \beta\mathcal{D}_{jk}$.
This is no longer true out of equilibrium. With (\ref{ttt})
we can give the explicit modification.

\newpage
We substitute $V = x_j$ and $Q= v_k$ into the integrated version of (\ref{ttt}) for constant $h$ to obtain
\begin{eqnarray*}
 \left.\frac{\partial}{\partial h}\left<v^k_t\right>^h_{\mu_0}\right|_{h=0} &=& \frac{1}{2}\beta_j\int_0^tds\left<v^j_sv^k_t\right>_{\mu_0}\\
&& -  \frac{1}{2D_j}\int_0^tds\left< [f_j(x_s)-
\frac{\partial U}{\partial x_j}(x_s)]v^k_t\right>_{\mu_0}\\
&& + \frac{1}{2D_j}\int_0^tds\frac{\partial}{\partial s}\left< v_j^s v^k_t\right>_{\mu_0}
\end{eqnarray*}
The integrand of the full right-hand side is zero for $s>t$ due to causality,
so we can as well integrate $s$ from $0$ up to $u>t$. Integrating then $t$ gives us
\begin{eqnarray*}
  \mathcal{X}_{jk} &=& \beta_j \mathcal{D}_{jk} - \frac{1}{2uD_j}\int_0^udt\int_0^uds\left< [f_j(x_s)-
\frac{\partial U}{\partial x_j}(x_s)]v^k_t\right>_{\mu_0}\\
&& + \frac{1}{2uD_j}\int_0^udt\left< [v^j_t-v^j_0]v^k_t\right>_{\mu_0}
\end{eqnarray*}
This relation reduces to the familiar Einstein relation if the unperturbed system is in equilibrium. Furthermore
very formally when the overdamped limit is taken, i.e. we neglect changes in momenta, then the last term in the relation
drops and we recover the relation found in (\ref{er}).

\subsection{Coupled oscillators}

We now consider coupled one-dimensional oscillators at different temperatures $T_i$
in the stationary or in a transient regime.
This means that each index $i$ represents a particle which can move in one dimension.
In (\ref{ud}) we then set a conservative potential $U$ 
that is the sum $U=\sum_{i=0}^n \varphi(x_{i+1}-x_i)$
of local couplings between the oscillators $i$ and $i+1$,
with $\varphi(x) = \frac{1}{2} x^2 + \frac{1}{4} x^4$.
Boundary conditions are imposed by keeping $x_0 = x_{n+1} = 0$.
For simplicity we take all $D_i=D$. 
A basic perturbation  is given by switching on an external 
field on the $j$-th particle $V(x) = - E x_j$.
Taking the velocity $Q = v_k$ at site $k$ to be the observable,
the variable excess in entropy flux (\ref{eq:S}) reduces to
\[
 S_{ex}(\omega) = - \beta_j E \int\limits_{0}^{t} d s \:  v^j_s \,h_s
\]
while the traffic equals
\[
\mathcal{T}(\omega) = - \frac{E}{D} \int\limits_{0}^{t}  d s
\left[ \dot{v}^j_s + \frac{\partial U}{\partial v_j}(x_s) \right] h_s 
\]

From (\ref{ttt}) we have
\begin{eqnarray}
 \frac{\delta}{\delta h_s} \left.\left< v^{k}_t \right>_{\rho}^{h}\right|_{h=0} &=&
  - \beta_j E \left< v^{j}_s v^{k}_t \right>^0_{\rho} -
   \frac{E}{2D} \frac{d}{d s} \left< v^{j}_s v^{k}_t \right>^0_{\rho}\nonumber\\ 
&&- \frac{E}{2D} \left< \frac{\partial U}{\partial x_{j}}(x_s)\, v^{k}_t \right>^0_{\rho}\label{FDR}
\end{eqnarray}
This last relation is still valid for all times $s,t$ and is
automatically equal to zero for $s>t$ (causality).  Remembering that $\beta_j = \frac{\gamma_j}{D} = 1/T_j$ we can rearrange
formula (\ref{FDR}) for the situation where $s<t$ (like we did in (\ref{response3})):
\begin{eqnarray*}
 \frac{d}{d h_s} \left< v^{k}_t \right>_{\rho}^{h} & =
  & - E \left( \frac{\beta_j + \beta_k}{2} \right) \left< v^{j}_s v^{k}_t \right>^0_{\rho}\\
& & - \frac{E}{2 D} \left( \left< \frac{\partial
U}{\partial x_{j}}(x_s)\, v^{k}_t\right>^0_{\rho} + \left<
v^{j}_s \frac{\partial U}{\partial x_{k}} x_t \right>_{\rho}
\right)
\end{eqnarray*}
Note that the right-hand side now shows a formal space-time symmetry
for exchanging $j\leftrightarrow k, s \leftrightarrow t$. In
equilibrium the symmetry is true on spatial level alone,
$j\leftrightarrow k$, because time-symmetry is automatic.  That is
then an instance of Onsager reciprocity \cite{ons31a,ons31b}.\\

\begin{figure}[!tb]
\begin{center}
\includegraphics[angle=0,width=11.5cm]{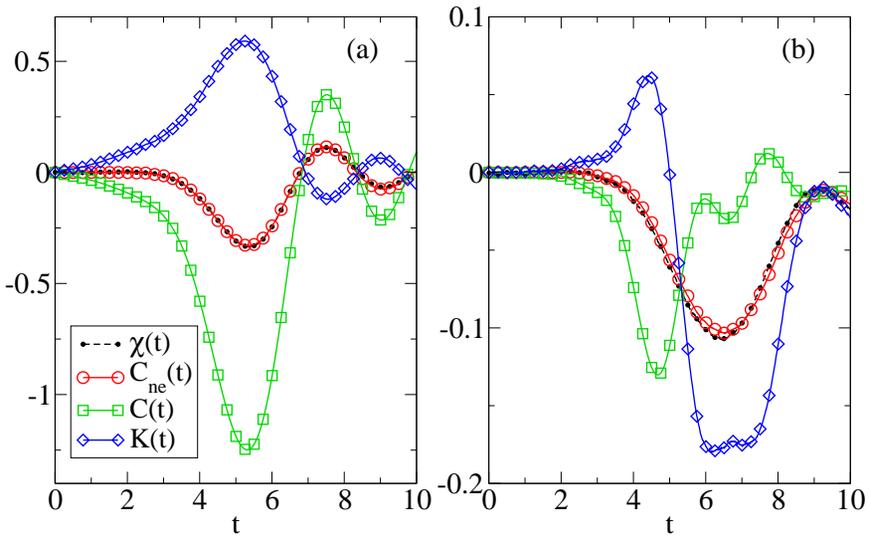}
\end{center}
\caption{
Visualization of the fluctuation-response relation $\chi(t) = C_{\textrm{ne}}(t)$,
for $n=11$ coupled oscillators, with parameters, $D=0.01$, $h=0.03$,
and (a) $T_i=i/10$, (b) $T_i=0.2$ and transient regime as described in the text.
\label{fig:co}}
\end{figure}

Choosing constant $h=0.03$,
in Fig.~\ref{fig:co}(a) we show the response in a system with $n=11$ oscillators,
with a linear gradient of temperature $T_i = i/10 = 1/\beta_i$, perturbation applied on site $j=1$
and response tested at central site $k=6$.
In Fig.~\ref{fig:co}(b) instead we have constant temperatures $T_i=0.2$, but we start
from a state out of equilibrium, by choosing $p_1(0)=10$ and the other momenta equal to zero.

\cleardoublepage
\chapter{Conclusions and literature}\label{fdrcon}

\textit{To conclude this part of the thesis we give an overview
of previous formulations of nonequilibrium fluctuation dissipation
relations, and present our conclusions of our own work.}

\section{Overview of previous formulations}

The literature on (extensions of) the fluctuation-dissipation
theorem is vast, and we can't list all
possible and even essential contributions. The equilibrium
formulation spans all of the previous century, while
nonequilibrium versions started to appear since the 1970's and
very much continue up to now. Early works include \cite{aga72,wei71}
and also \cite{dh75,ht75}, where we see a discussion within the
theory of stochastic dynamics.  In contrast to \cite{ht75,ht82}
our unperturbed process is time-homogeneous. Coming to more recent
times, violations of the equilibrium relation have been most often
discussed in transient regimes (equilibrium dynamics but not relaxed to equilibrium).  
For example, in the context of
ageing phenomena \cite{cr03,hp09}, much thought has been given to
making sense of an effective temperature as was discussed in
Section \ref{sec-efftemp}.

  However, recently much new
work has been also directed to find generic extensions of the fluctuation-dissipation theorem to
nonequilibrium steady states and to discussions of the dissipative
elements in relaxations to nonequilibrium. There is for example
the line starting from Sasa \textit{et al} \cite{hs05,hs01,ns08,op98} which treats
nonequilibrium heat effects.  In particular \cite{hs05} might also
be useful in real experiments because the response function is there
directly connected with the energy dissipation, which we also discussed in Section \ref{sec-diss}.  We do not know
yet how to relate that to the new ideas surrounding the traffic
term in our work.

For other recent extensions of the
FDR, we refer to \cite{bdgjl02,bslsb07,fiv90,ss06,ss09}. We
have also mentioned in \cite{bbmw09,bmw09b} (but not in this thesis) 
how our approach is related to the
co-moving frame interpretation of Chetrite, Gaw\c{e}dzki \textit{et al} \cite{cfg08,cg09,gpccg09}. 
There the equilibrium fluctuation-dissipation relation is recovered for systems out-of-equilibrium
by going to a description in a co-moving frame.
One disadvantage of this approach, and of the approaches in the previous references, is that one keeps the
stationary density $\rho$ (or its logarithm) as an observable in the
correlation functions. In our approach the largely unknown
distribution only enters in the statistical averaging.  
On the other hand, the other approaches come with a
new and still interesting interpretation (e.g. response in the time-reversed dynamics in
\cite{bdgjl02} and co-moving frame in
\cite{cfg08,cg09}).

A first generalized fluctuation-dissipation relation giving a
response formula {\it in our sense} appears in Section 2 of
\cite{ckp94} (Kurchan \textit{et al}). It treats an overdamped Langevin dynamics for a variable $y(t)$, used for soft spin
models, and the resulting equation (2.10) in \cite{ckp94} is exactly
our formula (\ref{response3}) for $Q=V$ equal to
$y$, where (\ref{trafgen}) can be used.  Another treatment for
systems of Ising spins is done in \cite{lcsz08,lcz05} (Lippiello \textit{et al}),
there the response is also investigated to higher orders in $h$. 
A more general treatment for jump
processes is offered in \cite{die05} (Diezemann), in particular its equations
(16)-(17), see also \cite{mw09}. In contrast, we have emphasized the
interpretation via entropy and traffic in nonequilibrium
fluctuation theory (more is to come on the meaning of traffic in the next part of this thesis). 
In fact, that interpretation is exactly what
makes a systematic generalization possible at all.  
The only study in which we
recognize some of the ideas related to the traffic term is in
\cite{rue09} (Ruelle), in the context of dynamical systems.

\section{Conclusions}
We have studied linear response relations under general
nonequilibrium conditions (stationary and not). 
From physical constraints on the probability of
trajectories, we have obtained in Chapter \ref{chap-response} a general fluctuation-dissipation relation for the response of a
driven system to the addition of a potential. Most generally in formula (\ref{response1a}),
and more specifically for systems in contact with an environment at one specific temperature in (\ref{response2}).
Our scheme lifts the nonequilibrium FDR
beyond formal first order perturbation theory applied to a specific dynamics,
by identifying in general physical terms the statistical quantities that
determine the response: entropy and traffic. 

Chapter \ref{chap-mjpod} has dealt with jump
processes and with overdamped diffusions, which is the usual
set-up for discussions on the violation of the
fluctuation-dissipation theorem and for the possible emergence of
an effective temperature. The general fluctuation-dissipation relation
for these classes of models is (\ref{fdr}). 
  As in this relation the correlations with entropy and with traffic are expressed in
terms of explicit averages, they constitute a formula ready to use
in a general context. For example, estimates of these correlations
can be obtained with usual averaging in simulations, without any
need to know further details or approximations of the stationary
density of states. On the theoretical side, several previous
approaches can be recovered or extended within the our
scheme. In many cases these results have been discussed in specific model
dynamics or for specific observables. It is interesting that there
is a unifying approach w\`{\i}th statistical interpretation behind
this very broad variety of previous results.

In Chapter \ref{chap-under}, we have applied
our scheme to underdamped diffusions, resulting in (\ref{ttt}).  It is seen that the correlation
with entropy involves dissipation over
different reservoirs, each one with its own equilibrium temperature.
The correlation with traffic is again written in terms of
observable quantities.

Generally, we like to stress the emergence of
traffic, as an important player out of equilibrium, complementary to the entropy flux.
In equilibrium entropy and traffic give the same contribution to the response function.
Out of equilibrium, they detach and the response needs
 to be evaluated in terms of correlation functions with both entropy and traffic.
So far traffic 
is in general not understood operationally for real experiments, but in the previous chapters 
it has been expressed as a
statistical average and correlation function of theoretically
known forces. The fluctuation-dissipation relations we derived also give observational significance to
the notion of traffic, which so far has mostly appeared as a theoretical concept in
fluctuation theories (see also the next part of this thesis).

In the ongoing research it will be interesting to consider more 
general and other models, e.g even non-Markovian
systems. Also within the models we considered, more general types of perturbation
are possible and interesting. For example, what will happen if the perturbation
is not of the potential type but actually changes the nonequilibrium part of the
dynamics? Think of changing one of the temperatures of several heat baths,
or adding an extra nonconservative forcing. In all these cases our results will have to be modified, of course, but 
we believe that the general method and framework we use is applicable to this wider set of 
questions. Specifically, the concepts of entropy and traffic will remain major
actors in nonequilibrium systems.

\cleardoublepage
\part{Dynamical fluctuations}

\vspace*{6cm}

``Large deviations estimates have proved to be the crucial tool
required to handle many questions in statistics, engineering, statistical mechanics,
and applied probability.''

\begin{flushright}
 Amir Dembo and Ofer Zeitouni, in their preface to \textit{Large Deviations Techniques and Applications} (1998), \cite{dz98}
\end{flushright}

\cleardoublepage
\chapter{Large deviations in statistical mechanics}\label{chap-ld}
\index{large deviations}
\textit{In 1910, Einstein proposed to invert the equation on Boltzmann's gravestone, to express probabilities of fluctuations
in terms of entropy:
\[ W = \exp\{ S/k_B \} \]
Such a formula shows that the probability of a fluctuation (a deviation of the macrostates from their equilibrium values)
is exponentially small and the entropy is the function that governs this exponential behaviour (rate function).
This idea is generalized in the application of large deviation theory to statistical mechanics.
This theory mathematically deals with the exponential decay of probabilities of large deviations in
stochastic processes, as a correction to the law of large numbers. 
It is very useful and forms a natural mathematical formalism for statistical mechanics, because thermodynamic
potentials as entropy and free energy naturally emerge as rate functions.
This chapter forms a brief introduction for the next chapters.}

\section{Introductory example: a discrete ideal gas}

Consider a discrete ideal gas \cite{ell99}. We mean by this a system consisting of $N$ independent particles, 
labelled by $i=1,\ldots,N$. The energy $\omega_i$ of each particle is a stochastic variable, drawn from the set $\{E_k| k=1,\ldots,m\}$.
The probability that particle $i$ has an energy $\omega_i = E_k$ is $\rho_k$,
where $\rho = (\rho_1,\ldots,\rho_m)$ is a probability vector, i.e. $\sum_k\rho_k=1$.
The total state of the system is given by $\omega = (\omega_1,\ldots,\omega_N)$.
Because the particles are assumed to be independent, the probability 
of the whole state $\omega$ is then 
\[\textrm{Prob}(\omega) = \prod_{i=1}^N\textrm{Prob}(\omega_i) =  \prod_{i=1}^N\left[\sum_{k=1}^m\delta_{\omega_i,E_k}\rho_k\right] \]
As a first step towards a more macroscopic description of such a system, we are not
interested in the exact energy of each particle, but rather in how many particles
have a certain energy. We define therefore the fraction of particles that have an energy $E_k$:
\begin{equation}\label{eov}
  p_{\omega}(k) = \frac{1}{N}\sum_{i=1}^N \delta_{\omega_i,E_k}
\end{equation}
where $\delta$ is the Kronecker delta. We call this fraction the empirical occupation
of $E_k$. The empirical occupation vector for the collection of energies is
\[ p_{\omega} = [ p_{\omega}(1),\ldots, p_{\omega}(m)] \]
Note that this (macrostate) is a stochastic variable, and at the same time defines a
probability distribution, because $\sum_kp_{\omega}(k) = 1$.

\paragraph{The law of large numbers}
By the law of large numbers, we expect this empirical occupation vector
to resemble $\rho = (\rho_1,\ldots,\rho_m)$ when the particle number $N$ is very large.
More precisely, we define the space of probability vectors
\[ \mathcal{X} = \{\mu = (\mu_1,\ldots,\mu_m) \in [0,1]^m |\sum_{k=1}^m \mu_k = 1\}\]
The collection of possible empirical occupation vectors for $N$ particles
form a subspace of $\mathcal{X}$, which we denote by $\Lambda_N$. The metric we consider on $\mathcal{X}$ is 
\[ d(\mu,\nu) = \sum_k(\mu_k-\nu_k)^2 \ \ \ \ \ \mu,\nu \in \Omega \]
Then the law of large numbers dictates that for any positive $\epsilon\in\mathbb{R}$
\[ \lim_{N\to\infty} \textrm{Prob}(d(p_{\omega},\rho)<\epsilon) = 1 \]
However, for finite $N$, fluctuations/deviations from this expected behaviour are always possible.
If we want to know more about these fluctuations, we have to go beyond
the law of large numbers.

\paragraph{Central limit theorem}

If we want to go beyond the law of large numbers, we could use the central 
limit theorem. 
Formally, the central limit theorem then says that in a distributional sense
\begin{equation}\label{clt}
  \textrm{Prob}(\frac{Np_{\omega}-N\rho}{\sqrt{N}} \sim z) \to \frac{1}{\mathcal{N}}\exp\{-\sum_{k=1}^m \frac{z_k^2}{2\rho_k}\}
\end{equation}
for $z \in \mathbb{R}^m$ and $\mathcal{N}$ a normalization factor.
To derive (\ref{clt}) is not trivial and we will not elaborate on this, because it is not very important
for our purposes. 
We do stress that for large $N$ this theorem only says something about
the probabilities of events for which $p_{\omega}-\rho$ is of the order of $1/\sqrt{N})$. We call these
small fluctuations. To know more, we have to consider the probabilities
of large fluctuations.

\paragraph{Beyond the central limit theorem}
In a simple system as this, it is not difficult to compute the probabilities
of arbitrary fluctuations.
Let us therefore take an arbitrary $\mu\in\Lambda_N$. This $\mu$
can be totally different from $\rho$ and is therefore called a (large) fluctuation.
Elementary combinatorics give us
\[\textrm{Prob}(p_{\omega} = \mu) = \frac{N!}{(N\mu_1)!\cdot\ldots\cdot(N\mu_m)!}\rho_1^{\mu_1}\cdot\ldots\cdot\rho_m^{\mu_m} \]
For systems with large $N$ this expression as such is not very convenient. We therefore use
Stirling's approximation for large $N$ to derive the following:
\begin{equation}\label{exex}
   \frac{1}{N}\log \textrm{Prob}(p_{\omega} = \mu) \approx -\sum_{k=1}^m \mu_k\log\left( \frac{\mu_k}{\rho_k} \right)
\end{equation}
The approximation becomes exact in the limit of $N\to\infty$.
For this reason, we write the probabilities in the following form:
\[ \textrm{Prob}(p_{\omega} = \mu) \approx e^{-NI(\mu)} \]
where $I(\mu) = \sum_{k=1}^m \mu_k\log \frac{\mu_k}{\rho_k}$ is minus the
relative entropy between the probability vectors $\mu$ and $\rho$. 
One can easily prove that $I(\mu)\geq 0$ in general. Moreover, it is only
zero for $\mu=\rho$.

Even though we have made an approximation, this result still gives more information
than the law of large numbers and even more than the central limit theorem: 
it tells us that the probability that the empirical occupation vector
to be equal to some $\mu\neq\rho$ is exponentially small in the number of particles
(only when $\mu=\rho$ there is no exponential decay, in accordance with the law of large numbers).
At least, this is the dominant behaviour for large $N$. 
It is not very surprising that there is an exponential behaviour in 
the number of particles, because all particles are independent.
However, also in many systems which do have interactions, this exponential behaviour is observed. 
This may not come as a surprise if we realize that many interactions are short-ranged,
and parts of the system that are sufficiently far apart are still approximately independent.

Such an exponential behaviour as in (\ref{exex}) is an instance
of a large deviation principle. As it turns out large deviation theory
forms a natural framework in which statistical mechanics can be embedded.
It incorporates in a natural way many key features, as the framework
of thermodynamic potentials and the variational principles that characterize
equilibrium.

In this chapter we
introduce the reader into some aspects of the theory of large deviations, by
using the example introduced above.
The introduction that follows here is based on \cite{ell99,tou09}, and for more explanation
we therefore refer to those texts. For a thorough treatment
of the mathematical theory, we refer to \cite{dz98}, 
while for the embedding of the theory of statistical mechanics in the large deviation formalism
we refer to \cite{ell85,ell95,lan73,tou09}.

\section{The large deviation principle}\label{sec-ldp}

In the example above, we defined the empirical occupation vector $p_{\omega}$,
and computed approximately its probability distribution for large $N$. 
Let us try to be mathematically more precise.
We consider the probability that the empirical occupation vector is close
to some given probability vector $\mu\in\mathcal{X}$.
We thus take an arbitrary (small) positive $\epsilon\in\mathbb{R}$ and consider the probability  $P(d(p_{\omega},\mu)<\epsilon)$.
Writing $B_N(\mu,\epsilon) = \{ \nu\in\Lambda_N| d(\mu,\nu)<\epsilon \}$, this probability becomes
\[ P(d(p_{\omega},\mu)<\epsilon) = \sum_{\nu\in B_N(\mu,\epsilon)}P(p_{\omega} = \nu)\]
Although it is not trivial, one can prove \cite{ell99} that this probability satisfies a large deviation principle,
stated here in a simplified way:
\[ -\lim_{N\to\infty}\frac{1}{N}\log P(d(p_{\omega},\mu)<\epsilon) = 
\sum_{k=1}^m \mu_k\log\left( \frac{\mu_k}{\rho_k} \right) + O(\epsilon)= I(\mu) + O(\epsilon) \]
As this can be done for arbitrary small $\epsilon$, one often denotes this by
\begin{equation}\label{ldnot1}
  -\lim_{N\to\infty}\frac{1}{N}\log P(p_{\omega}=\mu) = I(\mu)
\end{equation}
The functional $I$ is here the relative entropy between the probability vectors $\mu$ and $\rho$.
Generally, in large deviation theory, $I$ is called the rate function\index{rate function}, as it
quantifies the rate of exponential decay, or fluctuation functional\index{fluctuation functional}., as it
quantifies the probability of a fluctuation (deviation from the typical value).

To define the large deviation principle in general, one needs to go into
many mathematical details \cite{dz98}. This is not our goal. 
For our purposes, a definition as in (\ref{ldnot1}) is enough.
Moreover, we do not concern ourselves in this text
with proving the large deviation principle. We do assume that, for the systems
considered in the next chapters, those proofs can be done (in a physically
uninteresting way). Instead we usually just assume the large deviation principle to hold,
and refer to e.g. \cite{dz98,dv75a} for full mathematical proofs.

Following \cite{ell95,tou09}, we denote
\[ P(p_{\omega}=\mu) \asymp e^{-NI(\mu)} \]
to express a large deviation principle as in (\ref{ldnot1}).

We state here some properties of rate functions, without giving proofs, (for proofs we refer again to \cite{tou09,dz98}):
\begin{itemize}
 \item For some $\mu\in \mathcal{X}$, the limit in (\ref{ldnot1}) may not exist, because the probability density
decays to zero faster than exponentially (super-exponentially) in $N$. In that case the rate function is set to $+\infty$.
 \item In some cases $I(\mu) = 0$ for some $\mu$, meaning that the probability density decays slower than exponentially (sub-exponentially).
In the example, this only happens when $\mu=\rho$. In the systems we consider in the next chapters, we assume that the rate function
only has one global minimum where $I$ is zero. Actually this is equivalent with the assumption that there is a unique stationary state
of the system.
 \item Rate functions are positive: $I(\mu)\geq 0$. If they were not, the corresponding probabilities would
diverge instead of decay.
\end{itemize}

\section{The contraction principle}
\index{contraction principle}
Suppose we are given a new random variable, defined through a
continuous function of the old one: $q_{\omega} = h(p_{\omega})$. As we will see in the following,
such a function can be many-to-one. A heuristic argument gives the rate function
for $q$ in terms of the rate function for $p$.
Through the large deviation principle we write 
\[ p(q_{\omega}=\nu) \asymp \int_{\mu:h(\mu)=\nu} d\mu e^{-N I(\mu)}  \]
Laplace's saddle-point approximation gives an exact expression for the following limit:
\[ -\lim_{N\to\infty}\frac{1}{N}\log\int_{\mu:h(\mu)=\nu} d\mu e^{-N I(\mu)} = \inf_{\mu:h(\mu)=\nu} I(\mu) \]
One concludes that
\[ p(q_{\omega}=\nu) \asymp \exp\{ -N \inf_{\mu:h(\mu)=\nu} I(\mu) \} \]
defining the rate function for $q_{\omega}$ as $I_q(\nu) = \inf_{\mu:h(\mu)=\nu} I(\mu)$.
This way of computing one rate function as the infimum of another is called a contraction.

Let us consider the following case.
In absence of information about our system, we assume that the probabilities of all
the energy levels of the particles are the same, i.e. $\rho_k = 1/m$. The rate function
$I(\mu)$ then becomes $I(\mu) = \log m +\sum_k\mu_k\log \mu_k$.
Up to some constants this is minus the Shannon entropy of the probability vector $\mu$.  
Consider the probability
that the total energy per particle (average energy) is equal to a value $E$, giving
the following contraction: 
\[ I(E) = \inf_{\mu: \sum_k\mu_kE_k = E }I(\mu) \]
We can find such an infimum by using Euler-Lagrange equations with Lagrange multipliers
to ensure the restrictions. We get $k$ equations of the form:
\[ 0=\frac{\partial }{\partial \mu_k}[\log m +\sum_k\mu_k\log \mu_k +\beta(\sum_{k}\mu_kE_k - E) + \gamma(\sum_{k}\mu_k-1) ]  \]
where $\gamma$ makes sure that $\mu$ sums to one and $\beta$ makes sure that the average energy is $E$.
The solution of these equations is then a $\mu$ of the form
\[ \mu_k = \frac{1}{Z}e^{-\beta E_k} \]
where $Z$ is a normalization constant, and $\beta$ is determined by $\sum_{k}\mu_kE_k = E$.
This is the same form as the probability distribution of the canonical ensemble! 
It arises here as the most probable probability vector given an average energy $E$.
The
rate function becomes
\[ I(E) = \log m -\beta E - \log Z \]
One can check that $\frac{dI}{dE} = \beta$.
If we return to the idea that entropy and probability are related by $S = k_B\log P$,
then the rate function $I$ should represent minus the entropy per particle, divided by $k_B$.
As a result $dS/N = k_B\beta dE$. Indeed, the Lagrange multiplier $\beta$ is nothing but the
inverse temperature associated to this system. We also see that $-\frac{1}{\beta}\log Z$
is the free energy of the system. 

In this example, we clearly see that the ensemble theory of equilibrium statistical
mechanics naturally appears in large deviation theory,
at least for this example. The entropy plays
the role of rate function, and the canonical ensemble is found through
a contraction principle as the most probable probability vector given an average energy $E$.

\section{The G\"artner-Ellis theorem}

Let us take an arbitrary vector $M\in \mathbb{R}^m$. We denote the scalar product between
such a vector and the probability vector $\mu$ by $M\cdot \mu = \sum_k M_k\mu_k$.
The scaled cumulant generating function $\lambda(M)$ for our example is defined by
\[ \lambda(M) = \lim_{N\to\infty} \frac{1}{N}\log\left<e^{N M\cdot p_{\omega}}\right> \]
According to the G\"artner-Ellis theorem\index{G\"artner-Ellis theorem} \cite{ell84,gar77}, 
if $\lambda(M)$ exists and is differentiable for all $M$, then
$P(p_{\omega} = \mu)$ satisfies a large deviation principle with a rate function given by
\[ I(\mu) = \sup_{M\in\mathbb{R}^m}[M\cdot \mu -\lambda(M)] \]
This defines the rate function as the Legendre-Fenchel transform of $\lambda$. The relation can also be
inverted: $\lambda(M) = \sup_{\mu\in\mathcal{X}}[M\cdot\mu-I(\mu)]$. We will not prove this theorem here, but
heuristically, one can argue that
\[ \left<e^{N M\cdot p_{\omega}}\right>  \asymp \int d\mu e^{N[M\cdot\mu-I(\mu)]} \]
Similar as before, the saddle-point approximation of Laplace tells us that
\[ \left<e^{N M\cdot p_{\omega}}\right>  \asymp \exp\{ N\sup_{\mu\in\mathcal{X}}[M\cdot\mu-I(\mu)] \} \]
The G\"artner-Ellis theorem can help for explicit calculations of rate functions,
as it is sometimes easier to calculate the generating function $\lambda$.
Vice versa, sometimes it is easier to compute a rate function to get information about a generating function.

As an illustrative example, we assume again that $\rho_k=1/m$.
Consider the following generating function:
\[ \lambda({E_k}) = \lim_{N\to\infty} \frac{1}{N}\log\left<e^{-N\beta\sum_kE_k\mu_k}\right> \]
As we already know the rate function $I(\mu)$ in this case, we can compute the generating function
by
\[ \lambda({E_k}) = \sup_{\mu}[-\beta\sum_kE_k\mu_k - I(\mu)] \]
We find that the canonical distribution $\mu_k = \frac{1}{Z}e^{-\beta E_k}$ solves this
variational problem. As a consequence, the generating function $\lambda({E_k}) = \log Z$.
Remember that we interpret the rate function $I(\mu)$ as the entropy per particle $S/(Nk_B)$,
and $-\frac{1}{\beta}\log Z$ is the free energy $F$.
The G\"artner-Ellis theorem thus relates these two by the Legendre transform, schematically:
\[ F = \sup_{\mu}[\left<E\right>_{\mu}-TS(\mu) ] \]

\section{Typical behaviour and small fluctuations}

The large deviation theory incorporates the law of large numbers and the central
limit theorem in the following sense.

\paragraph{Law of large numbers}
\index{law of large numbers}

Suppose that the rate function $I(\mu)$ has a unique global minimum at $\mu=\rho$.
First of all, we can show that this minimum should satisfy $I(\rho) = 0$.
This follows from the fact that $\lambda(0) = 0$:
\[ 0 = \lambda(0) = \sup_{\mu}[-I(\mu)] = -I(\rho) \]
Secondly we see that for any (measurable) set $B\subset\mathcal{X}$:
\[ P(p_{\omega}\in B) \asymp e^{-N\inf_{\mu\in B}I(\mu)} \]
This tells us that $P(p_{\omega}\in B)\to 0$ whenever $\rho\notin B$, and  $P(p_{\omega}\in B)\to 1$
whenever $\rho\in B$. This is the law of large numbers, with $\rho$ the typical value
to which $p_{\omega}$ converges.

\paragraph{Central limit theorem}
\index{central limit theorem}
Given again a rate function $I(\mu)$ with a unique global minimum at $\rho$.
Suppose that the rate function is twice differentiable at $\rho$. We define
`small fluctuations' as values $\mu$ close enough to $\rho$ so that we can approximate the rate function by
\[ I(\mu) \approx \frac{1}{2}\sum_{k,l}\left.\frac{\partial^2 I}{\partial\mu_k\partial\mu_l}\right|_{\mu=\rho}(\mu_k-\rho_k)(\mu_l-\rho_l) \]
In this approximation, the probability density has the form of a Gaussian for large $N$:
\[ P(p_{\omega}=\mu) \approx \exp\{ -\frac{N}{2}\sum_{k,l}\left.\frac{\partial^2 I}{\partial\mu_k\partial\mu_l}\right|_{\mu=\rho}(\mu_k-\rho_k)(\mu_l-\rho_l) \} \] 
Which is (a form of) the central limit theorem. For this reason
fluctuations in this regime are also called small fluctuations,
Gaussian fluctuations or normal fluctuations.

For our specific example, where $I(\mu) = \sum_k\mu_k\log\frac{\mu_k}{\rho_k}$ we have that
\[ \left.\frac{\partial^2 I}{\partial\mu_k\partial\mu_l}\right|_{\mu=\rho} = \delta_{k,l}\frac{1}{\rho_k} \]
so that for small fluctuations
\[ P(p_{\omega}=\mu) \approx \exp\{ -\frac{N}{2}\sum_{k}\frac{(\mu_k-\rho_k)^2}{\rho_k} \} \]
which coincides with (\ref{clt}) as expected.

\section{Out of equilibrium}\label{sec-ldooe}

Because large deviation theory gives such a natural framework
for equilibrium statistical mechanics, it is a natural candidate
to give an extension to nonequilibrium. As we said in the introduction,
stochastic processes are
an important tool in nonequilibrium statistical mechanics.

The (mathematical) large deviation theory for Markov processes
was thoroughly founded by Donsker and Varadhan in \cite{dv75a,dv75b,dv76,dv83}.
In the context of physics, Onsager and Machlup \cite{om53} already used the large
deviation approach in the 1950's to examine relaxation to equilibrium in linear diffusion systems. 
A more general treatment of this problem, with a different emphasis, was done by Freidlin and Wentzell \cite{fw84,fw70} 

In more recent years advances have been made in what is called static fluctuation theory \cite{bdgjl01,bdgjl02}.
This is set in nonequilibrium stationary dynamics, and one imagines the system to be initially
in the stationary regime a long time ago (at time $-\infty$, the distribution of states
was the stationary distribution). The probability is then considered
that the distribution at time zero is equal to some other given distribution (a fluctuation).

In this thesis we discuss dynamical fluctuation theory, in which
time-integrated observables are considered, with time as the large parameter. 
This is mathematically not a very big step: it boils down to renaming the
particle number $N$ to the time $T$. For example: consider 
the vector $\omega=(\omega_1,\ldots,\omega_N)$ from the example above. It describes the energies
of the different particles. We could also see it as the consecutive energies of one
particle through time, so that $\omega$ becomes a trajectory like we have used throughout this text. 
The empirical occupation vector $p_{\omega}$ then gives
us the fractions of time that the particle has spent in each energy state. 
The only mathematical difference is that we work with continuous time.
This corresponds to measurements that are made continuously. We clarify this in the next chapters.
Apart from having a different interpretation, this switch to time as a large parameter also allows us to 
consider small systems, without having to make some hydrodynamic approximations.

In \cite{mnw08b,mnw08a,mnw09,sw10} we have reported the application of dynamical fluctuation
theory on various classes of models. The major goal in this line of research
is similar to that of fluctuation-dissipation relations: to find out what
thermodynamic quantities govern the rate functions of the large deviations.
Again, entropy and traffic turn out to be the basic ingredients
for constructing this out-of-equilibrium theory. 

We restrict ourselves in the next two chapters
to the work reported in \cite{mnw08b,mnw08a} in which respectively Markov jump processes
and overdamped diffusions are treated. However, the overall strategy in 
all models we have considered is the same: one starts by correctly
defining the observables (the time integrated occupations and currents,
as is explained there). Then, the rate functions (fluctuation functionals) are computed for
the joint probability of occupations and currents, which can be done
explicitly. The rate functions for occupations or currents alone
can then be computed by the contraction principle, which is unfortunately
not always possible. Throughout this calculation the physical aspect
is most important: we keep track of the influence of entropy and traffic
on the rate functions. We also show that the large
deviation theory provides a way of defining thermodynamic potentials,
although these potentials do not directly correspond to the ones defined
in equilibrium thermodynamics.
When only small deviations from the stationary behaviour are considered,
the rate functions are approximated to a quadratic dependence on the fluctuations
(the regime where the central limit theorem applies). When the dynamics is also close to equilibrium,
known variational principles such as the minimum entropy production principle
are recovered.

\cleardoublepage
\chapter{Driven overdamped diffusions}

\textit{We apply the theory of large deviations to time integrated observables
in general overdamped diffusion processes. These observables are empirical
occupation densities and empirical currents. Central is the joint rate function
for occupations and currents and the role of entropy and traffic in it.
This chapter represents work reported in \cite{mnw08a}.}

\section{Model}
We quickly recapitulate the equations describing overdamped diffusion,
which we discussed in Section \ref{sec-odiff}.

\subsection{Overdamped diffusions}
General overdamped diffusions in $d$ dimensions
are governed by the Langevin equation (\ref{langgencomponent}), interpreted 
in the It\^o way:
\begin{equation}
 dx^i_t = \sum_j[\chi_{ij}(x_t) F^j(x_t)dt +\frac{\partial  D_{ij}}{\partial x^j}(x_t)dt+\sqrt{2D(x_t)}_{ij}dB^j_t]
\end{equation}
where $x^i$ denotes a spatial component of the position $x$. In the shorter vector/matrix notation this gives:
\begin{equation}\label{langgen1}
 dx_t = \chi(x_t) [f(x_t)-\nabla U(x_t)]dt +\nabla \cdot D(x_t)dt+\sqrt{2D(x_t)}dB_t
\end{equation}
In this equation we have explicitly written the energy $U$ of the system, giving a conservative forcing, while $f_t$ 
represents a nonconservative
force (i.e. the driving). Furthermore $D$ and $\chi$ are $d\times d$ symmetric matrices
both depending on the position $x_t$, and there are $d$ Wiener processes $dB^j_t$.
We restrict ourselves to the case that the system is in contact with one heat bath
at a temperature $\beta$. Remember that the local detailed balance condition implies $\chi = \beta D$.

We restrict ourselves to two types of boundary conditions:\\
(1) \emph{periodic}---the particle moves on the unit torus
$[0,1)^d$ and the fields $U$, $f$,
and $\chi$ are smooth functions on the torus;\\
(2) \emph{decay at infinity}---the potential $U$ grows fast enough at infinity so that the
particle is essentially confined to a bounded
 region, i.e., the density and its derivative vanish at infinity.

Under either of the above boundary conditions we can simply ignore
boundary terms when performing integrations by parts.  The particles are essentially
 confined in their configuration space.\\
The probability density $\mu_t$ evolves according to the
Fokker-Planck equation (\ref{gsd1}):
\begin{equation}\label{fk1}
  \frac{\partial\mu_t}{\partial t} + \nabla\cdot j_{\mu_t} = 0,\qquad
  j_\mu = \chi \mu \,(F - \nabla U) - D\nabla\mu
\end{equation}
 The stationary condition reads
$\nabla \cdot j_\rho = 0$.

This class of models is still quite large. It is illustrative to
consider a simple example of these dynamics: a particle moving on a 
circle (one dimension). The advantage here is that we can explicitly
calculate the stationary distribution.

\subsection{Diffusion on a circle}\label{sec-circle}
We consider a particle undergoing an overdamped motion on the
circle with unit length:
\begin{equation}\label{gsd}
  d x_t = \chi(x_t)\big[ f(x_t) - U'(x_t)\big] d t + D'(x_t) d t
  + \sqrt{2D(x_t)}\, d B_t
\end{equation}
The prime as superscript is a shorthand notation for the spatial derivative.

The corresponding Fokker-Planck equation for the time-dependent probability density
$\mu_t$ is
\begin{equation}\label{fkcircle}
 \frac{\partial\mu_t(x)}{\partial t} + j_{\mu_t}'(x) = 0,\qquad
 j_\mu = \chi \mu (f-U') -  D\mu'
\end{equation}
The stationary density $\rho$ solves the
stationary equation $j'_\rho = 0$,
i.e.,
\begin{equation}\label{circle-stat}
  \chi \rho (f - U') - D \rho' = j_\rho
\end{equation}
is a constant. For $f = 0$, equation (\ref{circle-stat}) has the solution
\begin{equation}\label{bg}
  \rho(x) = \frac 1{Z}\,e^{-\beta U(x)},\qquad
  Z = \int_0^1 e^{ -\beta U} d x
\end{equation}
and the corresponding stationary current is $j_\rho = 0$; this is
a detailed balanced dynamics with $\rho$ the equilibrium density.\\
When adding a nongradient driving force,
$\int_0^1 f d x \neq 0$, (\ref{circle-stat}) can still be solved; the stationary density obtains the form
\begin{equation}
  \rho(x) = \frac{1}{\cal Z}
  \int_0^1 \frac{e^{\beta W(y,x)}}{D(y)}\,
   d y,\qquad {\cal Z} =
    \int_0^1\int_0^1 \frac{e^{\beta W(y,x)}}{D(y)}\, d y d x
\end{equation}
where
\begin{equation}
  W(y,x) = U(y) - U(x) +\left\{
  \begin{array}{lr}
    \int_y^x f\, d z &  \textrm{for } y \leq x \\
    \int_y^1 f\, d z + \int_0^x f d z  &  \textrm{for } y > x
  \end{array}\right.
\end{equation}
is the work performed by the applied forces along the positively
oriented path $y \rightarrow x$. In this model the stationary
current can be computed by dividing the stationary
equation (\ref{circle-stat}) by $\rho\chi$ and by integration over
the circle:
\begin{equation}\label{circle-current}
  j_\rho = \frac{\overline{W}}{\int_0^1 (\rho\chi)^{-1}d x}
\end{equation}
where $\overline{W} = \int_0^1 f d x$ is the work carried over a completed cycle.
The non-zero value of this stationary current indicates that time-reversibility is broken.
In the simplest nonequilibrium setting when $U = 0$ and $f, \chi > 0$ are some constants, the
steady state has the uniform density $\rho(x) = 1$ and the current is $j_\rho = \chi f$.

\section{Entropy and traffic}

Before going on to defining the observables for which we want to describe
large deviation principles, let us first discuss the basic ingredients that
play a role in the rate functions: entropy and traffic.

To properly define them, we need to define a reference process. We take this process
to be (see \ref{sec-ldb2}):
\[
 dx_t = \nabla \cdot D(x_t)dt+\sqrt{2D(x_t)}dB_t
\]
i.e. the overdamped diffusion (\ref{langgen1}) with the force $f-\nabla U$ put to zero.
The action that describes the probabilities of trajectories $\omega = (x_t)_{0\leq t \leq T}$ of
the original process with respect to the reference process is given by (\ref{actiongen}):
\begin{eqnarray}\label{ldaction}
 e^{-A(\omega)} &=&\frac{d\mathcal{P}_{\mu_0}}{d\mathcal{P}^0_{\mu_0}}(\omega)\\ &=& 
\exp\left\{ \frac{1}{4}\int_0^Tdt[2\dot{x}_t+2\nabla \cdot D - \chi (f - \nabla U) ]\cdot D^{-1}\chi (f - \nabla U) \right\}\nonumber
\end{eqnarray}

\subsection{Entropy}
The entropy flux into the environment is equal to the time-antisymmetric part of the action (by local
detailed balance, see (\ref{ldbdiff})):
\begin{eqnarray}
  S(\omega) = A(\theta\omega) - A(\omega) &=& \beta\int dx_t \circ (f(x_t) - \nabla U(x_t))  \nonumber\\
&=& \beta\int dx_t \circ f(x_t) - \beta[U(x_T)-U(x_0)]\label{entld}
\end{eqnarray}
which is equal to the work done by the nonconservative force (see (\ref{work})) minus the change of energy of the system.
It is useful for later calculations to compute the average of the entropy flux, started
from some initial distribution $\mu_0$. For that we need to compute the average of the Stratonovitch integral.
We rewrite the Stratonovitch integral to an It\^o integral using (\ref{itostrat}):
\[ \int_0^Tdx_t\circ f(x_t) = \int_0^T dx_t f(x_t) + \int_0^T dt [D(x_t) \nabla]\cdot f(x_{t}) \]
Using the Langevin equation, the average of the It\^o integral $\left<\int_0^T dx_t f(x_t)\right>_{\mu_0}$ gives:
\begin{eqnarray*}
&&\int_0^T \Bigl<f(x_t)\cdot\Bigl(\chi(x_t) [f(x_t)-\nabla U(x_t)]dt
 + \nabla \cdot D(x_t)dt+\sqrt{2D(x_t)}dB_t\Bigr) \Bigr>_{\mu_0}\\
&&= \int_0^T dt\Bigl<f(x_t)\Bigl(\chi(x_t) [f(x_t)-\nabla U(x_t)] +
\nabla \cdot D(x_t)\Bigr) \Bigr>_{\mu_0}
\end{eqnarray*}
In the last step we used that the average over the term with $dB_t$ drops, because $dB_t = B_{t+dt}-B_t$ is independent
of $x_t$ and has average zero. Then, by definition of the time-evolved distribution $\mu_t$
we get
\[
 \left<\int_0^Tdx_t\circ f(x_t)\right>_{\mu_0} = \int_0^T dt\int dx \mu_t\Bigl(f\cdot\chi [f-\nabla U] +
\nabla \cdot[Df]\Bigr)
\]
which can be rewritten succinctly, using partial integration, as
\begin{equation}\label{stratcurrent}
 \left<\int_0^Tdx_t\circ f(x_t)\right>_{\mu_0} = \int_0^T \int dx dt f\cdot j_{\mu_t}
\end{equation}
where $j_{\mu_t}$ is the probability current as in (\ref{fk1}).
The average entropy flux, starting from a density $\mu_0$ is thus equal to
\begin{equation}
 \left<S(\omega)\right>_{\mu_0} = \int_0^T dt\left[\int dx f\cdot j_{\mu_t} - \frac{d}{dt}\int dx U(x)\mu_t(x)\right]
\end{equation}
In Section \ref{traj-ae} we also discussed the average entropy. There we defined the entropy
of the system through the Shannon (or Gibbs) entropy:
\[ s(\mu) = -\int dx \mu(x) \log \mu(x) \]
The total average change of entropy (entropy production) in the world, given the initial distribution $\mu_0$ is then, see (\ref{avent}):
\begin{equation}
\left<S_{\mu_0}(\omega)\right>_{\mu_0} =\int_0^T dt\sigma(\mu_t)
\end{equation}
 where the instantaneous entropy production rate $\sigma(\mu)$ is easily computed to be
\begin{equation}\label{epr}
 \sigma(\mu) = \int j_\mu  \cdot(\mu D)^{-1} j_\mu\, d x
\end{equation}
It is very important to note that, although computed through the action relative to a reference process,
the average entropies computed here do not depend on that reference process (as long as the reference process
is an equilibrium process).

\subsection{Traffic}

The traffic is defined as the time-symmetric part of the action (\ref{ldaction}):
\begin{eqnarray}
\mathcal{T}(\omega) &=& A(\theta\omega) + A(\omega)\label{trafld}\\ &=&  \frac{\beta^2}{2}\int_0^Tdt(f-\nabla U)\cdot D (f-\nabla U) 
+\beta\int_0^Tdt \nabla \cdot [D(f-\nabla U)]\nonumber
\end{eqnarray}
The average value of the traffic, given an initial distribution $\mu_0$ is then equal to
\[ \left<\mathcal{T}(\omega)\right>_{\mu_0} = \int_0^T dt \tau(\mu_t) \]
with the instantaneous `traffic rate' given by
\begin{equation}\label{taumu}
 \tau(\mu) = \frac{\beta^2}{2}\int dx\mu(f-\nabla U)\cdot D (f-\nabla U) 
+\beta\int dx \mu\nabla \cdot [D(f-\nabla U)]
\end{equation}
The traffic, in contrast to entropy, is very much dependent on the choice of the reference process.
It is interesting to note that for overdamped diffusions, the entropy rate and traffic rate are closely related.
To see this, note that the entropy production rate for the reference process, can be found by simply putting $f-\nabla U$ to zero
in the explicit formula (\ref{epr}):
\[ \sigma^0(\mu) = \int dx \frac{(\nabla \mu)\cdot D (\nabla \mu)}{\mu} \]
One can then check directly that
\begin{equation}\label{ent-traffic}
 \tau(\mu) = \frac{\sigma(\mu) - \sigma^0(\mu)}{2}
\end{equation}
In following sections we often compare the traffic rates of different dynamics (determined by different forces).
This equality then tells us that these traffic differences can be replaced by entropy differences.
Hence, relation
(\ref{ent-traffic}) brings about a simplification in the structure
of fluctuations that is characteristic and restricted to
diffusions. See for example the next chapter, where 
there is no relation like (\ref{ent-traffic}).  We believe that this also indicates that a more
general nonequilibrium theory should reach beyond the
Langevin or diffusion approximation.

\section{Observables}
Here we define the observables for which we want to compute the rate functions.

A basic and time-symmetric dynamical
observable is the empirical distribution of the occupation times\index{empirical observable!occupation times}. It is
defined as the fraction of time spent in a subset $A$ of the state space $\Omega$ over
a fixed time interval $T$:
\[ p_{\omega}(A) = \frac{1}{T} \int_0^T \chi_A(x_t)\,d t \]
where $\chi_A(x)$ is the indicator function giving a value 1 if $x\in A$
and zero otherwise. Note that this resembles the definition of the empirical
occupation vector (\ref{eov}) in the example explained in the last chapter.
The difference is that the particle number $N$ is replaced by the duration time $T$
and the state space is continuous.

Having this in mind we formally write the empirical occupation density $p$:
\begin{equation}
  p_{\omega}(x) = \frac{1}{T} \int_0^T \delta(x_t - x)\, d t
\end{equation}
This is a path-dependent observable as it varies over the paths
$\omega = (x_t)_{0\leq t\leq T}$. Note that it is normalized (i.e. integration
over state space gives one).
It is also a useful observable, because we can write
time-integrated state-functions $g$ in terms of this empirical density:
\[ \frac{1}{T} \int_0^T dtg(x_t)\, d t = \int dx p_{\omega}(x)g(x) \]
Because we always assume that there exists a unique stationary distribution,
we have ergodicity:
\[ \int dx p_{\omega}(x)g(x) \to \left<g\right>_{\rho} \]
almost surely for $T \to \infty$ for any state function $g$
(this means that the probability to have a path $\omega$ for
which this convergence holds is equal to one).
 We conclude that for large times $p_{\omega} \to \rho$ almost surely.

The time-antisymmetric observable of special relevance is the
empirical current. We define the total time-averaged current\index{empirical observable!current}
in a set $A\subset \Omega$ as
\[ J_{\omega}(A) = \frac {1}{T} \int_0^Td x_t\circ\chi_A(x_t) \]
where the circle denotes a Stratonovitch stochastic integral, and $\chi_A$ is again an indicator function.
This observable is the sum of all displacements the particle makes when in a subset $A$ of $\Omega$ and
thus represents the time-integrated particle current (per unit of time). As usual, there can be an ambiguity 
in choosing the type of stochastic integral. 
The Stratonovitch integral is there to assure that the observable is time-antisymmetric. 
The current density is then formally defined as
\begin{equation}
   J_{\omega}(x) = \frac {1}{T} \int_0^T d x_t\circ\delta(x_t - x)  
\end{equation}
It depends again on the (random) path $\omega$ and
it measures the time-averaged current while in $x$. As for the empirical
occupations, this observable is useful because quantities like work and the entropy
flux can be written in terms of it: for an arbitrary state function $g$ we get 
\[ \frac {1}{T} \int_0^T d x_t\circ g(x_t) = \int dx J_{\omega}(x) g(x) \]
In the same way as for occupations and with (\ref{stratcurrent}) we see 
that $J_{\omega} \to j_\rho$ for $T \to \infty$ almost surely.

\section{Fluctuations of occupations and currents}

We already argued that the empirical densities of occupations and currents
will, in the large time limit, converge almost surely to the stationary densities $\rho$ and $j_{\rho}$.
In this section we compute the rate functions (fluctuation functionals) for deviations from these typical values.
This amounts to computing the probabilities that the empirical observables $p_{\omega}, J_{\omega}$ are close to some
given but arbitrary density $\mu$ and current $j$.

\subsection{Definitions and restrictions}

To be able to compute the probabilities of the empirical observables, 
we have to define the distance $d$ between $p_{\omega}$ and $\mu$ and between $J_{\omega}$ and $j$.
With such a definition, probabilities like $P(d(p_{\omega},\mu)<\epsilon)$ for some small number $\epsilon>0$ should
be computed. Actually proving that the large deviation principle holds, i.e.
\begin{equation}\label{ldpod}
  -\lim_{T\to\infty}\frac{1}{T}\log P(d(p_{\omega},\mu)<\epsilon) = I(\mu) + O(\epsilon)
\end{equation}
and under what restrictions, is not trivial. In this text we do not concern ourselves with this.
For mathematical details, see \cite{dz98,dv75a,dv75b,dv76,dv83}. Instead, we assume that the large deviation principle holds,
and concentrate on computing the rate functions using heuristic arguments. We begin by introducing the following
simplified notation for (\ref{ldpod}):
\[  -\lim_{T\to\infty}\frac{1}{T}\log P(p_{\omega} = \mu) = I(\mu)\ \ \ \ \ \ \textrm{or} \ \ \ \ \ \ P(p_{\omega} = \mu) \asymp e^{-TI(\mu)} \]
Similarly as for the occupations the probability of deviations for the currents are defined.
However, the natural starting point is the joint probability for currents and occupations:
\[ P(p_{\omega} = \mu, J_{\omega} = j) \asymp e^{-TI(\mu,j)} \]
as both $I(\mu)$ and $I(j)$ can be obtained in principle from the
joint fluctuations by the contraction principle.

A first observation that we can make, is that $I(\rho,j_{\rho})=0$, i.e. the stationary density and
current have a probability that does not decay to zero. Moreover, the stationary regime is characterized
by
\[ I(\rho,j_{\rho}) = \inf_{\mu,j}I(\mu,j) \]
A second observation is
that $I(\mu,j) = \infty = I(j)$ whenever $j$ is not stationary,
i.e.\ for $\nabla \cdot j \neq 0$. Indeed, for any smooth bounded
function $Y$ one has
\[
\int Y\, \nabla \cdot J_{\omega}\, d x =
  -\frac{1}{T} \int_0^T \nabla Y(x_t) \circ d x_t
  = -\frac{1}{T} [Y(x_T) - Y(x_0)] \to 0
\]
for $T\to\infty$. Hence, in a distributional sense, $\nabla \cdot J_{\omega} \to
0$ for $T \to \infty$ along \emph{any} particle trajectory, which
proves the above statement. That is why from now on we always
assume that $\nabla \cdot j = 0$, unless otherwise specified.

\subsection{Computation of the fluctuation functional}
The probability of the fluctuations can be written as
\[ P(p_{\omega} = \mu, J_{\omega} = j) = \int dP(\omega) \chi[p_{\omega} = \mu, J_{\omega}=j] \]
where $\chi$ is again an indicator function, giving one when its argument is satisfied and zero otherwise.
We have omitted initial conditions in the notation, as they are not important in the long-time limit.
(To be more rigorous, one has to define a neighbourhood of $\mu$ and $j$ using the distance defined on the function space.
The indicator function then gives one when the empirical densities are in that neighbourhood.)
To compute the rate function we exploit the knowledge we have of Radon-Nikodym derivatives. 
For this we define a new dynamics, by changing the force $f$ in (\ref{langgen1})
to a new force $g$. We take this $g$ such that the density $\mu$ and the current $j$ become typical,
i.e:
\[ j = \chi[g-\nabla U ]\mu - D\nabla \mu \]
This explicitly defines $g$. With this new dynamics we write:
\begin{equation}\label{probfluc}
  P(p_{\omega} = \mu, J_{\omega} = j) = \int dP_g(\omega) \frac{dP_f}{dP_g}(\omega)\chi[p_{\omega} = \mu, J_{\omega}=j]
\end{equation}
where we now have added subscripts $f$ and $g$ to denote respectively the old and the new dynamics.
The Radon-Nikodym derivative can be written in terms of the entropy production and traffic:
\[ \log\frac{dP_f}{dP_g}(\omega) = S_f(\omega) - S_g(\omega) + \mathcal{T}_g(\omega) - \mathcal{T}_f(\omega)  \]
which are defined in (\ref{entld}) and (\ref{trafld}). Again the quantities with a subscript $g$ are defined
in the dynamics with $f$ replaced by $g$. 
The excess entropy can be rewritten in terms of the empirical current:
\[ S_f(\omega) - S_g(\omega) = \beta\int dx_t \circ [f(x_t) - g(x_t)] = \beta T \int dx J_{\omega}(x)[f(x)-g(x)] \]
The excess traffic can be rewritten in terms of the empirical occupations:
\[ \mathcal{T}_g(\omega) - \mathcal{T}_f(\omega) = \beta T[\tau_g(p_{\omega})-\tau_f(p_{\omega})] \]
where $\tau_g$ is given by (\ref{taumu}), but with $f$ replaced by $g$.
We see that  the entropy production in both dynamics
only depends on the path $\omega$ through the empirical current, and the traffic in both dynamics
only depends on the empirical occupation density. When we substitute this in (\ref{probfluc}) we see that
the indicator function allows us to replace the empirical observables by the density $\mu$ and the current $j$.
We thus get
\begin{equation}\label{pfff}
  P(p_{\omega} = \mu, J_{\omega} = j) = e^{-TI(\mu,j)}\int dP_g(\omega) \chi[p_{\omega} = \mu, J_{\omega}=j]
\end{equation}
where the functional $I(\mu,j)$ is given by
\begin{equation}\label{imuj}
 I(\mu,j) = \frac{1}{2}\tau_f(\mu) - \frac{1}{2}\tau_g(\mu) + \frac{\beta}{2}\int dx j\cdot (g-f)
\end{equation}
This functional is exactly the rate function, because the last factor in (\ref{pfff}) goes to one
in the limit of large times, which is a consequence of the fact that $\mu$ and $j$ are the typical (stationary) values in the dynamics
determined by $g$.

Let us examine this joint rate function $I(\mu,j)$.
It is the sum of an excess instantaneous traffic rate
given the density $\mu$ alone, and an excess work (or, equivalently,
entropy flux) given stationary current $j$ alone. The excess is an excess
of the original dynamics with respect to the dynamics in which $\mu$ and $j$ are typical.
If we use (\ref{ent-traffic}), we can rewrite the functional in terms of the entropy production rate:
\begin{equation}
 I(\mu,j) = \frac{1}{4}\sigma_f(\mu) - \frac{1}{4}\sigma_g(\mu) + \frac{\beta}{2}\int dx j\cdot (g-f)
\end{equation}
Furthermore, using the explicit expressions of the traffic/entropy and the force $g$ we can rewrite
the fluctuation functional in the following explicitly positive form:
\begin{eqnarray}
  I(\mu,j) &=& \frac{1}{4}\sigma_f(\mu) - \frac{1}{4}\sigma_g(\mu) + \frac{\beta}{2}\int dx j\cdot (\mu D)^{-1}(j-j_{\mu})\nonumber\\
&=& \frac{1}{4} \int (j - j_{\mu}) \cdot (\mu D)^{-1} (j - j_{\mu})\,d x \label{imuj2}
\end{eqnarray}
(On the assumption $\nabla \cdot j = 0$; remember that $I(\mu,j) =
\infty$ otherwise.) This last formula resembles the Gaussian-like
expressions for the current distribution, typical for hydrodynamic
fluctuations of the diffusion-type.  Such expressions are
omnipresent in the works of e.g. \cite{bdgjl05,bd04}.  Although the
quadratic integrand in (\ref{imuj2}) resembles the (generalized
Onsager-Machlup) Lagrangian for macroscopic fluctuations in the
hydrodynamic limit, we have no spatial/temporal rescaling here. We
have started from a mesoscopic system as described by a diffusion
equation and the only large parameter is the time span $T$. 
The difference between our work and other work on dynamical large
deviations becomes visible from the formula (\ref{imuj}) and
(\ref{imuj2}). Our approach has nothing to do with hydrodynamic
rescaling or with macrostatistics.  It concerns the thermodynamic
interpretation of the fluctuation functional $I(\mu,j)$ for our
mesoscopic system: how it is shaped from quantities like traffic,
work and entropy production, and providing full account of the
steady dynamical fluctuations in both the time-symmetric and the
time-antisymmetric sectors.

\section{A notion of thermodynamic potentials}\label{sec-thpot}

\index{thermodynamic potential!overdamped diffusions}
In equilibrium systems it is useful to consider thermodynamic potentials (like free energy), as they have a clear physical
meaning and characterize equilibrium via variational principles. Moreover, to go from one potential to another,
Legendre transforms are used. Fluctuation functionals also bring with them variational principles. First of all, and mainly, minimizing them
characterizes stationarity.
In (\ref{imuj}) we gave an explicit
expression for the rate function of occupations and currents. Let us analyze this from the viewpoint of thermodynamic potentials 
and Legendre transforms. First of all,
the first two terms in (\ref{imuj}) constitute an excess traffic, or equivalently excess entropy production (\ref{ent-traffic}).
Let us consider the traffic in a dynamics governed by an arbitrary force $h$: 
\[\tau_h(\mu) = \frac{\beta^2}{2}\int dx\mu(h-\nabla U)\cdot D (h-\nabla U) 
+\beta\int dx \mu\nabla \cdot [D(h-\nabla U)] \]
This can be seen as a potential for the currents in the sense that the functional derivative
of it with respect to the force, gives
\begin{equation}\label{currentpotential}
  \frac{\delta \tau_h}{\delta h(x)} = \beta j^h_{\mu}(x)
\end{equation}
where $j^h_{\mu}$ is the probability current in a dynamics with a force $h$, see (\ref{fk1}). It is then natural to examine
the Legendre transform of $\tau_h$:
\[ G(\mu,j) =  \sup_{h}\left[ \beta \int  h\cdot j dx - \tau_h(\mu) \right] \]
We find the supremum by using Euler-Lagrange equations:
\[ j = j^h_{\mu} \]
so that $h = g$ is exactly the force needed to make the current $j$ typical, together with $\mu$:
\[ G(\mu,j) =  \beta\int  g\cdot j dx- \tau_g(\mu) \]
On the other hand, if we take the functional derivative of $G$, we get:
\begin{equation}\label{forcepotential}
  \frac{\delta G}{\delta j(x)} = \beta g(x)
\end{equation}
where $g$ is again the force that makes $\mu$ and $j$ typical. So $G$ is a potential for the forces, just
like $\tau$ was a potential for the currents, and by Legendre transforms we can switch between the two.

We can easily rewrite $I(\mu,j)$ in terms of $G$ and $\tau$:
\begin{equation}\label{res2}
  I(\mu,j) = \frac{1}{2}\left[G(\mu,j) + \tau_f(\mu) -\beta\int dx f\cdot j\right]
\end{equation}
Note that $\tau_0(\mu) = 0$, so that $\frac{1}{2}G(\mu,j)$ is exactly equal to the fluctuation functional
in the case that $f=0$. In other words:
\[ I_{f}(\mu,j) = I_{0}(\mu,j) + \frac{1}{2}\left[\tau_f(\mu) -\int dx f\cdot j\right] \]
This is nice, because the left-hand side is a fluctuation functional for a nonequilibrium dynamics, while on
the right-hand side, the first term is a fluctuation functional in an equilibrium dynamics. The rest is thus
`the correction to equilibrium.'

We can still rewrite (\ref{res2}) in the form
\begin{equation}
 4I_f(\mu,j)=  \sup_f\bigl\{ 2\beta\int dx f\cdot j -\sigma_f(\mu) \bigr\} - 2\beta\int dx f\cdot j+\sigma_f(\mu)
\end{equation}
fully in terms of entropic quantities, due to
(\ref{ent-traffic}).  A similar structure, cf.
(\ref{currentpotential}) and (\ref{forcepotential}), has been
established already before in the framework of jump processes, see
\cite{mn08} and also the next chapter. 

\section{Contractions}
\index{contraction principle}

Now that we have a fluctuation functional for both symmetric and
antisymmetric variables, we can compute the statistics of
empirical averages of arbitrary physical quantities.  In
particular, we can try to find the fluctuation functionals for
density $I(\mu)$ and for current $I(j)$  separately. To start we look at the
fluctuations of the occupation density.

\subsection{Occupation statistics}\label{occ}

As $I(\mu) = \inf_j
I(\mu,j)$, we have to compute the minimizing current $j$ for any
given density $\mu$. Since the minimization is constrained via the
stationary condition $\nabla\cdot j = 0$, we get the Euler-Lagrange equation from (\ref{imuj2})
\begin{equation}\label{eq-V}
 j = \chi\mu\bigl[ f-\nabla\cdot (U+V)\bigr] - D\nabla\mu
\end{equation}
where $V$ is a Lagrange multiplier (function of $x$). Not
surprisingly, we see that the minimizer is the stationary current
for a modified dynamics that makes $\mu$ stationary. This modified
dynamics is achieved here by adding to the imposed potential $U$
an extra potential $V$. We
therefore call the minimizing current in (\ref{eq-V}) $j_{\mu}^V$,
and the fluctuation functional becomes:
\begin{equation}\label{Imu-first}
  I(\mu) =  \frac{1}{4} \int (j_\mu^V - j_\mu)
  \cdot (\mu D)^{-1} (j_\mu^V - j_\mu)\,d x
\end{equation}
For some explicit examples of solutions to~(\ref{eq-V}),
see further down, in equations (\ref{V-circle}) and (\ref{minep}).

The fluctuation functional $I(\mu)$ obtains other equivalent forms
by substituting $g = f-\nabla V$ into (\ref{imuj}):
\begin{equation}\label{Imu-main}
  I(\mu) = \frac{\tau(\mu) - \tau_V(\mu)}{2} = \frac{\sigma(\mu) - \sigma_V(\mu)}{4}
\end{equation}
where the second equality follows again from~(\ref{ent-traffic}).
In this way we have recognized the excess traffic (or here also:
the excess entropy production) as governing the large time
statistics of the occupation times. Excess is here excess of the original with
respect to the modified process in which $\mu$ is typical.
The fluctuation functional $I(\mu)$ thus exactly equals
one quarter of a difference in entropy production rates when
having density $\mu$, these rates being computed respectively for
the original dynamics and for a modified dynamics that makes $\mu$
stationary.\\

In formul\ae\ (\ref{Imu-first}) and (\ref{Imu-main}) the potential
$V$ has to be determined from $\mu$ by solving the inverse
stationary problem~(\ref{eq-V}). We discuss two examples in
which this can be done explicitly: the equilibrium case and the example of diffusion
on the circle.

\paragraph{Equilibrium dynamics} Let us see what can
be said in general for equilibrium diffusions.  If $f = 0$ then
equation (\ref{eq-V}) has the solution $V = -\beta^{-1}\log\mu$, and
the corresponding current $j_\mu^V$ and the entropy production
$\sigma_V(\mu)$ are both zero. As a result,
\begin{equation}\label{I-eq}
  I(\mu) = \frac{\sigma(\mu)}{4}
\end{equation}
This exact relation between the equilibrium dynamical fluctuations
and the entropy production is solely true for diffusion processes.
In contrast, for jump processes (see the next chapter) $\sigma(\mu)$ gives only the leading
term in an expansion of $I(\mu)$ around the equilibrium density
$\rho \propto e^{-\beta U}$, and the relation (\ref{I-eq}) obtains
corrections when beyond small fluctuations; see~\cite{mn07} for
details.\\

\paragraph{Diffusion on the circle}
For the one-dimensional example of Section~\ref{sec-circle} the
inverse stationary problem~(\ref{eq-V}) allows for an explicit
solution. The current $j_\mu^V$ is immediately read off the
formula~(\ref{circle-current}),
\begin{equation}
  j_\mu^V = \frac{\beta\overline{W}}{\int_0^1 (\mu D)^{-1} d x},\qquad
  \overline{W} = \int_0^1 f\,d x
\end{equation}
and the potential $V$ obtains the form
\begin{equation}\label{V-circle}
  V(x) = -U(x) -\frac{1}{\beta}\log\mu(x) + \int_0^x
  \bigl( f - \frac{j_\mu^V}{\beta\mu D} \bigr)\,d y
\end{equation}
which is a nonlocal functional of the given density $\mu$. The
fluctuation functional is explicitly given as
\begin{equation}
  4 I(\mu) = \sigma(\mu) -
  \frac{\overline{W}^2}{\int_0^1 (\mu D)^{-1} d x}
\end{equation}
for $\mu\neq 0$.\\
 Observe that if $\mu = 0$ on some open set $A$
then the rate function equals $I(\mu) = \sigma(\mu)/4$. (That follows
also from the  equilibrium form (\ref{I-eq}) below as the circle
gets \emph{effectively} cut and the dynamics mimics a detailed
balance one.) The infimum of $I(\mu)$ over all densities $\mu$
that vanish on $A$ then
gives the escape rate\index{escape rate} from the complement $A^c = [0,1) \setminus A$.\\
As a simple example, assume that $U = 0$ and let $f$ and $D$ be
some constants. In this case the entropy production (\ref{epr}) reads
\begin{equation}\label{erescape}
  \sigma(\mu) = \beta^2 D f^2 + D\int_0^1 \frac{\mu'^2}{\mu}\,d x
\end{equation}
To compute the escape rate from $A^c$ (or, entrance rate to $A$)
we must take the infimum of (\ref{erescape}) over all $\mu$ that vanish
on $A$.
 Setting $A = (0,\delta)$ for some $0<\delta<1$, that infimum is reached
for the density $\mu^*(x) =
\frac{2}{1-\delta}\sin^2(\frac{\pi(x-\delta)}{1-\delta}), x\in
[\delta,1]$, and the escape rate is
\begin{equation}
  \inf_{\mu|_A = 0} I(\mu) = I(\mu^*)
  = \frac{\pi^2 D}{(1 - \delta)^2} + \frac{\beta^2 D f^2}{4}
\end{equation}
Even in equilibrium ($f=0$) the result is meaningful as it relates
the diffusion constant to an escape rate.   In the context of
dynamical systems, the analysis of the escape rates and of their
link to linear transport coefficients was initiated by Dorfman
and Gaspard, see~\cite{dor99,dg95} and references therein.

\subsection{Current statistics}
 The contraction to the current $j$ is
also possible. However, up to special examples, there is no
explicit solution to the associated variational problem and for
general models one has to resort to a perturbative or numerical
analysis.  In fact, often the calculation starting from the
generating function of the current appears more practical than to
do the contraction starting from $I(\mu,j)$, see e.g. \cite{der07}.
Through the G\"artner-Ellis theorem the generating function
is related to the fluctuation functional via a Legendre transform.

However, we can check in general a fluctuation theorem 
for currents: notice that
\[ I(\mu,-j) - I(\mu,j) = \frac{1}{2}\int dx j(\mu D)^{-1}j_{\mu} = \frac{\beta}{2}\int dx j \cdot f \] 
This remains true even when the contraction to currents alone is made.
We thus have that
\[ \frac{P(J_{\omega} = j)}{P(J_{\omega} = -j)} \asymp \exp\left\{ \frac{\beta T}{2}\int dx j \cdot f \right\} \]
We see that currents that have on average the same direction as the forcing, are exponentially
more probable than their reversed currents.

For explicit computations of the current rate function we restrict us here
to giving the result for a constant drift on the circle.

\paragraph{Constantly driven diffusion on the circle} We take
$U=0$ and $f,\chi$ constants. In this case, from (\ref{imuj2}) the
joined fluctuation functional reads:
\begin{equation}
 I(\mu,j) = \frac{1}{4D}\int\frac{1}{\mu}(j-\beta D f\mu -
 D\mu')^2\,d x
\end{equation}
and for all $j$, the infimum over $\mu$ is reached at the uniform
distribution, so that
\begin{equation}
 I(j) = \frac{(j-\beta D f)^2}{4D}
\end{equation}
and hence we see that here the current fluctuations are Gaussian.

\section{Small fluctuations and entropy principles}\label{sec:EP}

Considering small fluctuations (for which a quadratic approximation is valid) can
simplify the fluctuation functionals mathematically. It is also
experimentally a more accessible regime, as the fluctuations are more probable
than large fluctuations. The fluctuation
functionals can be
expanded in both the occupation densities and currents around their typical 
values and the strictly positive quadratic form obtained in
the leading order describes normal (Gaussian) fluctuations. From a physical
point of view, the structure of these normal fluctuations have been
first analyzed by Onsager and Machlup, \cite{om53}, for the case of
relaxation to equilibrium. Here we show a natural extension of the
original Onsager-Machlup formalism to nonequilibrium systems by
starting from the above fluctuation theory.

We
look here at the Gaussian approximation in a dynamics far from
equilibrium. Later we will also make the driving $f$ small, to be
close to equilibrium.\\
 As is clear from (\ref{imuj2}), current and occupations are coupled. It is
because of this coupling that contractions of $I(\mu,j)$ to
$I(\mu)$ and to $I(j)$ become rather complicated. Even for small
fluctuations this coupling remains: take $\mu =
\rho(1+\epsilon\mu_1)$ and $j = j_{\rho}+\epsilon j_1$, with
$\epsilon$ a small parameter. Because $j-j_{\mu}$ is then
$O(\epsilon)$, the fluctuation functional is $O(\epsilon^2)$:
\begin{eqnarray}
I(\mu,j) &=&  \frac{\epsilon^2}{4}\int d x
  \bigl[ j_1 \cdot (\rho D)^{-1} j_1 + \mu_1^2 j_\rho \cdot (\rho D)^{-1} j_\rho
\nonumber\\
  &&\hspace{15mm}+ \nabla\mu_1 \cdot \rho D\nabla\mu_1
  - 2\mu_1 j_1 \cdot (\rho D)^{-1} j_\rho \bigr] + o(\epsilon^2)
\label{smagen}
\end{eqnarray}
The last term in the integrand gives the coupling between
occupation and current fluctuations. It is proportional to the
stationary current, which is non-zero away from equilibrium. It is
only when we take a dynamics close to equilibrium, i.e. $f =
\epsilon f_1$, that the fluctuations decouple. In this
approximation we have that $j_{\rho} = O(\epsilon)$, and thus,
near equilibrium,
\begin{equation}\label{approx}
  I(\mu,j) =  \frac{\epsilon^2}{4} \int d x \bigl[ j_1 \cdot (\rho D)^{-1} j_1
  + \nabla\mu_1 \cdot \rho D\nabla\mu_1  \bigr] +  o(\epsilon^2)
\end{equation}
with, to leading order, a complete decoupling between the
time-symmetric (occupations) and the time-antisymmetric (current) sectors.

\paragraph{Occupations}
When close to equilibrium, the computation of $I(\mu)$ by
contraction is
    easy: we see from~(\ref{approx}) that the second term on its right-hand side
     is
    just $I(\mu)$.\\
In the same approximation of small fluctuations and
close-to-equilibrium, the entropy production becomes:
\begin{equation}\label{approx_ent}
  \sigma(\mu) =  \int d x \bigl[ j_{\rho} \cdot (\rho D)^{-1} j_{\rho}
  + \epsilon^2\nabla\mu_1 \cdot \rho D \nabla\mu_1 \bigr] + o(\epsilon^2)
\end{equation}
and thus we get
\begin{equation}\label{minep}
  I(\mu) = \frac{\sigma(\mu) - \sigma(\rho)}{4} + o(\epsilon^2)
\end{equation}
This reveals to be a special case of a general result, \cite{mn07},
according to which the entropy production governs the occupational
statistics in the linear irreversible regime. It provides a
fluctuation-based explanation for the minimum entropy production
principle\index{entropy production
principle!minimum} introduced by Prigogine to characterize stationarity via
an (approximate) variational principle, \cite{pri62}: the
stationary state has a minimal  entropy
production, which is not necessarily zero because the system is out of equilibrium.

\paragraph{Currents} For currents we have an analogue of the minimum
entropy production principle.  The starting point is
again~(\ref{approx}) from which we extract the current
fluctuations:
\begin{eqnarray}
I(j) &=& \frac{1}{4} \int (j - j_\rho) \cdot (\rho D)^{-1} (j - j_\rho)\,d x + o(\epsilon^2)
\nonumber\\
  &=& \frac{1}{4} \bigl[ {\cal D}(j_\rho) + {\cal D}(j) - 2\mathcal{S}(j) \bigr] + o(\epsilon^2)
\label{maxi}
\end{eqnarray}
with ${\cal D}(j) = \int j \cdot (\rho D)^{-1} j\,\,d x$ sometimes
called the Onsager dissipation function, and $\mathcal{S}(j)=\beta\int f\cdot jdx$ is the
entropy flux given a current $j$. In particular, this leads to a
variational characterization of the steady current $j_\rho$ which
can be written as the following maximum entropy production
principle: the $j_\rho$ maximizes the entropy flux $\mathcal{S}(j)$ under
the two stationary constraints
\begin{equation}
  \textrm{(1)  } \nabla \cdot j = 0,\qquad
  \textrm{(2)  }  {\cal D}(j) = \mathcal{S}(j)
\end{equation}
The second condition is indeed satisfied at $j = j_\rho$ (note
also that $\rho$ can with no harm in this order be replaced by the
equilibrium density $\rho_{f=0} = e^{-\beta U}/Z$.) Such a
variational principle, known as a maximum entropy production
principle\index{entropy production
principle!maximum}, is often used in applications and apparently even
beyond the linear irreversible regime.  As is however clear from
(\ref{maxi}) from our dynamical fluctuation theory, the validity
of the maximum entropy principle is restricted to
close-to-equilibrium.  Beyond that regime, we must refer to
contractions from (\ref{imuj2}), (\ref{res2}) or even from
(\ref{smagen}) for generally valid expressions with a general
thermodynamic meaning.

\cleardoublepage
\chapter{Markov jump processes}

\textit{Here, the large deviation theory is applied to Markov jump processes,
again with empirical occupation and current densities as observables.
The lines of reasoning are the same as in the last chapter,
and therefore the discussion in this chapter is less detailed. 
For more details, we refer to \cite{mnw08b}.}

\section{Entropy and traffic}

In this chapter we consider time-homogeneous Markov jump processes on a finite state space $\Omega$,
determined by the transition rates $k(x,y)$.
We imagine the system to be in contact with a single heat bath. The local detailed balance
assumption then restricts these rates (\ref{ldbmj}):
\[ \frac{k(x,y)}{k(y,x)} = e^{-\beta[U(y)-U(x)-W(x,y)]} \]
where $U(x)$ is the energy of the system in state $x$ and $W(x,y)=-W(y,x)$
is the work done on the system by a nonconservative force during the transition $x\to y$.

Again the main ingredients for the dynamical fluctuation theory are the
entropy and traffic, which are defined through the path-probability density,
relative with respect to a reference process. As the reference process we take 
an equilibrium process with rates $k_0(x,y)$ defined through 
\begin{equation}\label{eqreference}
  k(x,y) = k_0(x,y)e^{\frac{\beta}{2}W(x,y)}
\end{equation}
One can check that the rates $k_0(x,y)$ are detailed balanced with equilibrium distribution
$\rho(x)\propto \exp[-\beta U(x)]$. The action of the original process
with respect to this reference process is given by (\ref{girsanov}):
\begin{eqnarray*}
 -A(\omega) &=& \log \frac{d\mathcal{P}}{d\mathcal{P}_0}(\omega)\\
&=& \int_0^Tdt[\lambda_0(x_t)-\lambda(x_t)] +\sum_{t\leq T}
\log\left(\frac{k(x_{t^-},x_{t})}{k_0(x_{t^-},x_{t})}\right)\\
\end{eqnarray*}
where the sum is over all jump times of the path, $x_{t^-}$ is the state just before the
jump and $\lambda(x) = \sum_yk(x,y)$.
Using the relation between the original and the reference process, this becomes
\begin{equation}
 -A(\omega) = \int_0^Tdt\sum_yk_0(x_t,y)[1-e^{\frac{\beta}{2}W(x_t,y)}] +\frac{\beta}{2}\sum_{t\leq T}W(x_{t^-},x_{t})
\end{equation}

\subsection{Entropy}
The excess entropy flux (original system with respect to the reference) into the environment during a trajectory $\omega = (x_t)_{0\leq t\leq T}$ is 
given by
\[ S_{ex}(\omega) = A(\theta\omega) - A(\omega) = \beta\sum_{t\leq T}W(x_{t^-},x_{t}) \]
For later purposes we would like to compute the average of the entropy production.
We prove that for any function of two states $g(x,y)$ and for any initial condition $\mu$:
\begin{equation}\label{averageg}
  \left<\sum_{t\leq T}g(x_{t^-},x_{t})\right>_{\mu_0} = \int_0^Tdt\sum_{x,y}\mu_t(x)k(x,y)g(x,y)
\end{equation}
We do this by defining yet another set of rates $k^*(x,y) = k(x,y)\exp[h\cdot g(x,y)]$,
with $h\in\mathbb{R}$
and using the identity $1 = \left<\frac{d\mathcal{P}^*}{d\mathcal{P}}(\omega)\right>_{\mu_0}$:
\[ 1 = 
\left<\exp\left\{ \int_0^Tdt\sum_yk(x_t,y)[1-e^{h g(x_t,y)}] + h\sum_{t\leq T}g(x_{t^-},x_{t}) \right\}\right>_{\mu_0}  \]
Taking the derivative with respect to $h$ in $h=0$ of this equation gives us then
\[ 0 = \left< -\int_0^Tdt\sum_yk(x_t,y) g(x_t,y) +\sum_{t\leq T}g(x_{t^-},x_{t})\right>_{\mu_0} \]
which proves (\ref{averageg}). The average of the excess entropy flux is thus
\[ \left<S_{ex}(\omega)\right>_{\mu_0} = \int_0^Tdt\sum_{x,y}\mu_t(x)k(x,y)W(x,y) = \frac{1}{2}\int_0^Tdt\sum_{x,y}j_{\mu_t}(x,y)W(x,y)\]
with $j_{\mu}(x,y) = \mu(x)k(x,y) - \mu_t(y)k(y,x)$ the probability current.
This entropy flux is an excess with respect to the reference equilibrium process. For the equilibrium process
we have, because of detailed balance:
\[ S_0(\omega) = \log\frac{\rho(x_T)}{\rho(x_0)} = -\beta[U(x_T)-U(x_0)] \]
The average of this can be written as
\[ \left<S_0(\omega)\right>_{\mu_0} = -\beta\int_0^T\sum_x \frac{d}{dt} \mu_t(x)U(x) = \frac{\beta}{2}\int_0^T \sum_{x,y}j_{\mu_t}(x,y)[U(x)-U(y)] \]

In Section \ref{traj-ae} we defined the entropy
of the system through the Shannon (or Gibbs) entropy:
\[ s(\mu) = -\int dx \mu(x) \log \mu(x) \]
The total average change of entropy (entropy production) in the world, given the initial distribution $\mu_0$ is then
the sum of the entropy flux of the reference process plus the excess entropy flux plus the entropy change of the system, 
see also (\ref{avent}):
\begin{equation}
\left<S_{\mu_0}(\omega)\right>_{\mu_0} = s(\mu_T)-s(\mu_0) + \left<[S_0(\omega)+S_{ex}(\omega)]\right>_{\mu_0} = \int_0^T dt\sigma(\mu_t)
\end{equation}
where the instantaneous entropy production rate $\sigma(\mu)$ is straightforwardly computed to be
\begin{eqnarray*}
  \sigma(\mu) &=& \frac{1}{2}\sum_{x,y}j_{\mu}(x,y)[\log\mu(x)-\log\mu(y)+\beta U(x)-\beta U(y)+W(x,y)]\\   
&=&\sum_{x,y}\mu(x)k(x,y)\log\frac{\mu(x)k(x,y)}{\mu(y)k(y,x)}
\end{eqnarray*}

\subsection{Traffic}

As for the entropy flux, we get the excess traffic from the action:
\begin{equation}\label{trafficldmj}
  \mathcal{T}_{ex}(\omega) = A(\theta\omega) + A(\omega) = 2\int_0^Tdt[\lambda(x_t)-\lambda_0(x_t)]
\end{equation}
The average of this, starting from an initial distribution $\mu_0$, is
\[ \left<\mathcal{T}_{ex}\right>_{\mu_0} = 2\int_0^T  dt [\tau(\mu_t)-\tau_0(\mu_t) ] \]
with
\begin{equation}\label{trafficrldmj}
  \tau(\mu) = \sum_x\mu(x)\lambda(x) = \sum_{x,y}\mu(x)k(x,y)
\end{equation}
which is equal to the average number of jumps per unit of time for a system in a distribution
$\mu$. We call it the activity of the system. Note that this activity can be seen as the instantaneous
traffic rate (only with a different reference than in (\ref{trafficldmj})).

\section{Observables}

Like in the last chapter our first observable is the empirical time-integrated occupation
density\index{empirical observable!occupation times}:
\[ p_{\omega}(x) = \frac{1}{T}\int_0^T dt \delta_{x_t,x} \]
where $\delta_{x_t,x}$ is the Kronecker delta, giving one when $x_t=x$ and zero otherwise.
Furthermore we define the empirical distribution that counts jumps:
\[ Q_{\omega}(x,y) = \frac{1}{T}\sum_{t\leq T} \delta_{x_{t^-},x}\delta_{x_t,y} \]
Again, as for the case of diffusions, ergodicity leads us to the conclusion that
in the long time limit $p_{\omega}\to\rho$ and $Q_{\omega}(x,y) \to \rho(x)k(x,y)$
almost surely. From $Q_{\omega}$ we can define the empirical current\index{empirical observable!current}:
\[ J_{\omega}(x,y) = Q_{\omega}(x,y) - Q_{\omega}(y,x) \]
which converges in the long time limit to $j_{\rho}$ almost surely.

\section{Joint fluctuations of occupations and currents}

The easiest starting point for deriving fluctuation functionals is
the joint fluctuations of the occupations and jumps:
\[ P(p_{\omega}=\mu,Q_{\omega} = q) \asymp e^{-TI(\mu,q)} \]
Again, one has to take care in correctly defining such a large deviation
principle, and prove when it is valid. We will not do this here. Instead
we concern ourselves with computing the rate functions. We refer
to \cite{dz98} for mathematical details.

The heuristic argument to compute the rate function $I(\mu,q)$
is essentially the same as the one in the last chapter. We first
define a new dynamics in which $q$ and $\mu$ are typical, i.e.
we take a process with transition rates $q(x,y)/\mu(x)$. Then, we compute the action
for a path $\omega$ for the original process with respect to the new one:
\[ A(\omega) = \int_0^Tdt\sum_y[k(x_t,y)-\frac{q(x_t,y)}{\mu(x_t)}] +\sum_{t\leq T}
\log\frac{q(x_{t^-},x_{t})}{\mu(x_{t^-})k(x_{t^-},x_{t})} \]
Using the definitions of our empirical observables, we rewrite this to
\[ A(\omega) = T\sum_{x,y}\left\{p_{\omega}(x)[k(x,y)-\frac{q(x,y)}{\mu(x)}] + 
Q_{\omega}(x,y)\log\frac{q(x,y)}{\mu(x)k(x,y)}\right\} \]
We then see that the probability of the fluctuation becomes
\[ P(p_{\omega}=\mu,Q_{\omega} = q) = \int d\mathcal{P}^q(\omega)e^{-A(\omega)}\chi[p_{\omega}=\mu,Q_{\omega} = q] \]
And, similarly as for diffusions we can replace the empirical observables in the action
by $\mu$ and $q$, so that in the end the rate function becomes
\begin{equation}
 I(\mu,q) = \sum_{x,y}\left\{\mu(x)k(x,y) - q(x,y) + 
q(x,y)\log\frac{q(x,y)}{\mu(x)k(x,y)}\right\}
\end{equation}
However, we would like to find the rate function for the joint fluctuations
of occupations $\mu$ and currents $j$. By the contraction principle, we see that this rate function
is the infimum of $I(\mu,q)$ over all $q$ for which $q(x,y)-q(y,x) = j(x,y)$.
Note that for any path $\omega$ and any (bounded) state function $Y$, we have that
\[ \sum_x Y(x)\left[\sum_yJ_{\omega}(x,y)\right] = \frac{Y(x_0)-Y(x_T)}{T} \]
which goes to zero for the long time limit. Therefore we only consider fluctuations
for which $\sum_yj(x,y) = 0$.
The $q^*$ that makes the infimum can be explicitly computed.
For that we compute the Euler-Lagrange equations:
\begin{equation}\label{modification}
  q(x,y) = \mu(x)k(x,y)e^{\frac{\beta F(x,y)}{2}}
\end{equation}
where $F(x,y) = -F(y,x)$ is a Lagrange multiplier, which makes sure that
\begin{equation}\label{Fstat}
  j(x,y) = q(x,y) - q(y,x) = \mu(x)k(x,y)e^{\frac{\beta F(x,y)}{2}} - \mu(y)k(y,x)e^{-\frac{\beta F(x,y)}{2}}
\end{equation}
This equation can be explicitly solved, giving the minimizer $q^*$: 
\[ q^*(x,y) = \frac{j(x,y)}{2} + \frac{1}{2}\sqrt{[j(x,y)]^2+ 4\mu(x)\mu(y)k(x,y)k(y,x)} \]
This can then be substituted into $I(\mu,q = q^*)$ to get $I(\mu,j)$.
Going beyond the mathematics, we compare (\ref{modification}) to the local detailed balance assumption
which says that
\[ \frac{q(x,y)}{q(y,x)} = \frac{\mu(x)}{\mu(y)}e^{-\beta[U(y)-U(x)-W(x,y)-F(x,y)]} \]
This gives an interpretation to the Lagrange multiplier $F$: it is a modification
of the work $W(x,y)$ originating from a nonconservative forcing. We therefore call $F$ a force. Moreover, the equation
(\ref{Fstat}) tells us that this force makes the fluctuation $\mu,j$ typical. This is the same kind
of modification as we had in the last chapter. There, to find the joint rate function
of occupations and currents we also had to introduce a new force which made the fluctuation
typical. It is illustrative to write $I(\mu,j)$ in terms of this $F$ by substituting (\ref{modification})
into $I(\mu,q)$:
\begin{eqnarray}
  I(\mu,j) &=& \sum_{x,y}\mu(x)k(x,y)[1-e^{\frac{\beta}{2}F(x,y)}] + \frac{\beta}{4}\sum_{x,y}j(x,y)F(x,y)\nonumber\\
&=& \tau(\mu)-\tau_F(\mu) + \frac{\beta}{4}\sum_{x,y}j(x,y)F(x,y)\label{imujmj1}
\end{eqnarray}
In the last line we have written the excess activity of the system, excess meaning the modified dynamics
with respect to the original one. Similarly, the last term in the rate function is an excess entropy
flux, given a current $j$. This last expression coincides physically with (\ref{imuj}),
and is as such the most general and physical form of writing the rate function.

We can also use the reference equilibrium process to rewrite the fluctuation functional
into a form even more clearly similar to (\ref{imuj}):
\begin{eqnarray*}
  I(\mu,j) &=& \sum_{x,y}\mu(x)k_0(x,y)[e^{\frac{\beta}{2}W(x,y)}-e^{\frac{\beta}{2}G(x,y)}]\\ 
&&+ \frac{\beta}{4}\sum_{x,y}j(x,y)[G(x,y)-W(x,y)]\\
&=& \tau_W(\mu)-\tau_G(\mu) + \frac{\beta}{4}\sum_{x,y}j(x,y)[G(x,y)-W(x,y)]
\end{eqnarray*}
where $G = W + F$ is the nonconservative forcing in the modified dynamics,
and we have denote subscripts $W$ and $G$ to denote the traffic in the original and modified dynamics.

\section{A notion of thermodynamic potentials}\label{sec-thpot2}
\index{thermodynamic potential!Markov jump processes}
Like in Section \ref{sec-thpot}, we can use the traffic of a process to construct a potential.
The activity, as defined in (\ref{trafficrldmj}), has the following property: define
a dynamics with rates $k_h(x,y) = k_0(x,y)\exp\{ \frac{\beta}{2}h(x,y) \}$, where $h(x,y)=-h(y,x)$.
then the function $\mathcal{H}(\mu,h) = \tau_h(\mu)-\tau_0(\mu)$ satisfies
\[ \frac{\delta \mathcal{H}(\mu,h)}{\delta h(x,y)} = \frac{\beta}{2}[\mu(x)k_h(x,y)-\mu(y)k_h(y,x)] =\frac{\beta}{2}j^h_{\mu}(x,y) \]
meaning that the activity (traffic) is a potential for the currents, like the traffic was in the
overdamped diffusion case. Analogous to that case we can define a Legendre transform:
\[ G(\mu,j) = \sup_{h}[\frac{\beta}{4}\sum_{x,y}j(x,y)h(x,y) - \mathcal{H}(\mu,h)] \]
This is a potential for forces: the minimizer $h$ is found to be the force $G = W+F$ that makes
$j=j^G_{\mu}$ typical, and
\[ \frac{\delta G}{\delta j(x,y)} = \frac{\beta}{2}G(x,y) \]
The fluctuation functional $I(\mu,j)$ can then be written as
\[ I(\mu,j) = \mathcal{H}(\mu,W) + G(\mu,j) - \frac{\beta}{4}\sum_{x,y}j(x,y)W(x,y) \]
Exactly like for the diffusion case (\ref{res2}). Again, $G(\mu,j)$
is equal to $I_{W=0}(\mu,j)$, i.e. the equilibrium fluctuation functional.

\section{Contraction to occupations}

To find the fluctuation functional for occupations alone, we make the contraction
\[ I(\mu) = \inf_{j}I(\mu,j)\ \ \ \textrm{or}\ \ \ I(\mu) = \inf_{q}I(\mu,q)  \]
The infimum has to be taken with the restriction that $\sum_y[q(x,y)-q(y,x)] = \sum_yj(x,y) = 0$.
If we do the contraction using $I(\mu,q)$, we get Euler-Lagrange equations of the following form:
\begin{eqnarray*}
 q(x,y) &=& \mu(x)k(x,y)e^{\frac{\beta}{2}[V(y)-V(x)]}\\
 0 &=& \sum_y \left[ \mu(x)k(x,y)e^{\frac{\beta}{2}[V(y)-V(x)]} - \mu(y)k(y,x)e^{\frac{-\beta}{2}[V(y)-V(x)]} \right]
\end{eqnarray*}
where $V$ is a Lagrange multiplier making sure that $\sum_y[q(x,y)-q(y,x)] = \sum_yj(x,y) = 0$.
Although we can not solve this in general, we see that $V$ is a potential to be added to the dynamics
that makes $\mu$ typical. This is analogous to the overdamped diffusion case. Substituting
this form of $q$ in $I(\mu,q)$ we can write $I(\mu)$ in terms of this potential:
\begin{equation}\label{Imu-main2}
 I(\mu) = \sum_{x,y}\left\{\mu(x)k(x,y) -\mu(x)k(x,y)e^{\frac{\beta}{2}[V(y)-V(x)]}\right\} = \tau(\mu) - \tau_V(\mu)
\end{equation}
Indeed, the fluctuation functional for occupations alone is defined in terms of an excess traffic,
just as for the case of overdamped diffusions. In contrast to that case, however, this excess is 
not equal to an excess of entropy production. It seems therefore, that the excess traffic is a more
general quantity than entropy to characterize fluctuation functionals.

\section{Small fluctuations}\label{sec:EPmj}

Consider the regime of small fluctuations, i.e. where the quadratic approximation
of the fluctuation functionals is valid. We write $\mu(x) = \rho(x)[1+\epsilon \mu_1(x)]$
and $j(x,y) = j_{\rho}(x,y) + \epsilon j_1(x,y)$ with $\epsilon \in \mathbb{R}$ a small number.
Note that $j_1$ has to satisfy the constraint $\sum_yj_1(x,y)=0$.
To write down $I(\mu,j)$ up to second order in $\epsilon$, we expand first (\ref{Fstat}),
giving $F(x,y) = \epsilon F_1(x,y) + o(\epsilon)$, with
\begin{eqnarray*}
  F_1(x,y) &=& \frac{1}{\tau_{\rho}(x,y)}\Bigl[ j_1(x,y) - \frac{1}{2}\tau_{\rho}(x,y)[\mu_1(x)-\mu_1(y)]\\ && 
\ \ \ \ \ \ \ \ \ \ \ \ \ \ \ \ \ \ \ \ \ \ \ \ \ \ \ \ \ \ -\frac{1}{2}j_{\rho}(x,y)[\mu_1(x)+\mu_1(y)]  \Bigr]
\end{eqnarray*}
where we have defined the symmetric counterpart of the current:
\[ \tau_{\mu}(x,y) = \mu(x)k(x,y) + \mu(y)k(y,x) \]
which is the expected number of jumps between states $x$ and $y$ per unit of time, when the system is in a distribution
$\mu$. This quantity is closely related to the activity defined in (\ref{trafficrldmj}),
because 
\[\sum_{x,y}\tau_{\mu}(x,y) = 2\tau(\mu)\] 
The factor of two in this equality stems from the fact that every bond $x,y$ is counted twice in the sum.
We therefore also call $\tau_{\mu}(x,y)$ traffic (or activity).
The fluctuation functional $I(\mu,j)$ in terms of the added force $F$ becomes
\[ I(\mu,j) = \frac{\beta^2\epsilon^2}{8}\sum_{x,y}\tau_{\rho}(x,y)[F_1(x,y)]^2 + o(\epsilon^2) \]
and substituting the explicit form of $F$, we get schematically:
\begin{equation}\label{imujapprox}
 I(\mu,j) = \frac{\beta^2\epsilon^2}{8}\sum_{x,y}\frac{1}{\tau_{\rho}}
[j_1 - \tau_{\rho}\nabla^-\mu_1 - j_{\rho}\nabla^+\mu_1 ]^2(x,y) + o(\epsilon^2)
\end{equation}
with the notation $\nabla^{\pm}\mu_1(x,y) = \frac{1}{2}[\mu_1(x)\pm\mu_1(y)]$.
In this quadratic approximation, we
see that the stationary traffic $\tau_{\rho}$ plays the role of a variance.
  
We can write (\ref{imujapprox}) in
the form
\begin{eqnarray}
  I(\mu,j) &=& \frac{\beta^2\epsilon^2}{4} \sum_{x,y} \Bigl[
  \frac{j_1^2}{2\tau_{\rho}} + \frac{\tau_{\rho}}{2} (\nabla^- \mu_1)^2
  - \frac{j_{\rho}j_1}{\tau_{\rho}}  \nabla^+ \mu_1
  + \frac{j_{\rho}^2}{2 \tau_{\rho}} (\nabla^+ \mu_1)^2
  \Bigr](x,y)\nonumber\\
&&+ o(\epsilon^2)\label{eq: L-far1}
\end{eqnarray}
which demonstrates that there is an occupation-current coupling, which is
proportional to the stationary current. It is this coupling that
makes contractions to occupation densities or currents alone difficult.

\paragraph{The close-to-equilibrium regime}

Let us assume that the dynamics of the process is close to equilibrium, meaning that
the transition rates are of the form (\ref{eqreference}):
\[ k(x,y) = k_0(x,y)e^{\frac{\beta\epsilon}{2}W(x,y)} \] 
with $k_0(x,y)$ the equilibrium reference process.
In such a dynamics, the stationary current $j_{\rho} = O(\epsilon)$.
In (\ref{eq: L-far1}) we can therefore drop the terms containing the stationary current to obtain
\begin{equation}\label{eq: L-close}
  I(\mu,j) = \frac{\beta^2\epsilon^2}{4} \sum_{x,y} \Bigl[
  \frac{j_1^2}{2\tau_{\rho}}\,  + \frac{\tau_{\rho}}{2} (\nabla^- \mu_1)^2
  \Bigr](x,y) + o(\epsilon^2)
\end{equation}
which gives us the same conclusions as for the case of overdamped diffusions:
the current and occupation fluctuations decouple in this close-to-equilibrium regime.
As a consequence a minimum entropy production principle can be derived for the occupation
fluctuations and a maximum entropy production principle for currents.

A question arises here especially for the occupation fluctuations,
which was not visible for overdamped diffusions:
why does entropy production govern the close-to-equilibrium regime?
The occupation fluctuation functional is expressed as an excess
traffic. For overdamped diffusions this was equal to an excess
entropy production, but here it is not.

However, in the close-to-equilibrium regime, the excess traffic and
excess entropy production merge. We see this by inserting
$\mu = \rho_0[1+\epsilon \mu_1]$ and $ k(x,y) = k_0(x,y)e^{\frac{\beta\epsilon}{2}W(x,y)}$
into the definitions of traffic and entropy production. Note that $\rho_0$ is the equilibrium
distribution for the rates $k_0$. Using subscripts $W$ to denote that we work in the dynamics with $W$ we get
\begin{eqnarray*}
 \sigma_W(\mu) &=& \frac{\epsilon^2}{2}\sum_{x,y}\rho_0(x)k_0(x,y)\Bigl[ \mu_1(x)-\mu_1(y)+ W(x,y) \Bigr]^2 + o(\epsilon^2)\\
 \tau_W(\mu) &=& \sum_{x,y}\rho_0(x)k_0(x,y)\Bigl[1 + \epsilon\mu_1(x) + \frac{\epsilon}{2}W(x,y) 
\\
&& \ \ \ \ \ \ \ \ \ \ \ \ \ \ \ \ \  +\frac{\epsilon^2}{4}[\mu_1(x)-\mu_1(y)]W(x,y) + \frac{\epsilon^2}{8}W(x,y)^2 \Bigr] + o(\epsilon^2)
\end{eqnarray*}
Up to second order in $\epsilon$, the excess entropy production and traffic with respect to the equilibrium
dynamics thus give the same:
\begin{equation}\label{mjenttraf}
  \sigma_W(\mu) - \sigma_0(\mu) = \tau_W(\mu) -\tau_0(\mu) + o(\epsilon^2)
\end{equation}
The reason that entropy production principles appear close to equilibrium is thus a result
of two facts:  The first is that the occupation and current fluctuations decouple, so that the joint fluctuation
functional becomes a sum of the functionals for occupations and currents alone. Second, the traffic and entropy production
become indistinguishable, so that we do not need traffic do describe these physical results.

\cleardoublepage
\chapter{Conclusions}\label{dfcon}

The questions and the methods covered in this part are not
entirely original.  They have appeared in the mathematical
literature in a systematic way since the theory of large
deviations was introduced in the framework of Markov processes, 
see the references in Chapter \ref{chap-ld} and especially in Section \ref{sec-ldooe}. The relevance to physics and to
statistical mechanics in particular is obvious, but the
thermodynamic interpretation of the resulting dynamical
fluctuation functionals has not been systematically investigated.
A first study can be found in \cite{mn08}. The research discussed
in this part and in \cite{mnw09,sw10} has added to that. There have of course been
many other studies of dynamical fluctuation theory in the
literature. 

We mention in particular the works of Derrida and
Bodineau, see \cite{bd04,der07} and of Bertini {\it et al}, e.g. in
\cite{bdgjl01,bdgjl02}.  With respect to diffusion
processes, our approach is especially similar to what one is doing
for a macrostatistical theory where a hydrodynamic limit is taken by
rescaling of certain Markov jump processes. The hydrodynamic fluctuations
can then be viewed as a solution of some infinite-dimensional diffusion
process.  Possible differences with the existing work are first of
all that the problems related to the diffusion-approximation or to
a hydrodynamic rescaling do not enter in our work.  We just start from a
finite-dimensional diffusion process or a Markov jump process as such,
and without extra rescaling.  In other words, we prefer to split the problem of
hydrodynamical scaling with possible diffusion approximation from
the problem of studying the dynamical fluctuations.  

The emphasis of our work is on the distinction between the time-symmetric
and time-antisymmetric sector of the fluctuations. On the one hand the
action of relative path-probability densities is split in entropy flux
and traffic. On the other hand the observables are split into occupation
density (time-symmetric) and current (time-antisymmetric). It then comes as no surprise
that the joint fluctuation functional $I(\mu,j)$ can be written as an excess traffic, 
 and an excess entropy flux, which constitutes a first main conclusion, expressed in (\ref{imuj}) and (\ref{imujmj1}). 
The excess is an excess of the original process with respect to
a modified process (by adding a force to the dynamics) in which the fluctuations are typical.
Contractions to occupations or currents alone is hard to do.

Our second main result covers occupation fluctuations, see (\ref{Imu-main}) and (\ref{Imu-main2}). 
There one can see that the fluctuation functional can be expressed
in terms of only an excess traffic, and the modified dynamics is made by adding a potential.
Furthermore, we like to emphasize a structure in the joint fluctuations of density and
current which reminds of thermodynamic potentials, which can be found in Sections \ref{sec-thpot} and \ref{sec-thpot2}.

Finally, for small fluctuations one can clearly see that the coupling between occupation and current
fluctuations only vanishes when close to equilibrium. In that regime minimum and maximum
entropy production principles emerge.

In \cite{mnw09} we have investigated dynamical fluctuations for semi-Markov
jump processes. These are an extension of Markov jump processes where the waiting
times are not exponentially distributed. In \cite{sw10} we extended the analysis
of overdamped diffusion processes to time-dependent but periodic dynamics.
In those cases the general conclusions we have outlined do remain true.

However, it is clear that the story as presented in this part is far from finished.
First of all, we have not provided immediately applicable results. 
We have rather begun to examine the general physical structure of dynamical
fluctuations. To go on, other models should be studied to see which structures
remain in more general settings. For example, one could see what happens for
underdamped diffusions, as we did for fluctuation-dissipation relations.
Large deviations in this setting have already been discussed in \cite{bl08},
where a quantity similar to our traffic plays a major role.

\cleardoublepage
\part{Conclusions and appendices}

\vspace*{6cm}

``While working on my note, I realized that I had met these guys (=traffic) before, 
when I had been trying to prove some of my conjectures on the driven lattice gas.  
Those conjectures came mainly from my `intuition' that was unfortunately based on 
the equilibrium physics, and turned out to be wrong.  Always it looked like the `proof' 
went OK, but at some stage I saw something went wrong because of unwanted factors coming 
from waiting times.  That was traffic, and (in those days) I only regarded it as annoying noise.  
But your message probably is to make friends with them.  I will try.''

\begin{flushright}
 Hal Tasaki, Department of Physics, Gakushuin University, 171-8588, JAPAN, in a recent e-mail conversation (September 2009)
\end{flushright}

\cleardoublepage
\chapter{Overall conclusions}

\textit{This chapter is devoted to some general conclusions of this text.
For conclusions about fluctuation-dissipation relations and dynamical
fluctuations separately, we refer to Chapters \ref{fdrcon} and \ref{dfcon}.
Basically we have used the same method throughout this text: 
on the level of stochastic trajectories we distinguish entropy and traffic,
that completely specify the probabilities of those trajectories. Quantities
and physical relations that are computed through averages over possible trajectories
are thus governed by both entropy and traffic. Entropy is a known quantity: it has an 
operational definition. In this chapter we shortly discuss the status of the less-known
traffic and its possible interpretations.}

\section{Traffic for Markov jump processes and overdamped diffusions}

We have investigated both fluctuation dissipation-relations
and dynamical fluctuations for Markov jump processes and overdamped diffusions.
This warrants a comparison. Generally we have used the same strategy in both 
parts, namely using a description in terms of stochastic trajectories and more
importantly the splitting of the action in a time-symmetric and time-antisymmetric
part.

The time-antisymmetric part of the action is, under the local detailed balance assumption,
equal to the entropy flux into the environment. The restrictions under which this works is
that the parts of the environment connected to the system, each are and remain in equilibrium.
This entropy flux can therefore be expressed in terms of the heat dissipation into that environment.

The time-symmetric part of the action is called traffic. It is much more difficult to give this quantity
a general physical meaning than for entropy flux. One reason for that, is that the concept
only makes sense as an excess. This excess is a difference of the same quantity in two dynamics.
Our results for Markov jump processes and overdamped diffusions do give us a tool
to give more interpretation to traffic.

\subsection{Traffic as an escape rate}

We have seen that the rate function for the large deviations of the empirical
occupation density $I(\mu)$ can be expressed as an excess in traffic (a traffic rate actually), see (\ref{Imu-main}) and (\ref{Imu-main2}).

How can we interpret this rate function? For this, first consider the set of distributions
$\mu$ that vanish on a given subset $A$ of the state space $\Omega$. The contraction principle
then tells us that the infimum of $I(\mu)$ over this set gives the probability of the particle not visiting this $A$.
This infimum can thus be considered an escape rate from the subset $A$. The rate function $I(\mu)$, and thus the excess traffic
can then be called a generalized escape rate from the distribution $\mu$.

Another way of seeing this is by reversing our viewpoint. The original viewpoint is that,
given a distribution $\mu$ we look for a potential $V$ that makes it typical.
Then we compute the excess traffic of the original with respect to this new dynamics with the added potential.
The reverse viewpoint is that we start from some dynamics with a stationary density $\mu$. We then add
to the dynamics a potential $-V$. The change in the instantaneous traffic rate then gives the rate function
of $\mu$ in that new dynamics, in which it is no longer stationary. The bigger the rate function (excess traffic)
the faster $\mu$ relaxes to the new stationary distribution, thus `escaping from $\mu$.'
An interesting observation follows from this reverse viewpoint: when a potential is added to a stationary system,
the positivity of the rate function tells us that the traffic should increase. In what follows,
we interpret traffic as activity, meaning that adding a potential increases the activity of the system.

Also on the level of trajectories we can see this. For Markov jump processes we wrote (\ref{trafficldmj}),
which is a time-integral of the escape rates of the states the system has visited.
Moreover, overdamped diffusions
can be recovered from Markov jump processes when space and time are rescaled in a specific way
(see Appendix \ref{app-md}). Therefore we still interpret the traffic in terms of
`excess escape rates' even on the level of paths for overdamped diffusions.

\subsection{Traffic as activity}

The excess traffic appearing in the rate function for the large deviations of the empirical
occupation density $I(\mu)$ is, for Markov jump processes, equal to
\[ I(\mu) = \tau(\mu) - \tau_V(\mu) = \sum_{x,y}\mu(x)k(x,y) - \sum_{x,y}\mu(x)k(x,y)e^{\frac{\beta}{2}[V(y)-V(x)]} \]
This is an excess in the expected number of jumps per unit of time, when in the distribution $\mu$.
In this sense, the traffic can be called an activity. Again, because overdamped diffusions
are recovered as a scaling limit of Markov jump processes, we can still see the excess traffic as an excess activity
for overdamped diffusion processes.

When the fluctuation is small, i.e. $\mu$ is close to the stationary distribution,
the potential $V$ that makes $\mu$ typical, is also small. if we write $\mu = \rho[1+\epsilon\mu_1]$
and $V \to \epsilon V$, we get:
\begin{eqnarray}
  I(\mu) = \tau(\mu) - \tau_V(\mu) &=& -\frac{\beta\epsilon^2}{2}\sum_{x,y}\rho(x)\mu_1(x)k(x,y)[V(y)-V(x)]\\
 &=& -\frac{\beta\epsilon^2}{2}\sum_{x}\rho(x)\mu_1(x)LV(x)
\end{eqnarray}
and similarly for overdamped diffusions. Indeed, we recognize here the version of traffic $\tau(x) = \beta LV(x)$
that appeared in the fluctuation-dissipation relations (\ref{trafgen}).

\subsection{Unpopular traffic}

When comparing the fluctuation-dissipation relation out of equilibrium, but for one heat bath
 (\ref{response2}) with the equilibrium one (\ref{fdt3}), we see that the equilibrium fluctuation-dissipation
theorem can be rewritten:
\[ R_{QV}(t,s) = \beta\frac{\partial}{\partial s}\left<V(x_s)Q(x_t)\right>_{\mu_0}^0 = -\left<\tau(\omega,s)Q(x_t)\right>_{\mu_0}^0 \]
The first equality is the usual one. In our framework we see that it is the correlation between the observable
and the excess entropy flux of the perturbed with respect to the unperturbed process. 
The last equality states that we can also consider the correlation with the excess traffic. In equilibrium these are the same!

Similarly, for dynamical fluctuations we have seen that close to equilibrium excess entropy production
and excess traffic are the same, see (\ref{mjenttraf}). This may explain why traffic
has not appeared much earlier in statistical physics. Around equilibrium one does not see it as 
an independent new quantity, and all attention goes to its more popular brother entropy.
This explains results as minimum and maximum entropy production principles, (see Sections \ref{sec:EP} and \ref{sec:EPmj}) and the McLennan formula
(Section {\ref{sec:physres}}).

Celebrated recent results concerning a far-from-equilibrium regime, 
like the fluctuation-theorem and work relations (see Section \ref{sec:physres}),
only make statements about the time-antisymmetric part of the dynamics. Because of that
entropy is sufficient to describe the results, avoiding traffic completely.

It seems however, that for a full description of nonequilibrium systems, the time-symmetric
sector of the process described by the traffic is indispensable. 
This constitutes a major message of this text: traffic should not be avoided,
it should be investigated.

\section{Outlook}

\paragraph{Fluctuation-dissipation relations}
In part II of this text we have outlined the general strategy to
derive fluctuation-dissipation relations out of equilibrium, as well
as given several explicit examples, ready to be tested experimentally.
But of course the study is not finished. Apart from applying the strategy to
other models, several related problems seem very interesting and can be investigated in the
same framework. We suggest some examples:
\begin{itemize}
 \item The response of a system with respect to a perturbation other than a potential.
This could be for example an added nonconservative force, or a change in the chemical potential
of one of the particle reservoirs. This should be perfectly doable within the framework
outline in this text.
 \item Changes in temperature of the heat baths as perturbations. This is a very interesting topic for
applications in meteorology but also in nonequilibrium calorimetry. However, such
a perturbation can be tricky in our framework: for example for diffusion processes: if the
diffusion coefficient (which depends usually on the temperature) is perturbed, then the
path-probability densities of the perturbed and unperturbed process are no longer absolutely
continuous with respect to each other.
 \item A step towards thermodynamics through quasistatic processes: suppose that amplitude $h_t$
of the perturbation potential is not small, but changes very slowly. So slowly that the system
is always very close to the stationary distribution of the dynamics at that moment. In such
cases the response relations we have derived can be used to write expressions for the excess
heat, work and change of energy. The second law of thermodynamics for example can be investigated
in such processes.
 \item The use of our results in models of spin-glasses, where the notion of effective temperature 
often is useful. As our approach gives the response explicitly in terms of correlation functions
for Markov jump processes, which often model spin-glasses, it could be used to develop the knowledge
of effective temperatures. 
\end{itemize}

\paragraph{Dynamical fluctuations}
In part III we have discussed dynamical fluctuations
of occupations and currents in overdamped diffusions and Markov jump processes. 
Although some quite general conclusions were made in terms of entropy and traffic,
and this theory adds physical interpretation to traffic as escape rate and as a
`thermodynamic' potential, the story is far from finished.
Some suggestions for future work:
\begin{itemize}
 \item An investigation of the connection between dynamical and static
fluctuations, how one theory can be derived from the other, can give a more
systematic understanding of large deviations in nonequilibrium statistical mechanics.
 \item Dynamical fluctuations in underdamped diffusion processes can give more insight
into the concept of traffic in these models. As is clear from Markov jump processes and overdamped
diffusions, dynamical fluctuation theory is of great help here.
 \item It would be good to have some explicit results, i.e. some explicit rate functions,
which moreover relate to experimental accessible predictions. Simulations could also
help in visualizing the rate functions, at least for simple systems. 
\end{itemize}

\appendix
\cleardoublepage

\cleardoublepage
\chapter{The generator for Markov processes}\label{chap-gen}

\section{Different generators}

\paragraph{Backward generator}
When the system evolves according to a Markov dynamics, there exists an operator $L$ acting on state functions $f$, such that
\begin{equation}\label{backward}
 \frac{d}{d t}\left<f(x_t)\right>_{\mu_0} = \left<Lf(x_t)\right>_{\mu_0} \ \ \ \ \ \ \forall f
\end{equation}
This is a differential equation with the following solution
\[ \left<f(x_t)\right>_{\mu_0} = \int dx \mu_0(x)(e^{tL}f)(x)\]
where the integral should be replaced by a sum when the configuration space is discrete. 
In words, $e^{tL}$ ``pulls a function f back to the time of the initial measure.'' $L$ is therefore often called the backward generator.\\

\paragraph{Forward generator}
There also exists a forward generator $L^{\dag}$, defined in the following way: for any pair of state functions $f,g$
\begin{equation}\label{backfor}
  \int dx g(x)Lf(x) = \int dx f(x)L^{\dag}g(x)
\end{equation}
If we now consider the definition of the time-evolved distribution $\mu_t$:
\[ \int dx \mu_0(x)(e^{tL}f)(x) = \left<f(x_t)\right>_{\mu_0} = \int dx \mu_t(x)f(x) \]
we see that $\mu_t = e^{tL^{\dag}}\mu_0$. So $L^{\dag}$ pushes the distribution $\mu_0$ forward in time,
and is therefore called the forward generator. The forward generator can be found from
the Master equation for Markov jump processes or the Fokker-Planck equation for diffusions, because
\begin{equation}\label{foreward}
  \frac{\partial \mu_t}{\partial t} = L^{\dag}\mu_t
\end{equation}

\paragraph{Adjoint generator}

For time-independent dynamics, the adjoint generator $L^*$ is defined with the help of the stationary distribution $\rho$: 
for any pair of state functions $f,g$
\begin{equation}\label{adjoint}
 \int dx \rho(x)g(x)L^*f(x) = \int dx \rho(x)f(x)Lg(x)
\end{equation}
It is very important to realize that this adjoint generator depends on the stationary distribution, 
and is therefore often not explicitly known in nonequilibrium systems. 
This adjoint generator is related to the time-reversed dynamics.\\

To apply time-reversal, we must remember that there are possibly variables
that change sign under time-reversal, i.e. velocities. 
To simplify the analysis however, we restrict ourselves here
to the case that we have only time-symmetric variables (Markov jump processes and overdamped diffusions).

The stationary distribution for the time-reversed dynamics (with path-probability density $\mathcal{P}(\theta\omega)$) 
is also $\rho$, and for any $f,g$:
\begin{eqnarray*}
 \left<g(x_0)f(x_t)\right>_{\rho} &=& \int d\mathcal{P}_{\rho}(\omega)g(x_0)f(x_t)\\
 &=&\int d\mathcal{P}_{\rho}(\theta\omega)g(x_t)f(x_0)\\
 &=&\left<g(x_t) f(x_0)\right>^{\theta}_{\rho}
 \end{eqnarray*}
where the $\theta$ denotes that the average is taken in the time-reversed dynamics. 
Writing the generator of the time-reversed dynamics as $L^{\theta}$, we get
\[ \int dx\rho(x)g(x)e^{tL}f(x) = \int dx \rho(x)f(x)e^{tL^{\theta}}g(x) \]
Because of the definition (\ref{adjoint}), we see that $L^{\theta} = L^*$.
More generally, one can derive that $L^{\theta} = \pi L^*\pi$ in the case that the
configuration of the system contains velocities.
In other words, up to some reversal of signs of velocities, the adjoint generator is the generator of 
the time-reversed dynamics. This also means that an equilibrium dynamics must satisfy $L^* = \pi L\pi$.

Here we also see that the equilibrium distribution can be defined as the distribution that satisfies
\begin{equation}\label{eqadj}
 \int dx \rho(x)f(x)Lg(x) = \int dx\rho(x)g(x)\pi L\pi f(x)\ \ \ \ \ \forall f,g
\end{equation}
 
\section{Correlation functions}
With the definitions of generators, one can write correlation functions in several different ways as follows, for $s\leq t$
\begin{eqnarray*}
 \left<f(x_t)g(x_s)\right>_{\mu_0} &=& \int dx \mu_0(x)e^{sL}(ge^{(t-s)L}f)(x)dx\\
&=& \int dx g(x)e^{(t-s)L}f(x)e^{sL^{\dag}}\mu_0(x)dx\\
&=& \int dx \mu_s(x)g(x)e^{(t-s)L}f(x)dx\\
&=& \left<f(x_{t-s})g(x_0)\right>_{\mu_s}
\end{eqnarray*}
One can see from this, that the time derivative of such a correlation function can be quite complicated. For example, for $s<t$:
\begin{eqnarray}
 \frac{d}{d s}\left<f(x_t)g(x_s)\right>_{\mu_0} &=& - \left<Lf(x_t)g(x_s)\right>_{\mu_0} + \left<f(x_t)\frac{g(x_s)}{\mu_s(x_s)}
\frac{\partial \mu_s}{\partial s}(x_s)\right>_{\mu_0}\nonumber\\
&=& - \left<f(x_t)\frac{L^{\dag}(\mu_sg)}{\mu_s}(x_s)\right>_{\mu_0} + \left<f(x_t)\frac{g(x_s)}{\mu_s(x_s)}L^{\dag}\mu_s(x_s)\right>_{\mu_0}\nonumber\\
\frac{d}{d t}\left<f(x_t)g(x_s)\right>_{\mu_0} &=&  \left<Lf(x_t)g(x_s)\right>_{\mu_0}\label{corrder}
\end{eqnarray}
However, we see that in stationary averages, this simplifies to
\[ \frac{d}{d s}\left<f(x_t)g(x_s)\right>_{\rho} = - \left<Lf(x_t)g(x_s)\right>_{\rho} = - \frac{d}{d t}\left<f(x_t)g(x_s)\right>_{\rho}\]
When the generator occurs in a stationary correlation function, one can replace it by the adjoint generator in the following case, for $s\le t$:
\begin{eqnarray*}
 \left<Lf(x_t)g(x_s)\right>_{\rho} &=& \int dx \rho(x)g(x)e^{(t-s)L}Lf(x)dx\\
&=&  \int dx \rho(x)g(x)Le^{(t-s)L}f(x)dx\\
&=& \int dx \rho(x)L^*g(x)e^{(t-s)L}f(x)dx\\
&=& \left<f(x_t)L^*g(x_s)\right>_{\rho}
\end{eqnarray*}
where we used in the second equality that $L$ commutes with $e^{(t-s)L}$.
However, we can not apply a similar procedure for $\left<f(x_t)Lg(x_s)\right>_{\rho}$, because $L$ and $L^*$  do not commute in general.

\section{Generators for jump processes and diffusions}

As additional information, we give here the generators for the main models used in this
thesis. The procedure to find them is straightforward. We start from the evolution
equation for the distribution $\mu_t$, which is the Master equation for Markov jump processes
and the Fokker-Planck equation for diffusions, to find the forward generator $L^{\dag}$:
\[\frac{\partial \mu_t}{\partial t} = L^{\dag}\mu_t\]
After that we use (\ref{backfor}) to find the backward generator, and (\ref{adjoint}) to find the
adjoint generator.

\subsection{Markov jump processes}
The generators for a general Markov jump process with transition rates $k(x,y)$ are:
\begin{eqnarray}
 L^{\dag}\mu(x) &=& \sum_y[\mu(y)k(y,x) - \mu(x)k(x,y)] \label{fgmj}  \\ 
 Lf(x) &=& \sum_yk(x,y)[f(y)-f(x)]       \label{bgmj}    \\
 L^*f(x) &=&     \sum_y \frac{\rho(y)k(y,x)}{\rho(x)}[f(y)-f(x)]    \label{agmj}
\end{eqnarray}

\subsection{Overdamped diffusions in more dimensions}
The generators for a general overdamped Langevin dynamics (\ref{langgen}) are:
\begin{eqnarray}
 L^{\dag}\mu &=& -\nabla\cdot[\chi F\mu - D\nabla\mu]  \label{fgod} \\ 
 Lf &=&  F\chi\nabla f + \nabla\cdot (D\nabla f)   \label{bgod}       \\
 L^*f &=& -F\chi\nabla f + \nabla\cdot (D\nabla f) + 2 \frac{\nabla \rho}{\rho} D \nabla f      \label{agod}
\end{eqnarray}

\subsection{Underdamped diffusions}
The generators for underdamped diffusions of the form (\ref{ud}) are:

\begin{eqnarray}
 L^{\dag}\mu &=& \sum_i\left[ -v_i\frac{\partial \mu}{\partial x_i} - \frac{\partial }{\partial v_i}
[(f_i-\frac{\partial U}{\partial x_i}-m_i\gamma_iv_i)\mu] + D_i\frac{\partial^2 \mu}{\partial v_i^2} \right]   \\ 
 L g &=&   \sum_i\left[ v_i\frac{\partial g}{\partial x_i} + 
(f_i-\frac{\partial U}{\partial x_i}-m_i\gamma_iv_i)\frac{\partial g}{\partial v_i} + D_i\frac{\partial^2 g}{\partial v_i^2} \right]          \\
 L^*g &=& \sum_i\left[ -v_i\frac{\partial g}{\partial x_i} - 
(f_i-\frac{\partial U}{\partial x_i}-m_i\gamma_iv_i)\frac{\partial g}{\partial v_i} + D_i\frac{\partial^2 g}{\partial v_i^2} 
+ 2D_i\frac{1}{\rho}\frac{\partial \rho}{\partial v_i}\frac{\partial g}{\partial v_i} \right]       \nonumber 
\end{eqnarray}

\cleardoublepage
\chapter{A derivation of the Fokker-Planck equation}\label{app-c}

Starting from a Langevin equation it is possible to derive the corresponding Fokker-Planck
equation. We show it here for an overdamped Langevin equation in one dimension (\ref{langover}):
\[ dx_t = \chi F(x_t)dt + \sqrt{2 D}dB_t \]
We do this by deriving the generator $L$ of this process. For this, take an arbitrary
state function $g$, which is smooth. Take a small number $\epsilon\in\mathbb{R}$. We consider the average
\[ \left<g(x_{t+\epsilon})\right>_{\mu_0} = \int_{\mathbb{R}} dx \mu_t(x)\int_{\mathbb{R}} dy Prob(x_{t+\epsilon} = x+y|x_t=x)g(x+y) \]
where, for $\epsilon$ small enough we can write
\[ Prob(x_{t+\epsilon} = x+y|x_t=x) = \frac{1}{N}\exp\left\{-\frac{1}{4D\epsilon}[y-\chi F(x)\epsilon]^2\right\} \]
with the normalization $N = \sqrt{4\pi D\epsilon}$. We make a Taylor-expansion of $g$ around $x$:
\newpage
\begin{eqnarray*}
&& \int_{\mathbb{R}} dy Prob(x_{t+\epsilon} = x+y|x_t=x)g(x+y)\\
&& = \sum_{n=0}^{\infty}\frac{1}{n!}
\frac{\partial^n g}{\partial x^n}(x)\int_{\mathbb{R}} dy \frac{1}{N}e^{-\frac{1}{4D\epsilon}[y-\chi F(x)\epsilon]^2}y^n
\end{eqnarray*}
In this way we have reduced the problem to calculating Gaussian integrals. We find
\begin{eqnarray*}
  \int_{\mathbb{R}} dy Prob(x_{t+\epsilon} = x+y|x_t=x)g(x+y) &=& g(x) +\epsilon \chi F(x)\frac{\partial g}{\partial x}(x)\\
&&  + 
\epsilon D \frac{\partial^2 g}{\partial x^2}(x)+ o(\epsilon)
\end{eqnarray*}
So that in the limit of $\epsilon\to 0$ we get
\begin{eqnarray*}
  \left<Lg(x_t)\right>_{\mu_0} &=& \lim_{\epsilon\to 0}\frac{\left<g(x_{t+\epsilon})\right>_{\mu_0}-\left<g(x_{t})\right>_{\mu_0}}{\epsilon}\\
&=& \int_{\mathbb{R}} dx \mu_t(x)[\chi F(x)\frac{\partial g}{\partial x}(x) + 
 D \frac{\partial^2 g}{\partial x^2}(x)]
\end{eqnarray*}
This gives us the backward generator $L$. The forward generator $L^{\dag}$ then defines the Fokker-Planck
equation. It is easily found by (\ref{backfor}).

The generators and Fokker-Planck equations for other and more general diffusion equations can be found by the same reasoning.

\chapter{From Markov jump processes to overdamped diffusions}\label{app-md}

It is possible to derive overdamped diffusions as a certain limit of Markov jump processes. 
As an example, we consider the case of a single particle performing a Markov jump process on a ring with $N$ sites,
labelled $x=0,\epsilon,2\epsilon,\ldots,N\epsilon$ with $N\epsilon =1 \equiv 0$.
The particle can hop every time only one site to the left or to the right, with rates $k(x,x\pm\epsilon)$. 
A Markov process is fully determined by its generator. We therefore prove that the generator
of this jump process converges in a certain scaling for $\epsilon\to 0$ to the generator of a diffusion process.
The generator of the Markov jump process on a state function $g$ gives:
\begin{eqnarray*}
 L_{\epsilon}g(x)  &=& \sum_{y=x\pm \epsilon}k(x,x \pm \epsilon)[g(x\pm\epsilon) - g(x)]\\
 &=& \frac{1}{2}[k(x,x+\epsilon) + k(x,x-\epsilon)][g(x+\epsilon)+ g(x-\epsilon)-2g(x)]\\
&& + \frac{1}{2}[k(x,x+\epsilon) - k(x,x-\epsilon)][g(x+\epsilon)- g(x-\epsilon)]
\end{eqnarray*}
We expand the transition rates in orders of $\epsilon$ and rescale them as follows with $\epsilon\to 0$:
\[ k(x,x + \epsilon) = \frac{1}{\epsilon^2}[k_0(x) + \epsilon k_1(x) + o(\epsilon)] \]
for the rates for hopping to the right.
This corresponds to a scaling of space (the space between two sites is $1/\epsilon$) 
and of time (the escape rates of the process are multiplied by $1/\epsilon^2$ so that the
process goes faster and faster).

Furthermore, the local detailed balance condition tells us that
\[ \frac{k(x,x+\epsilon)}{k(x+\epsilon,x)} = \exp\left\{ \beta[W(x,x+\epsilon)+U(x)-U(x+\epsilon)] \right\} \]
where $W(x,x+\epsilon)$ is the work of the nonconservative forcing when the particle travels a distance
$\epsilon$ to the right. We write it therefore as $W(x,x+\epsilon) = \epsilon f(x) + o(\epsilon)$.
This allows us to express the rates for hopping to the left as:
\begin{eqnarray*}
  k(x+\epsilon,x) &=& k(x,x+\epsilon)\exp\left\{ -\beta\epsilon f(x) + \beta\epsilon U'(x) + o(\epsilon) \right\} \\
&=& \frac{1}{\epsilon^2}\left[k_0(x) + \epsilon [k_1(x)- \beta k_0(x)f(x) + \beta k_0(x)U'(x)] + o(\epsilon)\right]
\end{eqnarray*}
For our calculation, we need $k(x,x-\epsilon)$, which is thus equal to
\[ \frac{1}{\epsilon^2}\left[k_0(x) + \epsilon [k_1(x)-k'_0(x)- \beta k_0(x)f(x) + \beta k_0(x)U'(x)] + o(\epsilon)\right] \]
Substituting this into the generator, we get the following limit:
\[ \lim_{\epsilon\to 0} L_{\epsilon}g(x) = k_0(x)g''(x) +[\beta k_0(x)f(x) - \beta k_0(x)U'(x) + k_0'(x)]g'(x) \]
which exactly defines the generator for a diffusion process with the Langevin equation
\[ dx_t = \chi(x)[f(x_t)-U'(x_t)]dt + D'(x_t)dt + \sqrt{2D(x_t)}dB_t \]
with $D = k_0$ and $\chi = \beta k_0$. We therefore see the forcing with $D'$ naturally emerging.
Moreover, the local detailed balance condition for Markov jump processes, has the relation $\chi = \beta D$
as a consequence after the rescaling.

\cleardoublepage
\phantomsection
\markboth{{Bibliography}}{{Bibliography}}
\addcontentsline{toc}{chapter}{Bibliography}
\bibliographystyle{plain}
\bibliography{bibliografie}

\begin{thebibliography}{100}

\bibitem{aga72}
G.~S. Agarwal.
\newblock Fluctuation-dissipation theorems for systems in non-thermal
  equilibrium and applications.
\newblock {\em Z. Physik}, 252:25--38, 1972.

\bibitem{bbmw09}
M.~Baiesi, E.~Boksenbojm, C.~Maes, and B.~Wynants.
\newblock Nonequilibrium linear response for {M}arkov dynamics, {II}: Inertial
  dynamics.
\newblock {\em J. Stat. Phys.}, 139:492--505, 2010.

\bibitem{bmw09a}
M.~Baiesi, C.~Maes, and B.~Wynants.
\newblock Fluctuations and response of nonequilibrium states.
\newblock {\em Phys. Rev. Lett.}, 103:010602, 2009.

\bibitem{bmw09b}
M.~Baiesi, C.~Maes, and B.~Wynants.
\newblock Nonequilibrium linear response for {M}arkov dynamics, {I}: Jump
  processes and overdamped diffusions.
\newblock {\em J. Stat. Phys.}, 137(5-6):1094--1116, 2009.

\bibitem{bdgjl01}
L.~Bertini, A.~{De Sole}, D.~Gabrielli, G.~Jona-Lasinio, and C.~Landim.
\newblock Fluctuations in stationary nonequilibrium states of irreversible
  processes.
\newblock {\em Phys. Rev. Lett.}, 87:040601, 2001.

\bibitem{bdgjl02}
L.~Bertini, A.~{De Sole}, D.~Gabrielli, G.~Jona-Lasinio, and C.~Landim.
\newblock Macroscopic fluctuation theory for stationary non equilibrium states.
\newblock {\em J. Stat. Phys.}, 107:635--675, 2002.

\bibitem{bdgjl05}
L.~Bertini, A.~{De Sole}, D.~Gabrielli, G.~Jona-Lasinio, and C.~Landim.
\newblock Current fluctuations in stochastic lattice gases.
\newblock {\em Phys. Rev. Lett.}, 94:030601, 2005.

\bibitem{bslsb07}
W.~Blickle, T.~Speck, C.~Lutz, U.~Seifert, and C.~Bechinger.
\newblock Einstein relation generalized to nonequilibrium.
\newblock {\em Phys. Rev. Lett.}, 98:210601, 2007.

\bibitem{bd04}
T.~Bodineau and B.~Derrida.
\newblock Current fluctuations in non-equilibrium diffusive systems: an
  additivity principle.
\newblock {\em Phys. Rev. Lett.}, 92:180601, 2004.

\bibitem{bl08}
T.~Bodineau and R.~Lefevere.
\newblock Large deviations of lattice hamiltonian dynamics coupled to
  stochastic thermostats.
\newblock {\em J. Stat. Phys.}, 133(1):1--27, 2008.

\bibitem{bmn07}
S.~Bruers, C.~Maes, and K.~Neto\v{c}n{\'y}.
\newblock On the validity of entropy production principles for linear
  electrical circuits.
\newblock {\em J. Stat. Phys.}, 129(4):725--740, 2007.

\bibitem{cg05}
P.~Calabrese and A.~Gambassi.
\newblock Ageing properties of critical systems.
\newblock {\em J. Phys. A: Math. Gen.}, 38:R133--R193, 2005.

\bibitem{cw51}
H.~B. Callen and T.~A. Welton.
\newblock Irreversibility and generalized noise.
\newblock {\em Phys. Rev.}, 83:34--40, 1951.

\bibitem{cha03}
C.~Chatelain.
\newblock A far-from-equilibrium fluctuation-dissipation relation for an
  {I}sing-{G}lauber-like model.
\newblock {\em J. Phys. A}, 36:10739, 2003.

\bibitem{cfg08}
R.~Chetrite, G.~Falkovich, and K.~Gaw\c{e}dzki.
\newblock Fluctuation relations in simple examples of non-equilibrium steady
  states.
\newblock {\em J. Stat. Mech.}, page P08005, 2008.

\bibitem{cg09}
R.~Chetrite and K.~Gaw\c{e}dzki.
\newblock Eulerian and lagrangian pictures of non-equilibrium diffusions.
\newblock {\em J. Stat. Phys.}, 137:890--916, 2009.

\bibitem{cr03}
A.~Chrisanti and J.~Ritort.
\newblock Violation of the fluctuation-dissipation theorem in glassy systems:
  basic notions and the numerical evidence.
\newblock {\em J. Phys. A: Math. Gen.}, 36:R181--R290, 2003.

\bibitem{ckw04}
W.~T. Coffey, Y.~P. Kalmykov, and J.~T. Waldron.
\newblock {\em The Langevin Equation; With Applications to Stochastic Problems
  in Physics, Chemistry and Electrical Engineering}.
\newblock World Scientific Publishing Co., 2nd edition, 2004.

\bibitem{cro98}
G.~E. Crooks.
\newblock Nonequilibrium measurements of free energy differences for
  microscopically reversible {M}arkovian systems.
\newblock {\em J. Stat. Phys.}, 90:1481--1487, 1998.

\bibitem{cro99}
G.~E. Crooks.
\newblock Entropy production fluctuation theorem and the nonequilibrium work
  relation for free energy differences.
\newblock {\em Phys. Rev. E}, 60:2721--2726, 1999.

\bibitem{ckp94}
L.~Cugliandolo, J.~Kurchan, and G.~Parisi.
\newblock Off equilibrium dynamics and aging in unfrustrated systems.
\newblock {\em J. Phys. I}, 4:1641, 1994.

\bibitem{dm07}
W.~{De Roeck} and C.~Maes.
\newblock Symmetries of the ratchet current.
\newblock {\em Phys. Rev. E}, 76:051117, 2007.

\bibitem{dh75}
U.~Deker and F.~Haake.
\newblock Fluctuation-dissipation theorems for classical processes.
\newblock {\em Phys. Rev. A}, 11:2043, 1975.

\bibitem{dz98}
A.~Dembo and O.~Zeitouni.
\newblock {\em Large Deviations Techniques and Applications}.
\newblock Springer-Verlag, New York, 2nd edition, 1998.

\bibitem{der07}
B.~Derrida.
\newblock Non equilibrium steady states: fluctuations and large deviations of
  the density and of the current.
\newblock {\em J. Stat. Mech.}, page P07023, 2007.

\bibitem{die05}
G.~Diezemann.
\newblock Fluctuation-dissipation relations for {M}arkov jump processes.
\newblock {\em Phys. Rev. E}, 72:011104, 2005.

\bibitem{dv75a}
M.~D. Donsker and S.~R.~S. Varadhan.
\newblock Asymptotic evaluation of certain {M}arkov process expectations for
  large time. {I}.
\newblock {\em Comm. Pure Appl. Math.}, 28:1--47, 1975.

\bibitem{dv75b}
M.~D. Donsker and S.~R.~S. Varadhan.
\newblock Asymptotic evaluation of certain {M}arkov process expectations for
  large time. {II}.
\newblock {\em Comm. Pure Appl. Math.}, 28:279--301, 1975.

\bibitem{dv76}
M.~D. Donsker and S.~R.~S. Varadhan.
\newblock Asymptotic evaluation of certain {M}arkov process expectations for
  large time. {III}.
\newblock {\em Comm. Pure Appl. Math.}, 29:389--461, 1976.

\bibitem{dv83}
M.~D. Donsker and S.~R.~S. Varadhan.
\newblock Asymptotic evaluation of certain {M}arkov process expectations for
  large time. {IV}.
\newblock {\em Comm. Pure Appl. Math.}, 36:183--212, 1983.

\bibitem{dor99}
J.~R. Dorfman.
\newblock {\em An Introduction to Chaos in Nonequilibrium Statistical
  Mechanics}.
\newblock Cambridge University Press, 1999.

\bibitem{dg95}
J.~R. Dorfman and P.~Gaspard.
\newblock Chaotic scattering theory of transport and reaction-rate
  coefficients.
\newblock {\em Phys. Rev. E}, 51:28--35, 1995.

\bibitem{ein1905}
A.~Einstein.
\newblock {\"U}ber die von der molekularkinetischen theorie der w{\"a}rme
  geforderte bewegung von in ruhenden fl{\"u}ssigkeiten suspendierten teilchen.
\newblock {\em Annalen der Physik:}, 17, 549--560.

\bibitem{ell84}
R.~S. Ellis.
\newblock Large deviations for a general class of random vectors.
\newblock {\em Ann. Probab.}, 12(1):1--12, 1984.

\bibitem{ell85}
R.~S. Ellis.
\newblock {\em Entropy, Large Deviations and Statistical Mechanics}.
\newblock Springer, New-York, 1985.

\bibitem{ell95}
R.~S. Ellis.
\newblock An overview of the theory of large deviations and applications to
  statistical mechanics.
\newblock {\em Scand. Actuar. J.}, 1:97--142, 1995.

\bibitem{ell99}
R.~S. Ellis.
\newblock The theory of large deviations: from {B}oltzmann{’}s 1877
  calculation to equilibrium macrostates in 2{D} turbulence.
\newblock {\em Physica D}, 133:106--136, 1999.

\bibitem{ecm93}
D.J. Evans, E.~G.~D. Cohen, and G.~P. Morris.
\newblock Probability of second law violations in shearing steady states.
\newblock {\em Phys. Rev. Lett.}, 71:2401--2404, 1993.

\bibitem{es94}
D.J. Evans and D.~J. Searles.
\newblock Equilibrium microstates which generate second law violating steady
  states.
\newblock {\em Phys. Rev. E}, 50:1645--1648, 1994.

\bibitem{fiv90}
M.~Falcioni, S.~Isola, and A.~Vulpiani.
\newblock Correlation functions and relaxation properties in chaotic dynamics
  and statistical mechanics.
\newblock {\em Phys. Lett. A}, 144:341, 1990.

\bibitem{fw84}
M.I. Freidlin and A.~D. Wentzell.
\newblock {\em Random Perturbations of Dynamical Systems}.
\newblock Springer-Verlag, New York, 1984.

\bibitem{gc95}
G.~Gallavotti and E.~G.~D. Cohen.
\newblock Dynamical ensembles in nonequilibrium statistical mechanics.
\newblock {\em J. Stat. Phys.}, 80:2694--2697, 1995.

\bibitem{gar04}
C.~W. Gardiner.
\newblock {\em Handbook of Stochastic Methods for Physics, Chemistry and The
  Natural Sciences}.
\newblock Springer, Berlin, 2004.

\bibitem{gar77}
J.~G{\"a}rtner.
\newblock On large deviations from the invariant measure.
\newblock {\em Theory Probab. Appl.}, 22:24--39, 1977.

\bibitem{gpccg09}
J.~R. Gomez-Solano, A.~Petrosyan, S.~Ciliberto, R.~Chetrite, and
  K.~Gaw\c{e}dzki.
\newblock Experimental verification of a modified fluctuation-dissipation
  relation for a micron-sized particle in a non-equilibrium steady state.
\newblock {\em Phys. Rev. Lett.}, 103:075007, 2009.

\bibitem{gs01}
G.~Grimmett and D.~Stirzaker.
\newblock {\em Probability and Random processes}.
\newblock Oxford University Press, USA, 3rd edition, 2001.

\bibitem{ht75}
P.~H{\"a}nggi and H.~Thomas.
\newblock Linear response and fluctuation theorems for nonstationary stochastic
  processes.
\newblock {\em Z. Physik B}, 22:295--300, 1975.

\bibitem{ht82}
P.~H{\"a}nggi and H.~Thomas.
\newblock Stochastic processes: time-evolution, symmetries and linear response.
\newblock {\em Phys. Rep.}, 88:207--319, 1982.

\bibitem{hs05}
T.~Harada and S.~Y. Sasa.
\newblock Equality connecting energy dissipation with violation of
  fluctuation-response relation.
\newblock {\em Phys. Rev. Lett.}, 95:130602, 2005.

\bibitem{hs01}
T.~Hatano and S.~Y. Sasa.
\newblock Steady-state thermodynamics of {L}angevin systems.
\newblock {\em Phys. Rev. Lett.}, 86:3463, 2001.

\bibitem{hp09}
M.~Henkel and M.~Pleimling.
\newblock {\em Non-Equilibrium Phase Transitions: Volume 2: Dynamical Scaling
  far from Equilibrium}.
\newblock Canopus Academic Publishing Limited, 2009.

\bibitem{jar97}
C.~Jarzynski.
\newblock Nonequilibrium equality for free energy differences.
\newblock {\em Phys. Rev. Lett.}, 78:2690--2693, 1997.

\bibitem{jar99}
C.~Jarzynski.
\newblock Microscopic analysis of {C}lausius-{D}uhem processes.
\newblock {\em J. Stat. Phys.}, 96(1-2):415--427, 1999.

\bibitem{jar00}
C.~Jarzynski.
\newblock Hamiltonian derivation of a detailed fluctuation theorem.
\newblock {\em J. Stat. Phys.}, 98(1-2):77--102, 2000.

\bibitem{john28}
J.~Johnson.
\newblock Thermal agitation of electricity in conductors.
\newblock {\em Phys. Rev.}, 32:97--109, 1928.

\bibitem{kls83}
S.~Katz, J.~L. Lebowitz, and H.~Spohn.
\newblock Phase transitions in stationary nonequilibrium states of model
  lattice systems.
\newblock {\em Phys. Rev. B}, 28(3):1655--1658, 1983.

\bibitem{kl99}
C.~Kipnis and C.~Landim.
\newblock {\em Scaling Limits of Interacting Particle Systems}.
\newblock Springer-Verlag, Berlin, 1999.

\bibitem{kry95}
N.~V. Krylov.
\newblock {\em Introduction to the Theory of Diffusion Processes}.
\newblock American Mathematical Society, 1995.

\bibitem{kubo66}
R.~Kubo.
\newblock The fluctuation-dissipation theorem.
\newblock {\em Rep. Prog. Phys.}, 29:255--284, 1966.

\bibitem{kur05}
J.~Kurchan.
\newblock In and out of equilibrium.
\newblock {\em Nature}, 433:222, 2005.

\bibitem{lan73}
O.~E. Lanford.
\newblock {\em Entropy and Equilibrium States in Classical Statistical
  Mechanics}, pages 1--113.
\newblock Springer, Berlin, 1973.
\newblock in: Statistical Mechanics and Mathematical Problems, in: Lecture
  notes in Physics.

\bibitem{leb59}
J.~L. Lebowitz.
\newblock Stationary nonequilibrium {G}ibbsian ensembles.
\newblock {\em Phys. Rev.}, 114(5):1192--1202, 1959.

\bibitem{lcsz08}
E.~Lippiello, F.~Corberi, A.~Saraccino, and M.~Zanetti.
\newblock Nonlinear response and fluctuation dissipation relations.
\newblock {\em Phys. Rev. E}, 78:041120, 2008.

\bibitem{lcz05}
E.~Lippiello, F.~Corberi, and M.~Zanetti.
\newblock Off-equilibrium generalization of the fluctuation dissipation theorem
  for ising spins and measurement of the linear response.
\newblock {\em Phys. Rev. E}, 71:036104, 2005.

\bibitem{ls78}
R.~S. Lipster and A.~N. Shiryayev.
\newblock {\em Statistics of Random Processes}, volume I, II.
\newblock Springer-Verlag, New York, 1978.

\bibitem{maes99}
C.~Maes.
\newblock Fluctuation theorem as a {G}ibbs property.
\newblock {\em J. Stat. Phys.}, 95:367--392, 1999.

\bibitem{mn03}
C.~Maes and K.~Neto\v{c}n{\'y}.
\newblock Time reversal and entropy.
\newblock {\em J. Stat. Phys.}, 110:269--310, 2003.

\bibitem{mn07}
C.~Maes and K.~Neto\v{c}n{\'y}.
\newblock Minimum entropy production principle from a dynamical fluctuation
  law.
\newblock {\em J. Math. Phys.}, 48:053306, 2007.

\bibitem{mn08}
C.~Maes and K.~Neto\v{c}n{\'y}.
\newblock Canonical structure of dynamical fluctuations in mesoscopic
  nonequilibrium steady states.
\newblock {\em Europhys. Lett.}, 82:30003, 2008.

\bibitem{mnw08b}
C.~Maes, K.~Neto\v{c}n{\'y}, and B.~Wynants.
\newblock On and beyond entropy production: the case of {M}arkov jump
  processes.
\newblock {\em Markov Processes Relat. Fields}, 14:445--464, 2008.

\bibitem{mnw08a}
C.~Maes, K.~Neto\v{c}n{\'y}, and B.~Wynants.
\newblock Steady state statistics of driven diffusions.
\newblock {\em Physica A}, 387:2675--2689, 2008.

\bibitem{mnw09}
C.~Maes, K.~Neto\v{c}n{\'y}, and B.~Wynants.
\newblock Dynamical fluctuations for semi-{M}arkov processes.
\newblock {\em J. Phys. A: Math. Theor.}, 42:365002, 2009.

\bibitem{mrv00}
C.~Maes, F.~Redig, and A.~{Van Moffaert}.
\newblock On the definition of entropy production via examples.
\newblock {\em J. Math. Phys.}, 41:1528--1554, 2000.

\bibitem{mrv01}
C.~Maes, F.~Redig, and M.~Verschuere.
\newblock From global to local fluctuation theorems.
\newblock {\em Moscow Mathematical Journal}, 1:421--438, 2001.

\bibitem{mw09}
C.~Maes and B.~Wynants.
\newblock On a response formula and its interpretation.
\newblock {\em arXiv:0910.2320}, 2009.

\bibitem{mprv08}
U.~{Marini Bettolo Marconi}, A.~Puglisi, L.~Rondoni, and A.~Vulpiani.
\newblock Fluctuation--dissipation: Response theory in statistical physics.
\newblock {\em Phys. Rep.}, 461(4-6):111--195, 2008.

\bibitem{mbd09}
K.~Martens, E.~Bertin, and M.~Droz.
\newblock Dependence of the fluctuation-dissipation temperature on the choice
  of observable.
\newblock {\em Phys. Rev. Lett.}, 103:260602, 2009.

\bibitem{mar04}
P.~A. Martin.
\newblock Physique statistique des processus irreversibles.
\newblock Lecture notes of the DEA de Physique Theorique, notes by F. Coppex,
  2004.

\bibitem{mcl59}
J.~A. {McLennan Jr.}
\newblock Statistical mechanics of the steady state.
\newblock {\em Phys. Rev.}, 115:1405--1409, 1959.

\bibitem{mcl89}
J.~A. {McLennan Jr.}
\newblock {\em Introduction to Nonequilibrium Statistical Mechanics}.
\newblock Prentice-Hall, En- glewood Cliffs, NJ, 1989.

\bibitem{mor69}
R.~E. Mortensen.
\newblock Mathematical problems of modeling stochastic nonlinear dynamic
  systems.
\newblock {\em J. Stat. Phys.}, 1(2):271--296, 1969.

\bibitem{ns08}
T.~Nakamura and S.~Y. Sasa.
\newblock A fluctuation-response relation of many {B}rownian particles under
  nonequilibrium conditions.
\newblock {\em Phys. Rev. E}, 77:021108, 2008.

\bibitem{niko30}
O.~Nikodym.
\newblock Sur une g{\'e}n{\'e}ralisation des int{\'e}grales de {M}. {J}.
  {R}adon.
\newblock {\em Fundamenta Mathematicae}, 15:131--179, 1930.

\bibitem{nyq28}
H.~Nyquist.
\newblock Thermal agitation of electric charge in conductors.
\newblock {\em Phys. Rev.}, 32:110--113, 1928.

\bibitem{ons31a}
L.~Onsager.
\newblock Reciprocal relations in irreversible processes. {I}.
\newblock {\em Phys. Rev.}, 37:405--426, 1931.

\bibitem{ons31b}
L.~Onsager.
\newblock Reciprocal relations in irreversible processes. {II}.
\newblock {\em Phys. Rev.}, 38:2265--2279, 1931.

\bibitem{om53}
L.~Onsager and S.~Machlup.
\newblock Fluctuations and irreversible processes.
\newblock {\em Phys. Rev.}, 91(6):1505--1512, 1953.

\bibitem{op98}
Y.~Oono and M.~Paniconi.
\newblock Steady state thermodynamics.
\newblock {\em Prog. Theor. Phys. Suppl.}, 130:29, 1998.

\bibitem{pri62}
I.~Prigogine.
\newblock {\em Introduction to Non-Equilibrium Thermodynamics}.
\newblock Wiley-Interscience, New York, 1962.

\bibitem{rh08}
A.~R{\'a}kos and R.~J. Harris.
\newblock On the range of validity of the fluctuation theorem for stochastic
  markovian dynamics.
\newblock {\em J. Stat. Mech.}, page P05005, 2008.

\bibitem{ric03}
F.~Ricci-Tersenghi.
\newblock Measuring the fluctuation-dissipation ratio in glassy systems with no
  perturbing field.
\newblock {\em Phys. Rev. E}, 68:065104(R), 2003.

\bibitem{rit03}
F.~Ritort.
\newblock Universal dependence of the fluctuation-dissipation ratio on the
  transition rates in trap models.
\newblock {\em J. Phys. A: Math. Gen.}, 36:10791--10805, 2003.

\bibitem{rue09}
D.~Ruelle.
\newblock A review of linear response theory for general differential dynamical
  systems.
\newblock {\em Nonlinearity}, 22:855--870, 2009.

\bibitem{ssd82}
J.~M. Sancho, M.~{San Miguel}, and D.~D{\"u}rr.
\newblock Adiabatic elimination for systems of {B}rownian particles with
  nonconstant damping coefficients.
\newblock {\em J. Stat. Phys.}, 28(2):291--305, 1982.

\bibitem{set06}
J.~P. Sethna.
\newblock {\em Entropy, Order Parameters, and Complexity}.
\newblock Oxford Univ. Press, Oxford, 2006.

\bibitem{sw10}
N.~Singh and B.~Wynants.
\newblock Dynamical fluctuations for periodically driven diffusions.
\newblock {\em J. Stat. Mech.}, page P03007, 2010.

\bibitem{ss06}
T.~Speck and U.~Seifert.
\newblock Restoring a fluctuation-dissipation theorem in a nonequilibrium
  steady state.
\newblock {\em Europhys. Lett.}, 74:391--396, 2006.

\bibitem{ss09}
T.~Speck and U.~Seifert.
\newblock Extended fluctuation-dissipation theorem for soft matter in
  stationary flow.
\newblock {\em Phys. Rev. E}, 79:040102, 2009.

\bibitem{tit78}
U.~M. Titulaer.
\newblock A systematic solution procedure for the {F}okker-{P}lanck equation of
  a brownian particle in the high-friction case.
\newblock {\em Physica}, 91A:321--344, 1978.

\bibitem{tou09}
H.~Touchette.
\newblock The large deviation approach to statistical mechanics.
\newblock {\em Phys. Rep.}, 478:1--69, 2009.

\bibitem{vhbwb10}
G.~Volpe, L.~Helden, T.~Brettschneider, J.~Wehr, and C.~Bechinger.
\newblock Influence of noise on force measurements.
\newblock {\em arXiv:1004.0874v1}, 2010.

\bibitem{wei71}
W.~Weidlich.
\newblock Fluctuation-dissipation theorem for a class of stationary open
  systems.
\newblock {\em Z. Physik}, 248:234--243, 1971.

\bibitem{fw70}
A.~D. Wentzell and M.I. Freidlin.
\newblock On small random perturbations of dynamical systems.
\newblock {\em Russ. Math. Surveys}, 25:1--55, 1970.

\bibitem{wil76}
G.~Wilemski.
\newblock In the derivation of {S}moluchowski equations with corrections in the
  classical theory of {B}rownian motion.
\newblock {\em J. Stat. Phys.}, 14(2):153--169, 1976.

\end{thebibliography}

\printindex

%
%
%
%

\end{document}